%% file: MultiJets2013.tex
\documentclass[11pt,a4paper]{article}
\pdfoutput=1
\usepackage{jheppub}
\usepackage{arydshln}

\input{frontmatter}

\input{title}
\input{abstract}

\usepackage{preprintcover} 
\PreprintCoverPaperTitle{\multijettitle}
\PreprintIdNumber{CERN-PH-EP-2013-110}  
\PreprintCoverAbstract{\multijetabstract}
\PreprintJournalName{JHEP}

\begin{document}

\title{\multijettitle}
\author{The ATLAS Collaboration}

\abstract{\multijetabstract}

\collaborationImg{\includegraphics{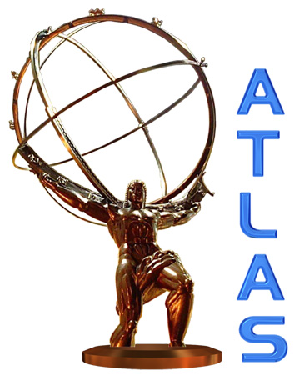}}

\maketitle

\newcommand\hepdatablurb{Acceptance times efficiency values are tabulated elsewhere~\cite{multijet_2013_auxiliary}.}

\newpage

\input{content}

\FloatBarrier

\section*{Acknowledgments}
\input{acknowledgements}


%

\providecommand{\href}[2]{#2}\begingroup\raggedright\endgroup

\FloatBarrier







\input{atlas_authlist}

\end{document}

%% file: title.tex
\newcommand{\ourintlumi}{{\ensuremath{20.3\,\ifb{}}}}
\newcommand{\mgluinoexcluded}{{1.1\,\TeV{}}}
\newcommand{\mlspexcluded}{{350\,\GeV{}}}
\newcommand{\multijettitle}{{
Search for new phenomena in final states with 
large jet multiplicities and missing transverse momentum
at ${\bf \sqrt{s}=8}$~TeV proton-proton collisions using the ATLAS experiment}}

%% file: abstract.tex
\newcommand\multijetabstract{{
A search is presented for new particles 
decaying to large numbers (7 or more) of jets, 
with missing transverse momentum and no isolated electrons or muons.
This analysis uses \ourintlumi{} of \pp{} collision data 
at \sqseight{} collected by the ATLAS experiment 
at the Large Hadron Collider.
The sensitivity of the search is enhanced by considering the number of $b$-tagged jets 
and the scalar sum of masses of large-radius jets in an event. No evidence is found for physics beyond the Standard Model.
The results are interpreted in the context of various simplified
supersymmetry-inspired models where gluinos are pair produced, as well
as an \msugra{} model.
}}

%% file: content.tex
\section{Introduction}
\label{intro}
Many extensions of the Standard Model of particle physics
predict the presence of TeV-scale strongly interacting particles
that decay to weakly interacting descendants.
In the context of R-parity-conserving supersymmetry
(SUSY)~\cite{Fayet:1976et,Fayet:1977yc,Farrar:1978xj,Fayet:1979sa,Dimopoulos:1981zb},
the strongly interacting parent particles are the partners of the
quarks (squarks, $\tilde{q}$)
and gluons (gluinos, $\tilde{g}$), and are
produced in pairs. The lightest supersymmetric particle (LSP) is
stable, providing a candidate that can contribute to the relic dark-matter density in the universe~\cite{Goldberg:1983nd,Ellis:1983ew}.
If they are kinematically accessible, the squarks and gluinos could be produced
in the proton--proton interactions at the Large Hadron Collider
(LHC)~\cite{Evans:2008zzb}.

Such particles are expected to decay in cascades,
the nature of which depends on the mass hierarchy within the model.
The events would be characterised by
significant missing transverse momentum from the unobserved weakly interacting
descendants, and by a large number of jets from emissions of quarks and/or gluons.
Individual cascade decays may include
gluino decays to a top squark (stop, $\tilde{t}$) and an anti-top quark,
\begin{subequations}\label{eq:gttonshell}
\begin{equation}\label{eq:gttonshell_1}
\gluino ~\to~ \stop + \tbar
\end{equation}
followed by the top-squark decay to a top quark and a neutralino LSP, \ninoone,
\begin{equation}
\stop \to t + \ninoone.
\end{equation}
\end{subequations}
Alternatively, if the top squark is heavier than the gluino, the three-body decay,
\begin{equation}\label{eq:gtt}
\gluino ~\to~ t + \tbar + \ninoone
\end{equation}
may result.
Other possibilities include decays involving intermediate charginos, neutralinos,
and/or squarks including bottom squarks.
A pair of cascade decays produces a large number 
of Standard Model particles, 
together with a pair of LSPs, one from the end of each cascade.
The LSPs are assumed to be stable and only weakly interacting,
and so escape undetected, resulting in missing transverse momentum.

In this \paper{} we consider final states with large numbers of jets
together with significant missing transverse momentum
in the absence of isolated electrons or muons, using the $\pp$ collision data 
recorded by the ATLAS experiment \cite{Aad:2008zzm} during 2012 
at a centre-of-mass energy of $\sqrt{s}=8\,\TeV$. 
The corresponding integrated luminosity is $\ourintlumi{}$.
Searches for new phenomena in final states with large jet multiplicities 
-- requiring from at least six to at least nine jets -- and missing
transverse momentum
have previously been reported by the ATLAS
Collaboration using LHC $pp$ collision data corresponding 
to $1.34\,\ifb$~\cite{Aad:2011qa} and to $4.7\,\ifb$~\cite{Aad:2012hm} at $\sqrt{s}=7\,\TeV$.
Searches with explicit tagging of jets from bottom quarks ($b$-jets) in multi-jet 
events were also performed by ATLAS~\cite{Aad:2012pq} and
CMS~\cite{Chatrchyan:1527115, Chatrchyan:1503581, Chatrchyan:2012lia}. These searches found no
significant excess over the Standard Model expectation and provide 
limits on various supersymmetric models, including decays such as
that in eq.~\eqref{eq:gtt} and an \mSUGRA{} \citemsugraandcmssm{} model that
includes strong production processes. The analysis presented in this \paper{} extends previous analyses
by reaching higher jet multiplicities and utilizing new sensitive variables.

Events are first selected with large jet multiplicities, 
with requirements ranging from at least seven to at least ten jets,
reconstructed using the anti-$k_t$
clustering algorithm~\cite{Cacciari:2008gp,Cacciari:2005hq} and jet
radius parameter $R=0.4$. 
Significant missing transverse momentum is also required in the event.  
The sensitivity of the search is further enhanced by the subdivision
of the selected sample into several categories using additional information.
Event clasification based on the number of 
$b$-jets gives enhanced sensitivity to 
models which predict either more or fewer 
$b$-jets than the Standard Model background.
In a complementary stream of the analysis,
the $R=0.4$ jets are clustered into large $(R=1.0)$ 
\composite{} jets to form an event variable, the sum of the masses of
the \composite{} jets,
which gives additional discrimination in models with a large number of 
objects in the final state~\cite{Hook:2012fd}.
Events containing isolated, high transverse-momentum (\pT{})
electrons or muons are vetoed in order to reduce backgrounds
involving leptonic $W$ boson decays.
The previous analyses~\cite{Aad:2011qa,Aad:2012hm} had signal regions with 
smaller jet multiplicities; those are now omitted
since the absence of significant excesses in earlier analyses places 
stringent limits on models with large cross sections.

Searches involving final states with many jets and missing transverse momentum 
have been confirmed to have good sensitivity to
decays such as those in eqs.~\eqref{eq:gttonshell} and \eqref{eq:gtt}~\cite{Aad:2012hm},
but they also provide sensitivity to any model
resulting in final states with large jet multiplicity
in association with missing transverse momentum. Such models include the pair production 
of gluinos, each of them decaying via an off-shell squark, as
\begin{subequations}\label{eq:onestep}
\begin{equation}\label{eq:onestep_1}
\tilde{g} \to \bar{q} + q' + \chinoonepm \to \bar{q} + q' + W^{\pm} + \ninoone,
\end{equation}
or alternatively 
\begin{equation}\label{eq:onestep_2}
\tilde{g} \to \bar{q} + q' + \chinoonepm \to \bar{q} + q' + W^{\pm} + \ninotwo \to
\bar{q} + q' + W^{\pm} + Z^{0} + \ninoone.
\end{equation}
\end{subequations}
Another
possibility is the pair
production of
gluinos which decay as in eq.~\eqref{eq:gttonshell_1} and the subsequent
decay of the \stop-squark via 
\begin{equation*} 
\stop \to b + \tilde{\chi}^{\pm}_{1},
\end{equation*} 
or via the R-parity-violating decay 
\begin{equation}\label{eq:RPVUDD} 
\stop \to \bar{b} + \bar{s}.
\end{equation} 

Several supersymmetric models are
used to interpret the analysis results: simplified models that include decays
such as those in eqs.~\eqref{eq:gttonshell}--\eqref{eq:RPVUDD},
and an \msugra{} model with parameters\footnote{A particular \msugra{} model point is specified by five parameters:
the universal scalar mass $m_0$,
the universal gaugino mass $m_{1/2}$,
the universal trilinear scalar coupling $A_0$,
the ratio of the vacuum expectation values of the two Higgs fields $\tan\beta$,
and the sign of the higgsino mass parameter $\mu$.}
$\tan{\beta}=30$, $A_0=-2m_0$ and $\mu>0$, which accommodates a
lightest Higgs boson mass compatible with the observed Higgs boson
mass at the LHC~\cite{Aad:2012tfa,Chatrchyan:1471016}.

\section{The ATLAS detector and data samples}\label{sec:samples}

The ATLAS experiment is a multi-purpose particle
physics detector with a forward-backward symmetric cylindrical
geometry and nearly 4$\pi$ coverage in solid angle.\footnote{ATLAS uses a right-handed coordinate system with its origin
  at the nominal interaction point (IP) in the centre of the detector
  and the $z$-axis along the beam pipe. The $x$-axis points from the
  IP to the centre of the LHC ring, and the $y$-axis points
  upward. Cylindrical coordinates $(r,\phi)$ are used in the
  transverse plane, $\phi$ being the azimuthal angle around the beam
  pipe. The pseudorapidity is defined in terms of the polar angle
  $\theta$ as $\eta=-\ln\tan(\theta/2)$, and the transverse energy $\ET$ by
$\ET=E\sin\theta$.}
The layout of the detector is
defined by four superconducting magnet systems, which comprise a
thin solenoid surrounding the inner tracking detectors (ID),
and a barrel and two end-cap toroids generating the magnetic field for
a large muon spectrometer.
The ID provides precision reconstruction of tracks in the region $|\eta|<2.5$.
The calorimeters lie between the ID and the muon system.  In the pseudorapidity
region $\left|\eta\right| < 3.2$, high-granularity liquid-argon (LAr)
electromagnetic (EM) sampling calorimeters are used.  An iron/scintillator-tile
calorimeter provides hadronic coverage for
$\left|\eta\right| < 1.7$.  The end-cap and forward regions, spanning
$1.5 < \left|\eta\right| < 4.9$, are instrumented with LAr calorimeters
for both EM and hadronic measurements.

The data sample used in this analysis was taken during
the period from March to December 2012
with the LHC operating at a $pp$ centre-of-mass energy of $\sqrt{s}=8~\TeV$.
Application of data-quality requirements results in
an integrated luminosity of $20.3 \pm 0.6\,\ifb$,
where the luminosity is measured
using techniques similar to those described in ref.~\cite{Aad:2013ucp},
with a preliminary calibration of the luminosity scale derived from
beam-overlap scans performed in November 2012.
The analysis makes use of dedicated multi-jet triggers, 
the final step of which required
either at least five jets with $\ET>55\,\GeV{}$ 
or at least six jets with $\ET>45\,\GeV{}$,
where the jets must have $|\eta|<3.2$.
The final level of the trigger selection 
is based on a jet algorithm and calibration method
closely matched to those used in the signal region selections.
In all cases the trigger efficiency is greater than 99\%
for events satisfying the jet multiplicity selection
criteria for the signal regions described in
\oursecref{sec:eventselection}.
Events selected with single-lepton triggers and prescaled multi-jet triggers are used for background determination in
control regions. 

\section{Physics object selection}\label{sec:objects}

Jets are reconstructed using the
anti-$k_t$ jet clustering algorithm with
radius parameter $R=0.4$. The inputs to this algorithm are the
energies and positions of clusters
of calorimeter cells, where the clusters are formed starting from
cells with energies significantly above the noise level~\cite{Aad:2011he}.
Jet momenta are constructed by performing a four-vector sum over
these clusters of calorimeter cells, treating each
as an $(E, {\bf p})$ four-vector with zero mass.
The local cluster weighting (LCW) calibration method~\cite{Issever:2004qh}
is used to classify clusters as being of either electromagnetic
or hadronic origin and, based on this classification, applies specific energy corrections
derived from a combination of Monte Carlo simulation and data~\cite{Aad:2011he}. A
further calibration
is applied to the corrected jet energies to relate the response of the calorimeter
to the true jet energy~\cite{Aad:2011he}.
The jets are corrected for energy from additional proton--proton
collisions (pile-up) using a method, proposed
in ref.~\cite{Cacciari:2007fd}, which estimates the pile-up
activity in any given event, as well as the sensitivity of any given
jet to pile-up. The method subtracts a contribution from the jet energy
equal to the product of the jet area 
and the average energy density of the event.
All jets are 
required to satisfy $\pT > 20$~GeV and $|\eta| < 2.8$.
More stringent requirements on $\pT{}$ and on $|\eta|$ are made when
defining signal regions as described in \oursecref{sec:eventselection}.

Jets with heavy-flavour content are identified using 
a tagging algorithm that uses both impact parameter and secondary
vertex information~\cite{ATLAS-CONF-2011-102}.
This $b$-tagging algorithm is applied to all jets that satisfy both
$|\eta| < 2.5$ and $\pT>40\,\GeV$. 
The parameters of the algorithm are chosen such that 70\% of $b$-jets
and about 1\% of light-flavour or gluon jets are selected in
\ttbar{} events in Monte Carlo simulations~\cite{ATLAS-CONF-2012-097}.
Jets initiated by charm quarks are tagged with about 20\% efficiency.

Electrons are required to have $\ourpt$ $>$ $10~\GeV$ and $|\eta|$ $<$
2.47. They must satisfy `medium' electron shower shape and track selection criteria based upon those described in
ref.~\cite{Aad:2011mk}, but modified to reduce the
impact of pile-up and to match tightened trigger requirements.
They must be separated by at least $\Delta R=0.4$ from any jet,
where $\Delta R = \sqrt{(\Delta \eta)^2 + (\Delta \phi)^2}$. Events containing electrons passing these criteria are vetoed when forming signal regions.
Additional requirements are applied to electrons when defining
leptonic control regions used to aid in the estimate of the SM
background contributions, as described in \Secref{sec:backgrounds:leptonic};
in this case, electrons must have $\ourpt>25\GeV$, 
must satisfy the `tight' criteria of ref.~\cite{Aad:2011mk},
must have transverse and longitudinal impact parameters 
within $5$ standard deviations and 0.4\,mm, respectively, of the primary vertex, 
and are required to be well isolated.\footnote{The electron isolation 
requirements are based on nearby tracks and calorimeter clusters, as follows.
The scalar sum of transverse momenta of tracks, other than the track
from the electron itself, 
in a cone of radius $\Delta R=0.3$ around the electron 
is required to be smaller than 16\% of the electron's \pT{}.
The scalar sum of calorimeter transverse energy around the electron 
in the same cone, excluding the electron itself, 
is required to be smaller than 18\% of the electron's \pT{}.} 

Muons are required to have $\ourpt>10\GeV$ and $|\eta|$ $<$ 2.5,
to satisfy track quality selection criteria, and 
to be separated by at least $\Delta R=0.4$ from the nearest jet candidate. Events containing muons passing these criteria are vetoed when forming signal regions.
When defining leptonic control regions, 
muons must have $\pT>25\GeV$, $|\eta|<2.4$, 
transverse and longitudinal impact parameters within $5$ standard deviations and
0.4\,mm, respectively, of the primary vertex 
and they must be isolated.\footnote{The scalar sum of the transverse
  momenta of the tracks, other than the track from the muon itself, within a cone of $\Delta R=0.3$ around the muon must be less than 12\% of the muon's \pT{},
and the scalar sum of calorimeter transverse energy in the same cone, excluding that from the muon, must be less than 12\% of the muon's \pT{}.  
}

The missing transverse momentum two-vector \ourvecptmiss{} is calculated from the negative 
vector sum of the transverse momenta of all calorimeter energy
clusters with $|\eta|<4.5$
and of all muons~\cite{Aad:2012re}. 
Clusters associated with either electrons or photons with $\pT>10\,\GeV$,
and those associated with jets with $\pT>20\,\GeV$ and $|\eta|<4.5$ make use of the calibrations
of these respective objects.
For jets the calibration includes the area-based pile-up correction described above. 
Clusters not associated with such objects are calibrated using both
calorimeter and tracker information. The magnitude of \ourvecptmiss{}, conventionally
denoted by \ourmagptmiss{}, is used to distinguish signal and background
regions.

\section{Event selection}\label{sec:eventselection}

Following the physics object reconstruction described in \oursecref{sec:objects},
events are discarded if they contain any jet
that fails quality criteria designed to suppress detector noise and
non-collision backgrounds, or if they lack a reconstructed primary
vertex with five or more associated tracks.
Events containing isolated electron or muon
candidates are also vetoed as described in section~\ref{sec:objects}.
The remaining events are then analysed in two complementary analysis streams,
both of which require large jet multiplicities and significant
\met. The selections of the two streams are verified to have good sensitivity to decays such as those in eqs.~\eqref{eq:gttonshell} -- \eqref{eq:RPVUDD}, but are kept
generic to ensure sensitivity in a broad set of models with large jet
multiplicity and \met{} in the final state. 

\subsection{The multi-jet + flavour stream}

In the multi-jet + flavour stream 
the number of jets with $|\eta|<2$ and \pT{} above   
the threshold $\pthreshold=50\,\GeV$ is determined.
Events with exactly eight or exactly nine such jets are selected,
and the sample is further subdivided according to the number of the 
jets (0, 1 or $\ge$2) with $\pt>40$\,GeV and $|\eta|<2.5$ which satisfy 
the $b$-tagging criteria.
The $b$-tagged jets may
belong to the set of jets with $\pT$ greater than $\pthreshold$,
but this is not a requirement.
Events with ten or more jets are retained in a separate category, 
without any further subdivision.

A similar procedure is followed for the higher 
jet-\pT{} threshold of $\pthreshold = 80\,\GeV$. Signal regions
are defined for events with exactly seven jets
or at least eight jets. Both categories are again subdivided
according to the number of jets (0, 1 or $\geq$2) that are $b$-tagged.
Here again, the $b$-tagged jets do not necessarily satisfy the $\pthreshold$ requirement.

In all cases the final selection variable is $\metsig$, the ratio
of the \met{}
to the square root of the scalar sum \HT{} of the transverse momenta
of all jets with $\pT>40\,\GeV$ and $|\eta|<2.8$.
This ratio is closely related to the significance of the \met{}
relative to the resolution due to stochastic variations
in the measured jet energies~\cite{Aad:2012re}.
The value of \metsig{} is required to be larger than 4\,\rootgev{}
for all signal regions.

\subsection{The multi-jet + ${\boldsymbol\MJ{}}$ stream}

Analysis of the multi-jet + \MJ{} stream proceeds as follows.
The number of ($R=0.4$) jets with \pT{} above 50\,GeV is determined, 
this time using a larger pseudorapidity acceptance of $|\eta|<2.8$.
Events with at least eight, at least nine or at least ten such jets are retained,
and a category is created for each of those multiplicity thresholds.
The four-momenta of the $R=0.4$ jets satisfying $\pT>20\,$GeV and $|\eta|<2.8$ are then used as inputs
to a second iteration of the anti-$k_t$ jet algorithm, 
this time using the larger distance parameter $R=1.0$.
The resulting larger objects are denoted as \composite{} jets.
The selection variable \MJ{} is then defined to be the sum 
of the masses $m_{j}^{R=1.0}$ of the \composite{} jets
\begin{equation*}
\MJ \equiv \sum_{\substack j} m_{j}^{R=1.0},
\label{eq:defMJ}
\end{equation*}
where the sum is over the \composite{} jets that satisfy 
$p_{\rm T}^{R=1.0} > 100 \GeV$ and 
${|\eta^{R=1.0}|<1.5}$.
Signal regions are defined for two different \MJ{} thresholds.
Again the final selection requires that $\metsig>4\,\rootgev$.

\subsection{Summary of signal regions}

The nineteen resulting signal regions are summarized in \tabref{tab:SR}.
Within the multi-jet + flavour stream
the seven signal regions defined with $\pthreshold=50\,\GeV$
are mutually disjoint. 
The same is true for the six signal regions defined with the threshold of 80\,GeV. 
However, the two sets of signal regions overlap; an event found in one of the $\pthreshold=80\,\GeV$ signal regions
may also be found in one of the $\pthreshold=50\,\GeV$ signal regions.
The multi-jet + \MJ{} stream has six inclusive signal regions;
for example an event which 
has at least ten $R=0.4$ jets with $\pt>50$~GeV, 
$\MJ>420$\,GeV and $\metsig>4\rootgev$
will be found in all six multi-jet + \MJ{} regions. 
These overlaps are treated in the results of the analysis as described in~\Secref{sec:results}.

\afterpage\clearpage
\begin{sidewaystable}[h]
\renewcommand\arraystretch{1.5}
\begin{center}
\begin{tabular}{| c || c | c | c | c | c | c | c || c | c | c | c | c | c || c | c | c |}
      \hline
 & \multicolumn{13}{|c||}{\bf Multi-jet + flavour stream} &
\multicolumn{3}{|c|}{\bf Multi-jet + ${\boldsymbol\MJ{}}$ stream} \\ \hline \hline
Identifier & \multicolumn{3}{|c|}{~\BSR{8j50}~} & \multicolumn{3}{|c|}{ ~\BSR{9j50}~} &
            { ~\BSR{\geq10j50}~} &
              \multicolumn{3}{|c|}{ ~\BSR{7j80}~} & \multicolumn{3}{|c||}{ ~\BSR{\geq8j80}~} &
            { ~\BSR{\geq8j50}~} & { ~\BSR{\geq9j50}~} & { ~\BSR{\geq10j50}~}\\ \hline
Jet $|\eta|$ &
  \multicolumn{7}{|c||}{$<2.0$} & 
  \multicolumn{6}{|c||}{$<2.0$} & 
  \multicolumn{3}{|c|}{$<2.8$} \\ \hline
Jet \pt{} &
  \multicolumn{7}{|c||}{$>50$\,\GeV} & 
  \multicolumn{6}{|c||}{$>80$\,\GeV} &
  \multicolumn{3}{|c|}{$>50$\,\GeV} \\ \hline
Jet count &
  \multicolumn{3}{|c|}{$=8$} &   \multicolumn{3}{|c|}{$=9$} & $\geq10$ & 
  \multicolumn{3}{|c|}{$=7$} &   \multicolumn{3}{|c||}{$\geq8$} &
  $\geq8$ & $\geq 9$ & $\geq10$ \\ \hline
$b$-jets & \multirow{2}{*}{0} & \multirow{2}{*}{1}  & \multirow{2}{*}{$\geq 2$} &
                             \multirow{2}{*}{0} & \multirow{2}{*}{1} & \multirow{2}{*}{$\geq 2$} & \multirow{2}{*}{---} & 
                             \multirow{2}{*}{0} & \multirow{2}{*}{1} & \multirow{2}{*}{$\geq 2$} &
                             \multirow{2}{*}{0} & \multirow{2}{*}{1} & \multirow{2}{*}{$\geq 2$} &
                             \multicolumn{3}{|c|}{\multirow{2}{*}{---}} \\ 
$(\pt> 40\,\GeV, |\eta|<2.5)$ & & & &
                              & &  &  & 
                              & &  &
                              & &  &
                              \multicolumn{3}{|c|}{}\\ \hline
\MJ{} [GeV] & \multicolumn{7}{|c||}{---} & \multicolumn{6}{|c||}{---} 
            & \multicolumn{3}{|c|}{$>340$ and $>420$ for each case}  \\ \hline
\metsig &  \multicolumn{7}{|c||}{$>4 \rootgev$} 
        &  \multicolumn{6}{|c||}{$>4 \rootgev$} 
        & \multicolumn{3}{|c|}{$>4 \rootgev$}\\
\hline
\end{tabular}
\end{center}
\caption{\label{tab:SR}
Definition of the nineteen signal regions.
The jet $|\eta|$, \pt{} and multiplicity all refer to the $R=0.4$ jets.
Composite jets with the larger radius parameter $R=1.0$ are used in the 
multi-jet + \MJ{} stream when constructing \MJ{}.
A long dash `---' indicates that no requirement is made.
}
\end{sidewaystable}

\section{Standard Model background determination}\label{sec:bg}

Two background categories are considered in this search: (1) multi-jet production,
including purely strong interaction processes and
fully hadronic decays of \ttbar, and hadronic decays
of $W$ and $Z$ bosons in association with jets, and (2) processes with
leptons in the final states, collectively referred to as leptonic backgrounds. 
The latter consist of
semileptonic and fully leptonic decays of \ttbar{}, including \ttbar{}
production in association with a boson; leptonically decaying $W$ or $Z$ bosons 
produced in association with jets; and single top quark production.

The major backgrounds 
(multi-jet, \ttbar{}, $W$\,+\,jets, and $Z$\,+\,jets)
are determined with the aid of control regions, 
which are defined such that they are enriched in the background 
process(es) of interest,
but nevertheless remain kinematically close to the signal regions. The
multi-jet background determination is fully data-driven, and the most
significant of the other backgrounds use data control regions to
normalise simulations. 
The normalisations of the event yields predicted by the simulations are adjusted simultaneously
in all the control regions using a binned fit described in \oursecref{sec:results}, 
and the simulation is used to extrapolate
the results into the signal regions.
The methods used in the determination of the multi-jet and \leptonic{}
backgrounds are described in sections~\ref{sec:backgrounds:multijet}
and \ref{sec:backgrounds:leptonic}, respectively.

\subsection{Monte Carlo simulations}\label{sec:mc}

Monte Carlo simulations are used as part of the \leptonic{} background determination process,
and to assess the sensitivity to specific SUSY signal models.
Most of the \leptonic{} backgrounds
are generated using \sherpaVER{1.4.1}~\cite{Gleisberg:2008ta}
with the \generator{CT10}~\cite{Lai:2010vv} set of parton
distribution functions (PDF).
For \ttbar{} production, up to four additional partons are modelled in
the matrix element. Samples of 
$W$~+~jets and $Z$~+~jets events are generated with up to five
additional partons in the matrix element,
except for processes involving $b$-quarks for which up to four additional partons
are included.
In all cases, additional jets are generated via parton showering.
The leptonic $W$~+~jets, $Z$~+~jets and \ttbar{} backgrounds are normalised
according to their inclusive theoretical cross sections~\cite{Aliev:2010zk,Catani:2009sm}. In the case of \ttbar{}
production, to account for higher-order terms which are not present in
the \sherpa\, Monte Carlo simulation, the fraction of events initiated
by gluon fusion, relative to other processes, is modified to improve the agreement with data in
\ttbar-enriched validation regions described in
\oursecref{sec:backgrounds:leptonic}. This corresponds to applying a
scale factor of 1.37 to the processes initiated by gluon fusion and a
corresponding factor to the other processes to keep the total \ttbar\,
cross section the same.
The estimation of the \leptonic{} backgrounds in the signal regions is described in detail in \oursecref{sec:backgrounds:leptonic}.

Smaller background contributions are also modelled for the following processes:
single top quark production in association with a $W$ boson in
the $s$-channel (\generator{MC@NLO} 4.06\,\cite{Frixione:2002ik,Frixione:2010wd,Frixione:2005vw,Frixione:2008yi}
 / \herwig\,6.520\,\cite{Corcella:2002jc} /  \jimmy\,4.31\,\cite{Butterworth:1996zw}), 
$t$-channel single top quark production (\acermcVER{3.8} \cite{acerMC} / \pythiaVER{6.426} \cite{Sjostrand:2006za}), and
\ttbar{} production in association with a $W$ or $Z$ boson
(\madgraphVER{5.1.4.8} \cite{Alwall:2011uj} / \pythiaVER{6.426}).

Supersymmetric production processes are generated using
\herwigppVER{2.5.2}~\cite{Bahr:2008pv} and \madgraphVER{5.1.4.8} with
the PDF set \generator{CTEQ6L1} \cite{Pumplin:2002vw}.
The cross sections are calculated to next-to-leading order in the strong coupling constant $\alpha_{\rm{S}}$, including the resummation of soft gluon emission at next-to-leading-logarithmic accuracy (NLO+NLL)~\cite{Beenakker:1996ch,Kulesza:2008jb,Kulesza:2009kq,Beenakker:2009ha,Beenakker:2011fu}.

For each process, the nominal cross section and its uncertainty are taken from an 
envelope of cross-section predictions using different PDF sets and factorisation and renormalisation scales,
as described in ref.~\cite{Kramer:2012bx}. 
All Monte Carlo simulated samples also include simulation of pile-up and employ a
detector simulation~\cite{Aad:2010ah} based on
GEANT4~\cite{Agostinelli:2002hh}. The simulated events are reconstructed with the same algorithms as the data.

\subsection{Multi-jet background}\label{sec:backgrounds:multijet}

The dominant background at intermediate values of
\met{} is multi-jet production
including purely strong interaction processes and
fully hadronic decays of \ttbar.
The contribution from these processes is
determined using collision data and the selection criteria were designed such 
that multi-jet processes can be accurately 
determined from supporting measurements.

The background determination method is based on the observation that
the \met{} resolution of the detector is approximately proportional to $\sqrt{\HT}$ and almost independent of the jet multiplicity in events dominated by jet activity, including hadronic decays of top quarks and gauge bosons~\cite{Aad:2011qa,Aad:2012hm}.
The distribution of the ratio \metsig{} therefore has a shape that 
is almost invariant under changes in the
jet multiplicity.
The multi-jet backgrounds can be determined 
using control regions with lower \metsig{} and/or lower 
jet multiplicity than the signal regions. The control regions are
assumed to be dominated by Standard Model processes, and that assumption
is corroborated by the agreement 
with Standard Model predictions of multi-jet cross-section
measurements for up to six jets~\cite{Collaboration:2011tq}.

Events containing heavy quarks show a different \metsig{} distribution
than those containing only light-quark or gluon jets, 
since semileptonic decays of heavy quarks contain neutrinos.
The dependence of \metsig{} on the number of heavy quarks
is accounted for in the multi-jet + flavour signal regions by using a consistent set of control regions with 
the same $b$-jet multiplicity as the target signal distribution.
The \metsig{} distribution is also found to be approximately 
independent of the \MJ{} event variable,
so a similar technique is used to obtain
the expected multi-jet background contributions to 
the multi-jet + \MJ{} signal regions.

\afterpage\clearpage
The leading source of variation in \metsig{} under changes in the jet
multiplicity
comes from a contribution to \met{} 
from calorimeter energy deposits not associated with jets 
and hence not contributing to \HT. 
The effect of this `soft' energy is corrected for by 
reweighting the \metsig{} distribution separately for each jet
multiplicity in the signal region, 
to provide the same \softoHT{} distribution, 
where \softet{} is the scalar sum of $\ET$ over all clusters 
of calorimeter cells not associated with jets having 
$\pT>20\,\GeV$ or electron, or muon candidates.

For example, to obtain the multi-jet contribution to the 
multi-jet + flavour stream \SR{9j50} signal region 
with exactly one $b$-jet, the procedure is as follows.
A template of the shape of the \metsig{} distribution is formed 
from events which have exactly six jets with $\pT>50$\,GeV,
and exactly one $b$-jet (which is not required to be one of the six previous jets). 
The expected contribution from \leptonic{} backgrounds is then subtracted,
so that the template provides the expected distribution resulting 
from the detector resolution, together with any contribution 
to the resolution from semileptonic $b$-quark decays.
The nine-jet background prediction for the signal region ($\metsig>4\rootgev$)
with exactly one $b$-jet is then given by 
\begin{eqnarray}
N^{\textrm{multi-jet}}_{\textrm{predicted}} &=& \left(N^{A,
  ~n^{\textrm{jet}}=9}_{\textrm{data}}-N^{A, ~n^{\textrm{jet}}=9}_{\textrm{~\leptonic\,MC}}\right)
\times \left(\frac{N^{B, ~n^{\textrm{jet}}=6}_{\textrm{data}}-N^{B,
    ~n^{\textrm{jet}}=6}_{\textrm{~\leptonic\,MC}}}{N^{A,
    ~n^{\textrm{jet}}=6}_{\textrm{data}}-N^{A, ~n^{\textrm{jet}}=6}_{\textrm{~\leptonic\,MC}}}
\right),
\label{Eq.QCDmethod}
\end{eqnarray}
where $A \equiv \metsig<1.5~\rootgev$, $B \equiv
\metsig>4~\rootgev$, and each of the counts $N$ is determined after requiring the same $b$-jet multiplicity as 
for the target signal region (i.e. exactly one $b$-jet in this
example). 
Equation~\ref{Eq.QCDmethod} is applied separately to each of ten bins (of width 0.1) in  
\softoHT{} to find the prediction for that bin, and then the contributions of the ten bins 
summed to provide the \softoHT{}-weighted multi-jet prediction.

An analogous procedure is used to obtain the expected multi-jet contribution
to each of the other multi-jet + flavour stream signal regions 
by using the appropriate \pthreshold,
jet multiplicity, and $b$-jet multiplicity as required by the target signal region.
In each case the shape of the \metsig{} distribution is obtained from 
a `template' with exactly six (five) jets for signal regions 
with $\pthreshold=50\,(80) \GeV$. The distributions of \metsig{} for
multi-jet + flavour stream control regions are shown in
figure~\ref{fig:metsig_qcd_vr}.

\begin{figure}\centering
\subfloat[~No $b$-jets]{\includegraphics[width=0.49\textwidth, keepaspectratio]{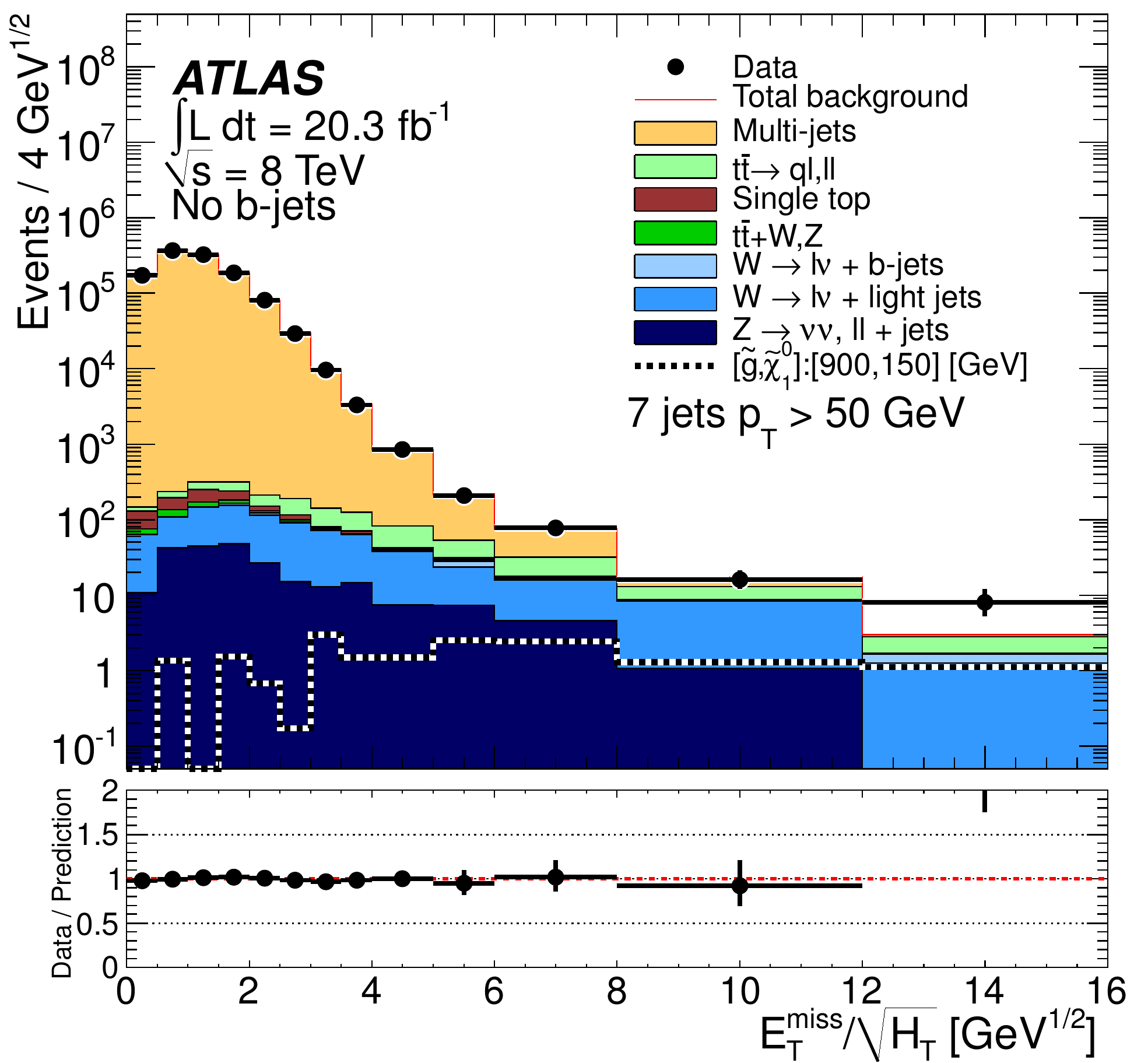}}
\subfloat[~Exactly one $b$-jet]{\includegraphics[width=0.49\textwidth, keepaspectratio]{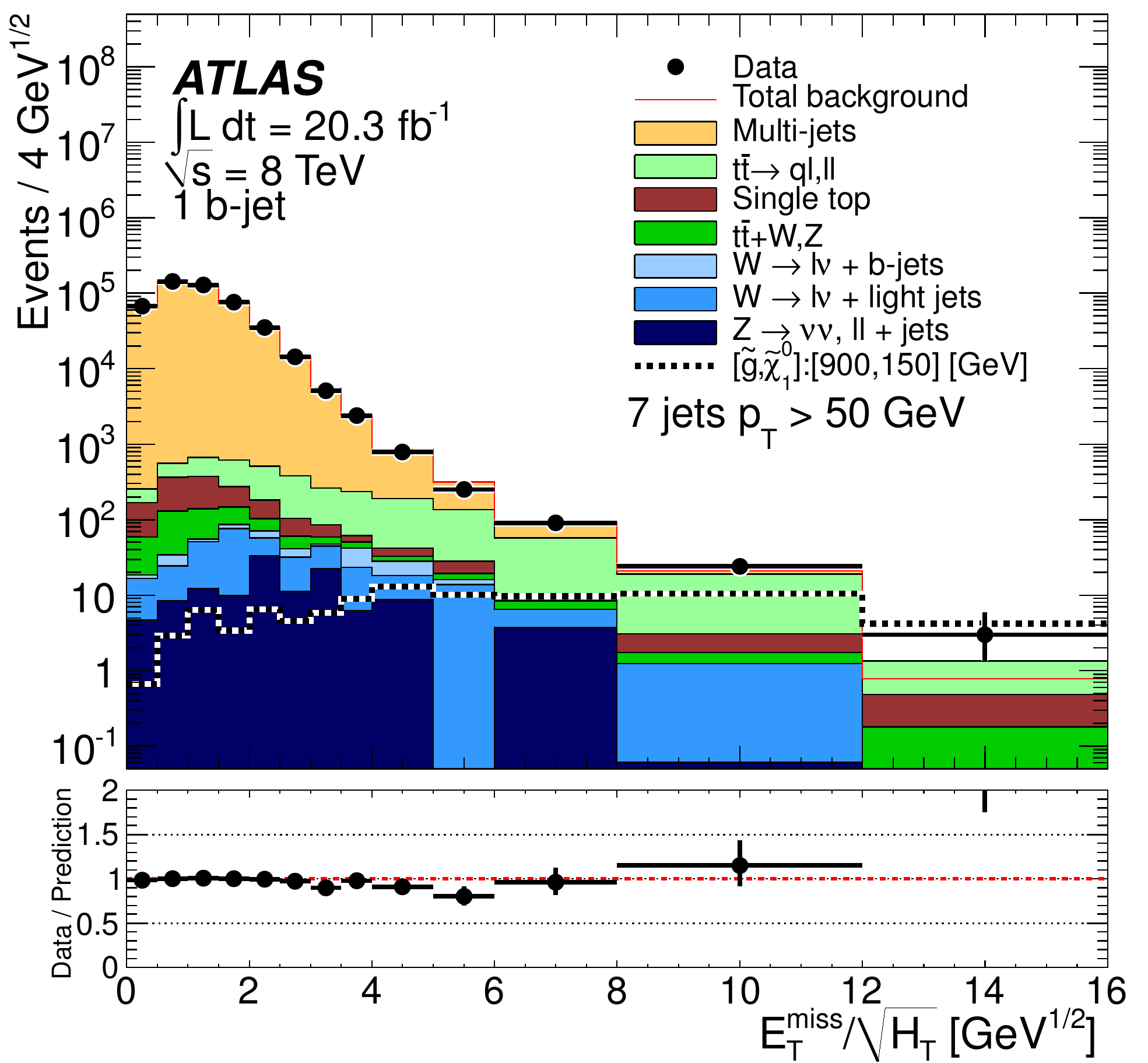}}\\
\subfloat[~At least two $b$-jets]{\includegraphics[width=0.49\textwidth, keepaspectratio]{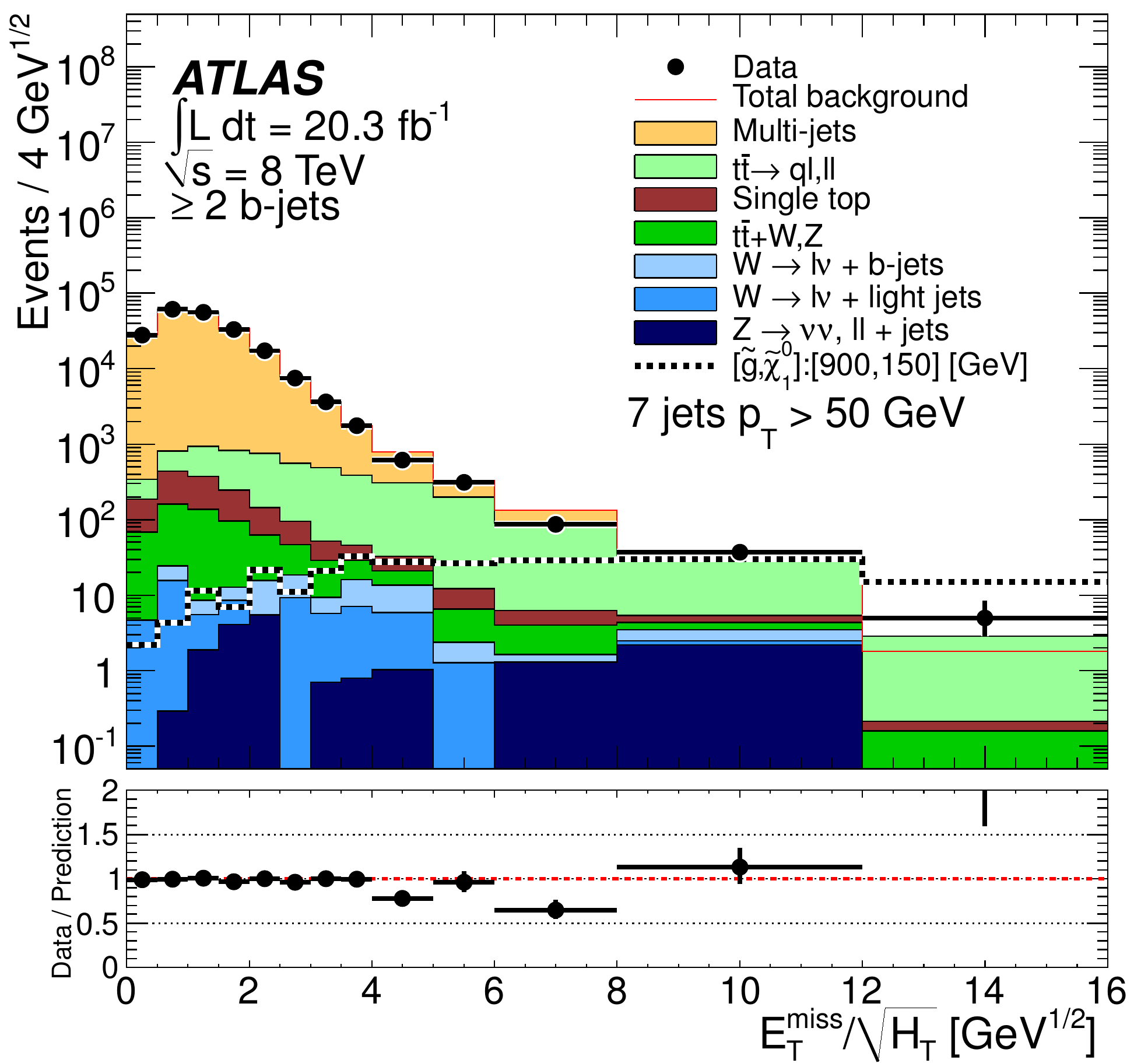}}
\caption{\label{fig:metsig_qcd_vr}
Distribution of \metsig{} for the control regions 
with exactly seven jets with 
$\pt \geq 50$~GeV and $|\eta|<2.0$, for different $b$-jet multiplicities.
The multi-jet prediction is determined from an \metsig{} 
template obtained from events with exactly six jets. It is normalised
to the data in the region $\metsig<1.5\,\rootgev$ after subtraction
of the \leptonic{} backgrounds. The most important \leptonic{}
backgrounds are also shown, based on Monte Carlo simulations. 
Variable bin sizes are used with bin widths (in units of~\rootgev) of 0.5 (up to $\metsig=4\,\rootgev$), 1 (from 4 to 6), 2 (from 6 to 8) and 4 thereafter. 
For reference and comparison, a supersymmetric model is used where gluinos of mass 900~GeV are 
pair produced and each decay as in eq.~\eqref{eq:gtt} to 
a \ttbar{} pair and a $\ninoone$ with a mass of 150~GeV. The model is referred to as `$[\tilde{g},\ninoone]:[900,150]$~[GeV]'.} 
\end{figure}

The procedure in the multi-jet + \MJ{} stream is similar:
the same jet \pthreshold, jet multiplicity and \MJ{} criteria 
are used when forming the template and
control regions that are required for the target signal
region. \metsig{} distributions for control regions with exactly
seven jets with \pt$>$50~GeV and additional \MJ{} selection criteria applied are shown
in figure~\ref{fig:METsigCR_MJ}.
\Leptonic{} backgrounds are subtracted, and \softoHT{} weighting 
is applied.
For all cases in the multi-jet + \MJ{} stream the \metsig{} template shape is determined 
from a sample which has exactly six jets with
$\pT>50$\,GeV.

\begin{figure}\centering
\subfloat[~$\MJ\geq340$ GeV]
    {\includegraphics[width=0.49\textwidth, keepaspectratio]{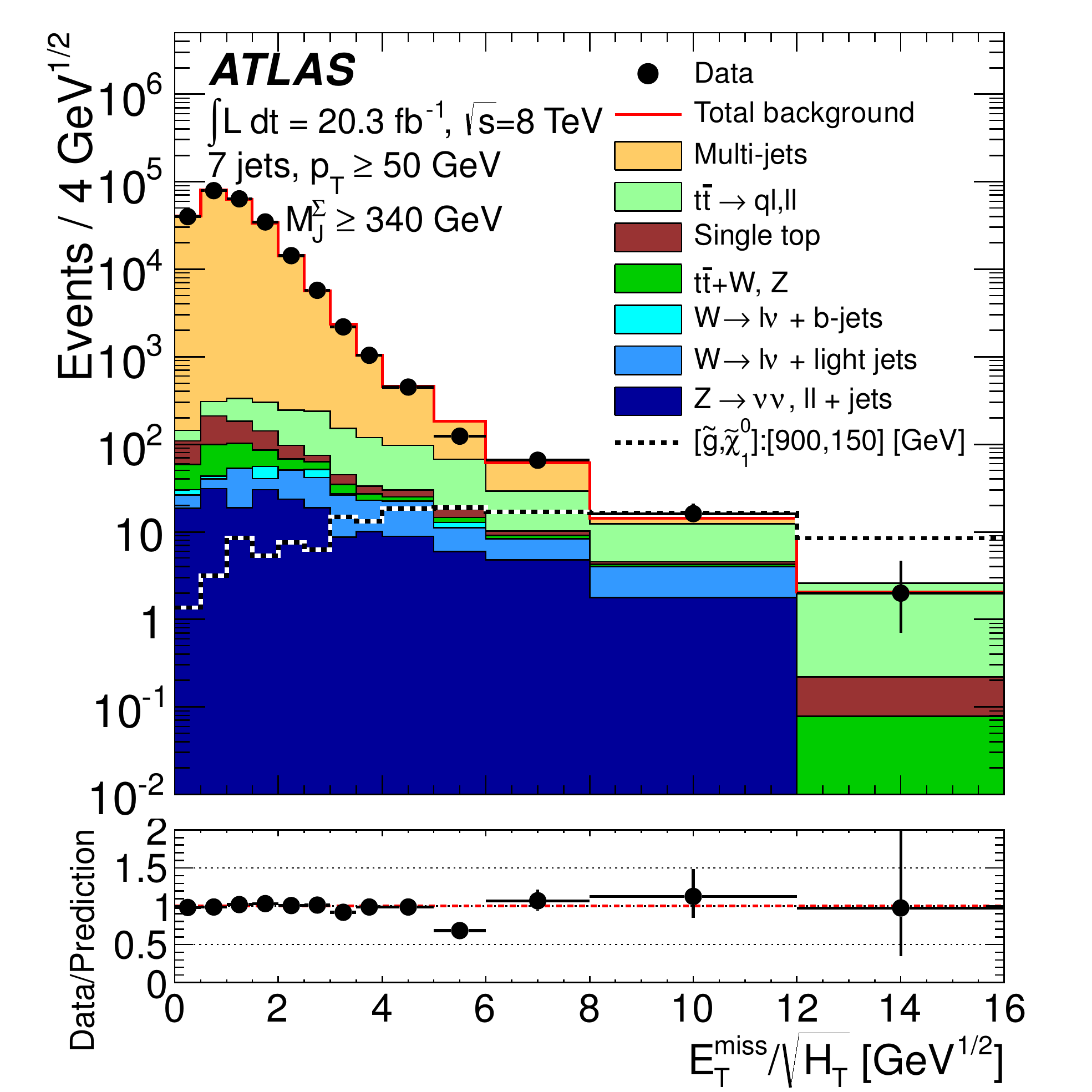}}
      \subfloat[~$\MJ\geq420$ GeV]
    {\includegraphics[width=0.49\textwidth, keepaspectratio]{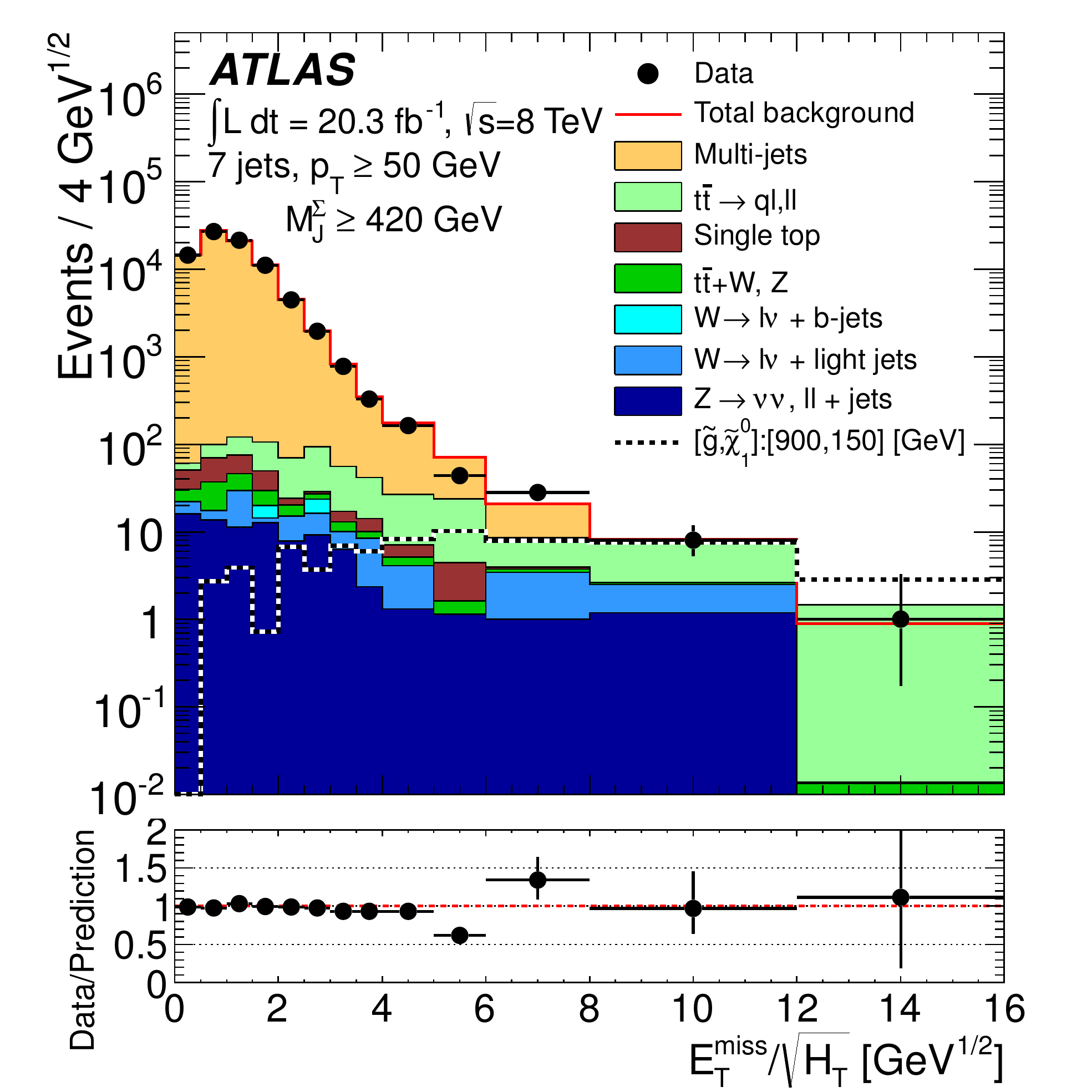}}
\caption{\label{fig:METsigCR_MJ}
Distribution of $\metsig$ for control regions with 
exactly seven jets with $\pt \geq 50$ GeV, and satisfying the same requirements
as the multi-jet + \MJ{} stream signal regions, other than that on \metsig{} itself. 
The multi-jet prediction was determined from an \metsig{} 
template obtained from events with exactly six jets. Other details are as
for figure~\ref{fig:metsig_qcd_vr}.
}
\end{figure}

Variations in the shape of the \metsig{} distribution under changes in the jet multiplicity are later used 
to quantify the systematic uncertainty associated with the method, 
as described in \oursecref{sec:backgrounds:multijet:systematic}.

\subsection{Systematic uncertainties in the multi-jet background determination}
\label{sec:backgrounds:multijet:systematic}

The multi-jet background determination method is 
validated by measuring the accuracy of the predicted \metsig{} template
for regions with jet multiplicities and/or \metsig{} smaller than those chosen for the signal regions.
The consistency of the prediction with the number of observed events (\textit{closure}) is tested in regions with
$\metsig$ [\rootgev] in the ranges $(1.5,\,2.0)$, $(2.0,\,2.5)$, and $(2.5,\,3.5)$ 
for jet multiplicities of exactly seven, eight and nine, 
and in the range $(1.5,\,2.0)$ and $(2.0,\,3.5)$ for 
$\geq$10 jets. 
The tests are performed separately for 0, 1 and $\geq 2$ $b$-tagged jets.
In addition, the method is tested for events with exactly six (five)
jets with $\pthreshold=50$~GeV (80~GeV)
across the full range of \metsig{} 
in this case using a template obtained from events with exactly five (four) jets.
The five-jet (four-jet) events are obtained using a prescaled trigger 
for which only a fraction of the total luminosity is available.
Agreement is found both for signal region jet multiplicities at intermediate values of \metsig{} and 
also for the signal region \metsig\, selection at lower multiplicity.
A symmetrical systematic uncertainty on each signal region is constructed 
by taking the largest deviation in any of the closure regions 
with the same jet multiplicity or lower, for the same $b$-tagging requirements. 
Typical closure uncertainties are in the range 5\% to 15\%;
they can grow as large as $\sim$50\% for the tightest signal regions,
due to larger statistical variations in the corresponding control regions.

Additional systematic uncertainties result from modelling of the 
heavy-flavour content (25\%), which is assessed by using combinations of the templates of
different $b$-tagged jet multiplicity to vary the purity of the different samples. The closure in simulation of samples with high heavy-flavour content is also tested.
The \leptonic{} backgrounds that are subtracted when forming the
template have an uncertainty associated
with them (5--20\%, depending on the signal region). Furthermore,
other uncertainties taken into account are due 
to the scale choice of the cutoff for the soft energy term, \softet{}, (3--15\%)
and the trigger efficiency
($<$1\%) in the region where the template is formed.

\subsection{\Leptonic{} backgrounds}\label{sec:backgrounds:leptonic}

The \leptonic{} backgrounds are defined to be those which
involve the leptonic decays $W\to\ell\nu$ or $Z\to\nu\nu$.
Contributions are determined for
partly hadronic (i.e.~semileptonic or dileptonic) $\ttbar$,
single top, $W$ and $Z$ production, and diboson production,
each in association with jets.
The category excludes semileptonic decays of charm and bottom quarks, 
which are considered within the 
multi-jet category (\oursecref{sec:backgrounds:multijet}).
The \leptonic{} backgrounds which contribute most to the 
signal regions are \ttbar{} and $W$\,+\,jets.
In each case, events can evade the lepton veto, either via hadronic $\tau$ decays 
or when electrons or muons are produced but not reconstructed.

The predictions employ the Monte Carlo simulations described in \oursecref{sec:mc}.
When predictions are taken directly from the Monte Carlo simulations,
the \leptonic{} background event yields are subject to large theoretical
uncertainties associated with the use of a leading-order
Monte Carlo simulation generator. These include scale variations as well as changes in  
the number of partons present in the matrix element calculation, and
uncertainties in the response of the detector.
To reduce these uncertainties the
background predictions are, where possible, 
normalised to data using control regions and cross-checked 
against data in other validation regions.
These control regions and validation regions are designed 
to be distinct from, but kinematically close to, the signal regions,
and orthogonal to them by requiring an identified lepton candidate.

\begin{figure}
\centering
\subfloat[~No $b$-jets]{\label{fig:leptonic_cr_0b}\includegraphics[width=0.42\linewidth]{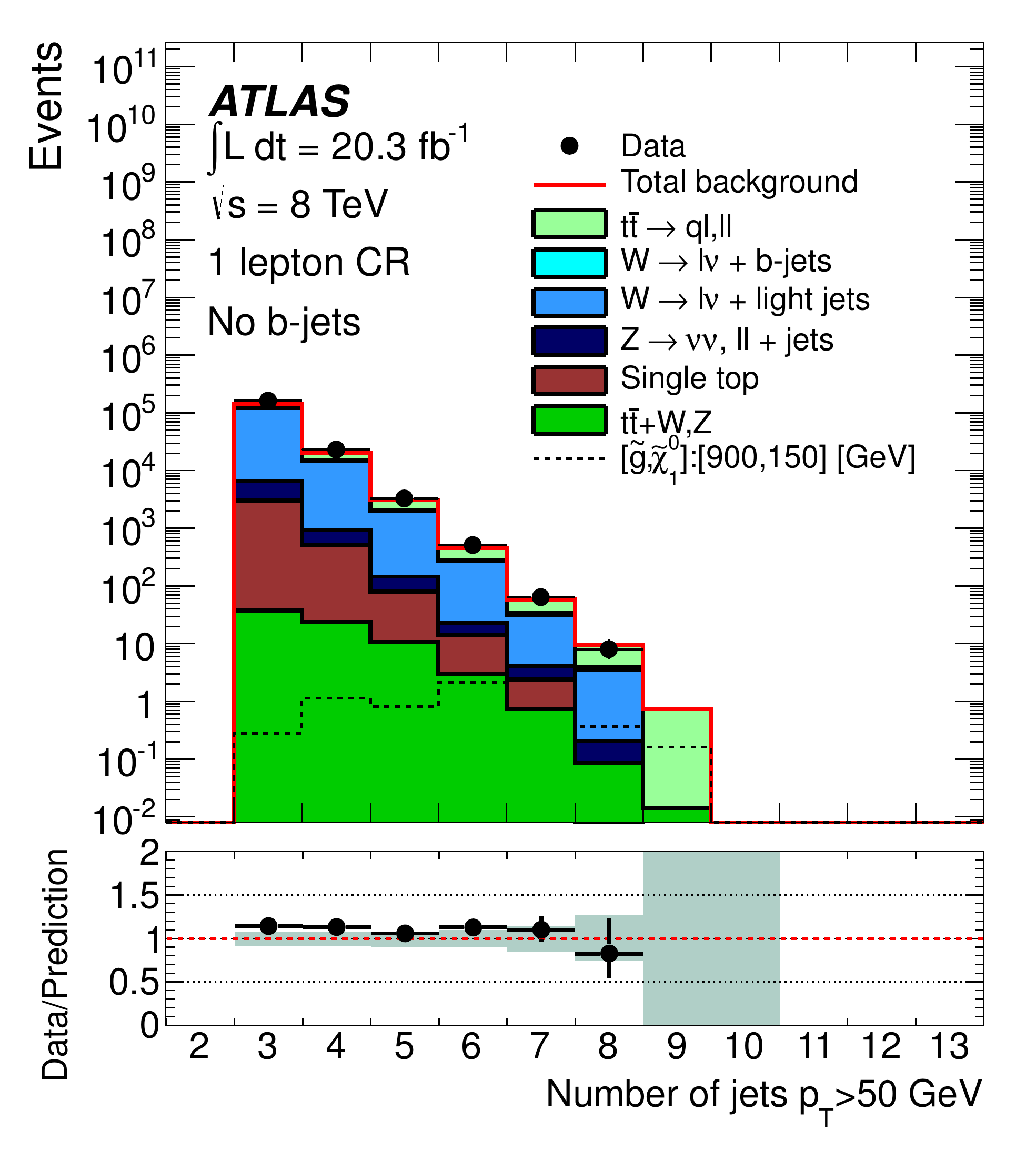}}
\subfloat[~Exactly one $b$-jet]{\label{fig:leptonic_cr_1b}\includegraphics[width=0.42\linewidth]{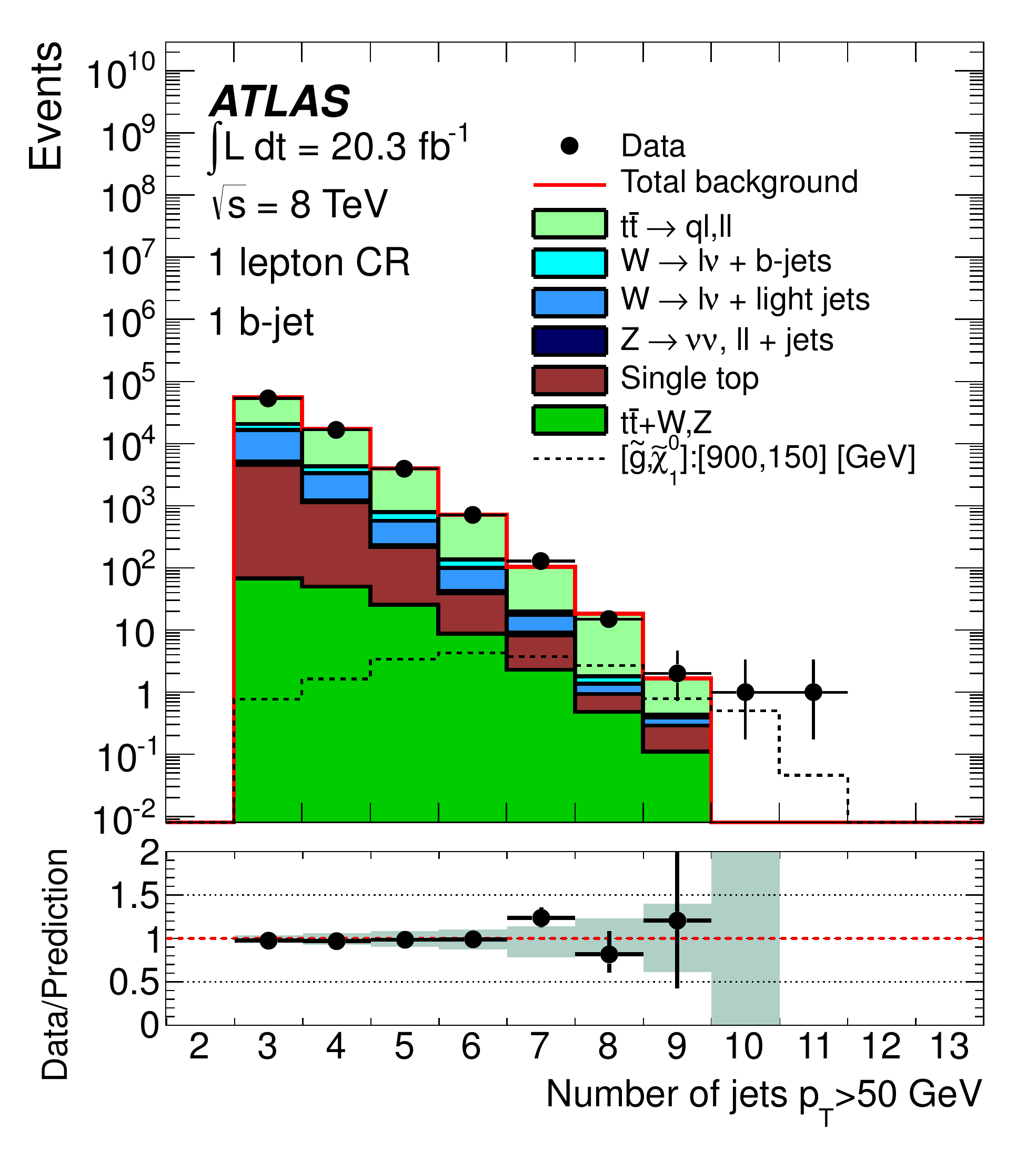}}\\
\subfloat[~$\geq$2 $b$-jets]{\label{fig:leptonic_cr_2b}\includegraphics[width=0.42\linewidth]{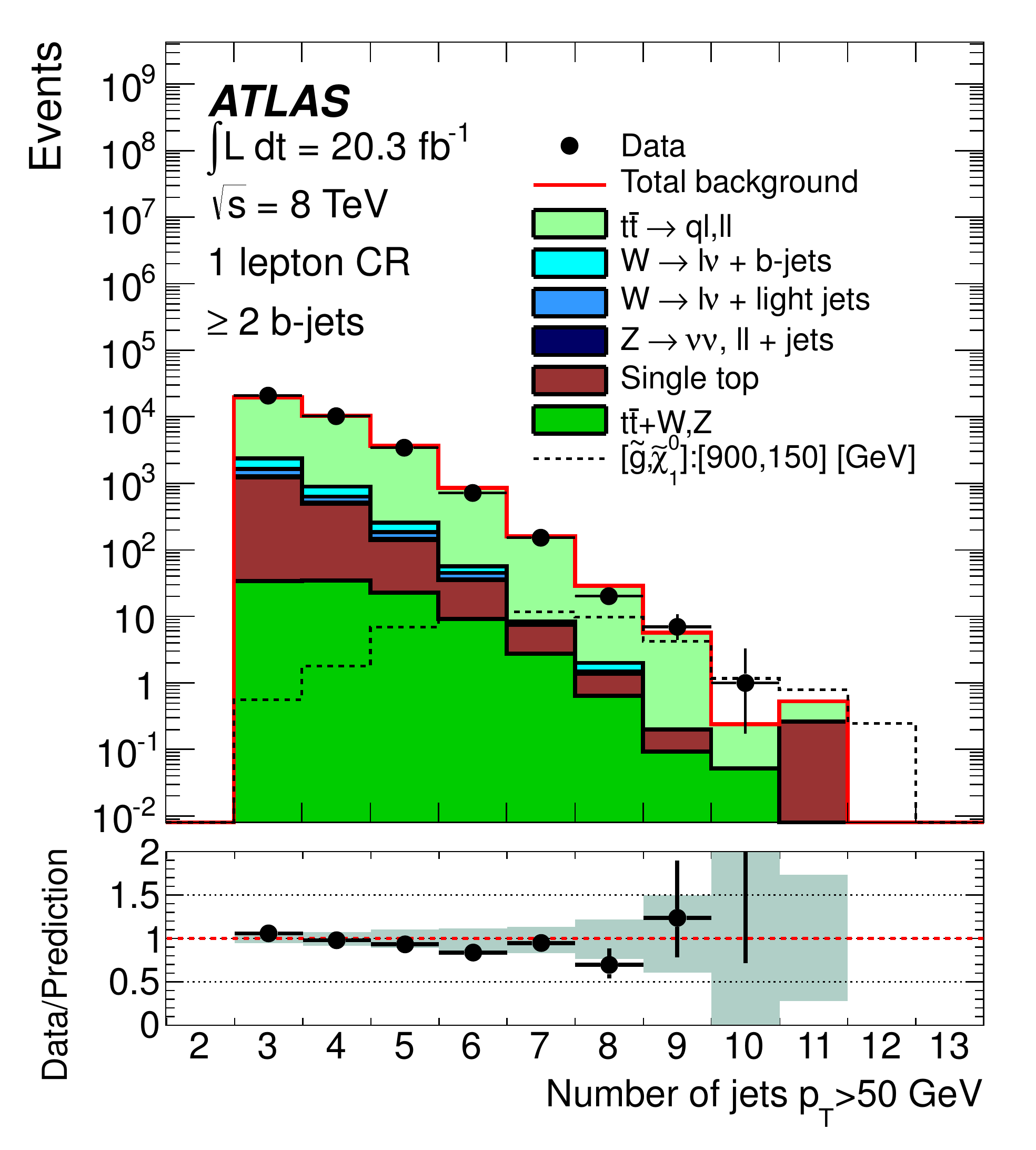}}
\caption{\label{fig:leptonic_cr_mult_1}
Jet multiplicity distributions for $\pthreshold=50$~GeV jets in the 
one-lepton \ttbar{} and  $W$\,+\,jets control regions (CR)
for different $b$-jet multiplicities. Monte Carlo simulation predictions are before fitting to data.
Other details are as for figure~\ref{fig:metsig_qcd_vr}. The band in the ratio plot indicates the 
experimental uncertainties on the Monte Carlo simulation prediction
and also includes the Monte Carlo simulation statistical uncertainty. Additional theoretical uncertainties are not shown.  }
\end{figure}

\begin{figure}
\centering
\subfloat[~\MJ$>$340~GeV, no
  $b$-jets]{\label{fig:leptonic_cr_0b_mj}\includegraphics[width=0.42\linewidth]{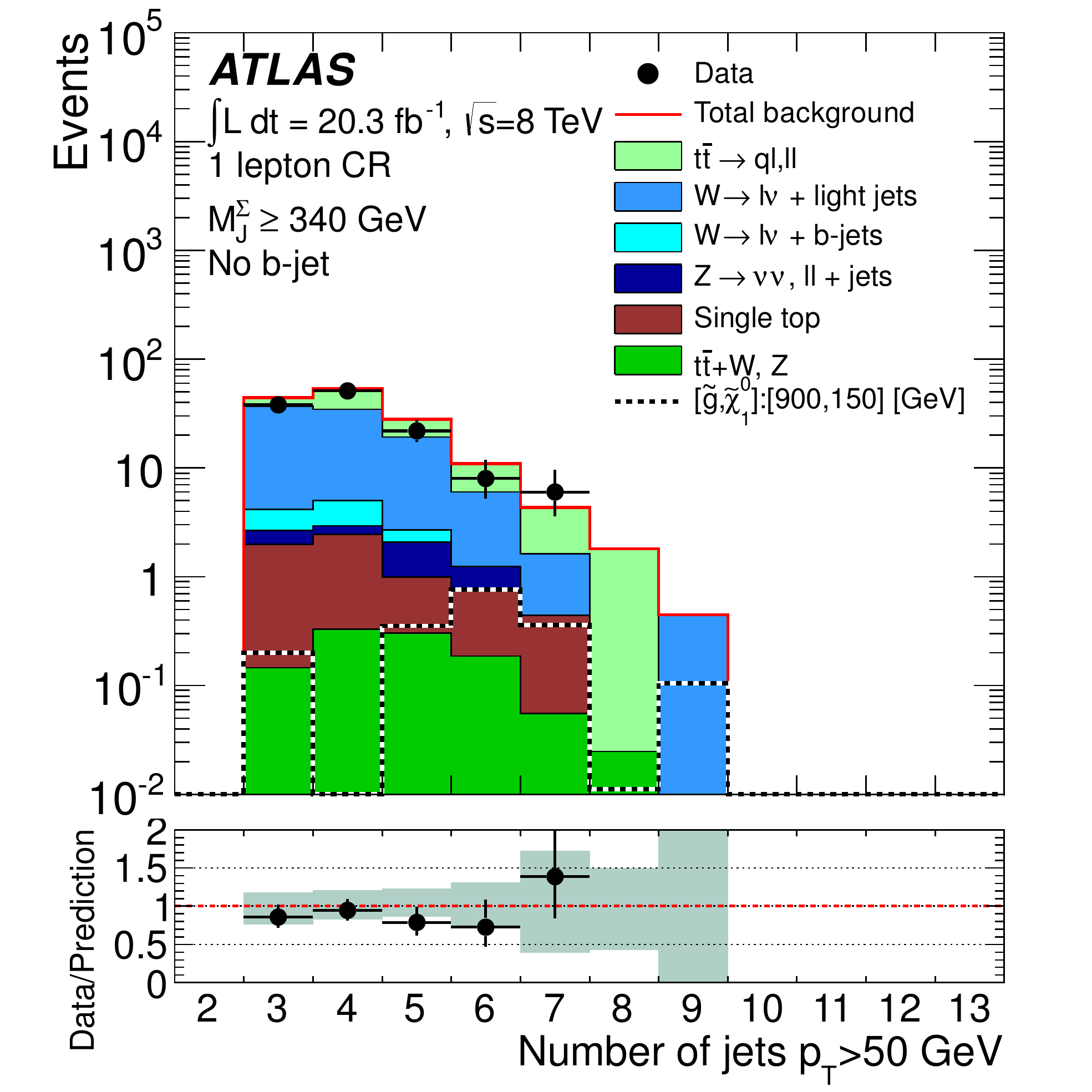}}
\subfloat[~\MJ$>$340~GeV, $\geq$1 $b$-jets]{\label{fig:leptonic_cr_1b_mj}\includegraphics[width=0.42\linewidth]{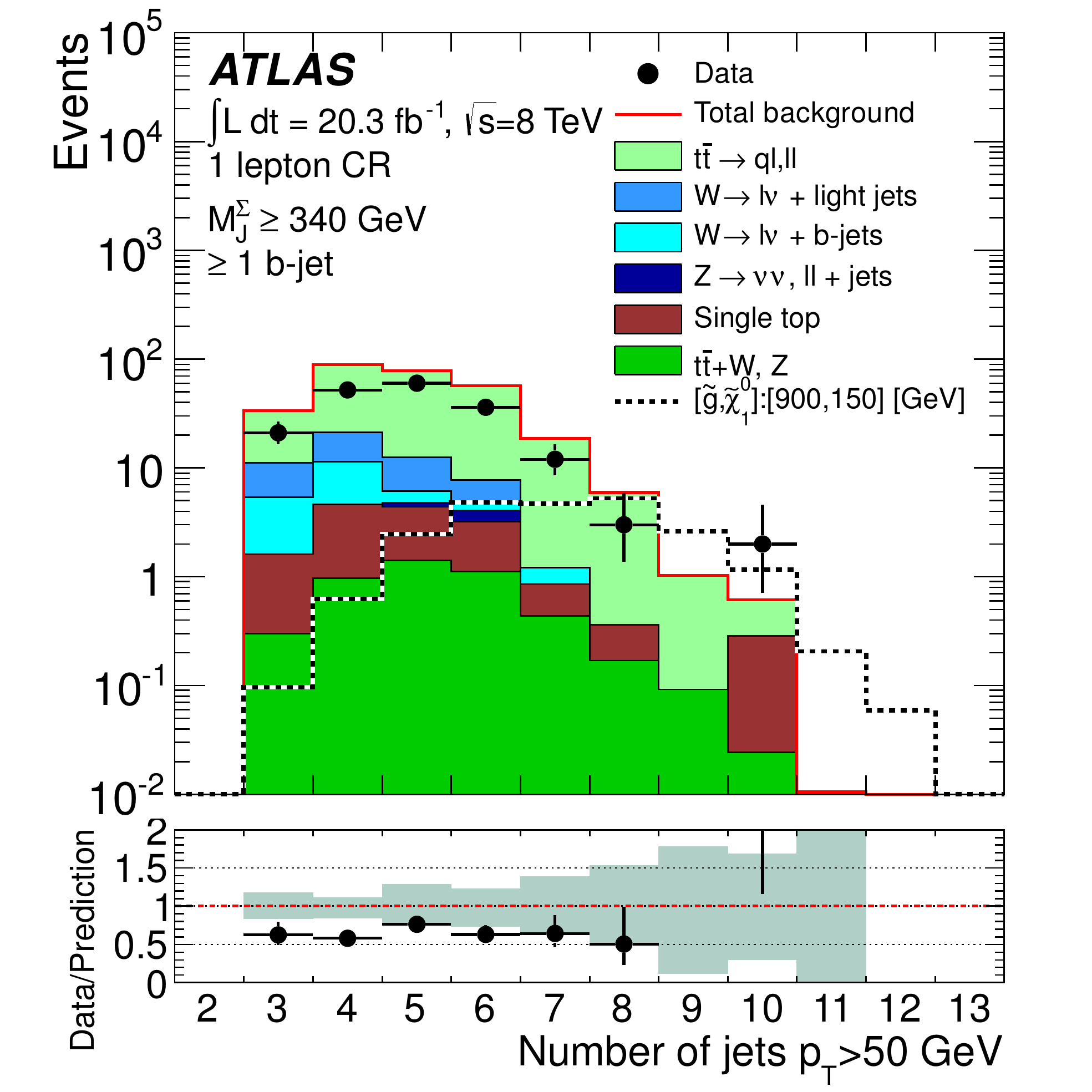}}
\\
\subfloat[\MJ{} distribution, $\geq$7j50 selection applied]{\label{fig:leptonic_cr_mj}\includegraphics[width=0.42\linewidth]{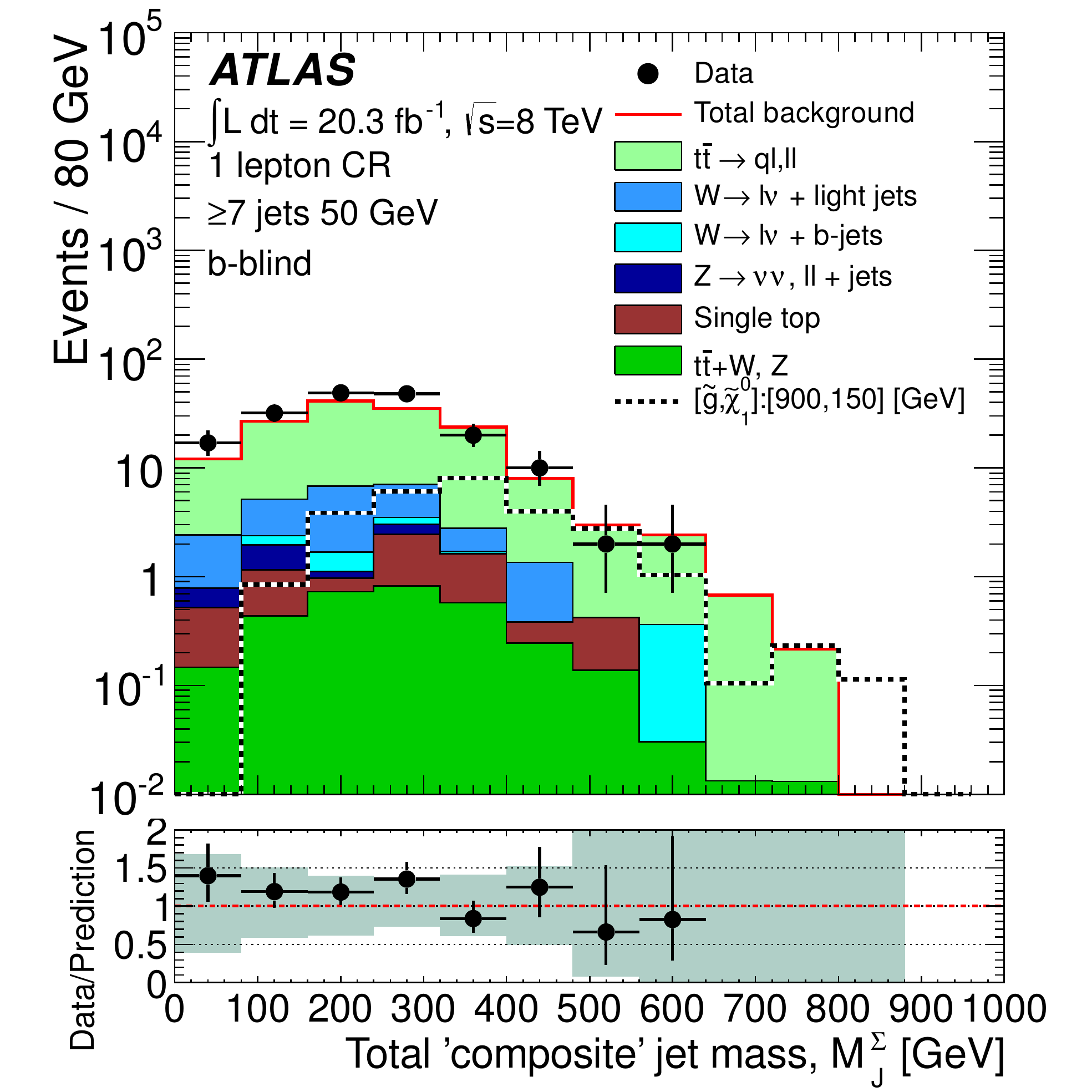}}
\caption{\label{fig:leptonic_cr_mult_2}
Jet multiplicity distributions for $\pthreshold=50$~GeV jets in the 
one-lepton \ttbar{} and  $W$\,+\,jets control regions (CR)
for different $b$-jet multiplicities
and a
selection on $\MJ>340$~GeV (\ref{fig:leptonic_cr_0b_mj}) - (\ref{fig:leptonic_cr_1b_mj}), and the \MJ{} distribution for an
inclusive selection of seven jets with $\pthreshold=50$~GeV (\ref{fig:leptonic_cr_mj}). Other details are as
for figure~\ref{fig:leptonic_cr_mult_1}.}
\end{figure}

The validation and control regions for the \ttbar{} and $W$\,+\,jets backgrounds
are defined in \tabref{tab:one_lepton_cr}.
In single-lepton regions, a single lepton ($e$ or $\mu$) is required,
with sufficient \pt{} to allow the leptonic trigger to be employed.
Modest requirements on \met{} and \metsig{} 
reduce the background from fake leptons.
An upper limit on 
\[
\mt = \sqrt{2 \left(|\ourvecptmiss||{\bf p}_{\rm T}^\ell| 
                     - \ourvecptmiss \cdot {\bf p}_{\rm T}^\ell\right)},
\]
where ${\bf p}_{\rm T}^\ell$ is the transverse momentum vector of the lepton,
decreases possible contamination from non-Standard-Model processes.

\begin{table}[h]
\small
\centering
\renewcommand\arraystretch{1.3}
\begin{tabular}{| l | c |}
\hline
\multicolumn{2}{|c|}{\textbf{Single-lepton validation region}} \\ \hline
Lepton \pT & $>25\,$GeV \\ \hline
Lepton multiplicity & Exactly one, $\ell \in \{e,\,\mu\}$ \\ \hline
\met{} & $>30$\,GeV \\ \hline
\metsig{} & $>2.0$\rootgev \\ \hline
$\mt$ & $<120\,$GeV \\ \hline
Jet \pT{} & \\ \cline{1-1}
Jet multiplicity & \multirow{2}{*}{\begin{minipage}{4cm}\begin{center}As for signal regions\\ (\tabref{tab:SR})\end{center}\end{minipage}} \\ \cline{1-1}
$b$-jet multiplicity & \\ \cline {1-1}
\MJ & \\ \cline{1-1}
\hline\hline
\multicolumn{2}{|c|}{\textbf{Control region (additional criteria)}} \\ \hline
Jet multiplicity & Unit increment if $p_{\rm T}^\ell > \pthreshold$ \\ \hline
$\met/\sqrt{\HT~(+p_{\rm T}^\ell)}$ & $>4.0$\rootgev \\ \hline
\end{tabular}
\caption{\label{tab:one_lepton_cr}
The selection criteria for the validation and control regions for the 
\ttbar\, and $W$\,+\,jets backgrounds.
In the control region the lepton is recast as a jet
so it contributes to \HT{} if $p_{\rm T}^\ell > 40$\,GeV 
and to the jet multiplicity count if $p_{\rm T}^\ell > \pthreshold$.
}
\end{table}

Since it is dominantly through hadronic $\tau$ decays that $W$ bosons and $\ttbar$ pairs
contribute to the signal regions, 
the corresponding control regions are created by recasting 
the muon or electron as a jet.
If the electron or muon has sufficient \pt{}\, (without any additional calibration), it is considered as an
additional `jet' and 
it can contribute to the jet multiplicity count, 
as well as to \HT{} and hence to the selection variable \metsig{}. The same
jet multiplicity as the signal region is required for the equivalent control 
regions. Additionally, the same criteria for \metsig, \MJ{}\, and the number of
$b$-tagged jets are required. For the \MJ{}\, stream these control regions
are further split into regions with no $b$-tagged jets and those with $b$-tagged jets 
to allow separation of contributions from $W+$jets and \ttbar\, events.
Provided the expected number of Standard Model events in the corresponding control region 
is greater than two, the number of observed events in that control region is used in a fit to determine the
Standard Model background as described in \oursecref{sec:results}. Distributions of jet multiplicity for the leptonic control regions
can be found in
figures~\ref{fig:leptonic_cr_mult_1}--\ref{fig:leptonic_cr_mult_2}. In
figure~\ref{fig:leptonic_cr_mult_2} the \MJ{} distribution for a
leptonic control region is also shown.

The $Z+$jets control regions require two same-flavour leptons with an invariant mass 
consistent with that of the $Z$ boson.
To create control regions that emulate the signal regions, 
the lepton transverse momenta are added to the missing momentum two-vector 
and then the requirement $\metsig>4\rootgev\,$ is applied. 
This emulates the situation expected for the $Z\rightarrow\nu\nu\,$ background. 
The details of the selection criteria are given in table
\ref{tab:leptonic_vr}. This selection, but with relaxed jet
multiplicity criteria, is used to validate the Monte Carlo simulation description of this process;
however, insufficient events remain at high jet multiplicity, so the estimation of this background is taken from Monte Carlo
 simulations.

\begin{table}[h]
\small
\centering
\renewcommand\arraystretch{1.3}
\begin{tabular}{| l | c |}
\hline
\multicolumn{2}{|c|}{\textbf{Two-lepton validation region}} \\ \hline
Lepton \pT & $>25\,$GeV \\ \hline
Lepton multiplicity & Exactly two, $e\,e$ or $\mu\,\mu$ \\ \hline
$m_{\ell\ell}$ & 80\,GeV to 100\,GeV  \\ \hline
Jet \pT{} & \\ \cline{1-1}
Jet multiplicity & \multirow{2}{*}{\begin{minipage}{4cm}\begin{center}As for signal regions\\ (\tabref{tab:SR})\end{center}\end{minipage}} \\ \cline{1-1}
$b$-jet multiplicity & \\ \cline {1-1}
\MJ & \\ \cline{1-1}
\hline\hline
\multicolumn{2}{|c|}{\textbf{Control region (additional criteria)}} \\ \hline
$|\ourvecptmiss + {\bf p}_{\rm T}^{\ell_1} + {\bf p}_{\rm T}^{\ell_2}| / \sqrt{\HT}$ 
                 & $>4.0$\rootgev \\ \hline
\end{tabular}
\caption{\label{tab:leptonic_vr}
The selection criteria for the validation and control regions for the $Z$\,+\,jets 
background.}
\end{table}

\subsection{Systematic uncertainties in the \leptonic{} background determination}
\label{sec:backgrounds:leptonic:systematic}

Systematic uncertainties on the \leptonic{} backgrounds originate from both detector-related
and theoretical sources from the Monte Carlo simulation modelling. Experimental uncertainties
are dominated by those on the jet energy scale, jet energy resolution and,
in the case of the flavour stream, $b$-tagging efficiency. Other
less important uncertainties result from the modelling of the pile-up,
the lepton identification and the soft energy term in the \met{}
calculation; these make negligible contributions to the
total systematic uncertainty.

The ATLAS jet energy scale and resolution are determined using in-situ
techniques~\cite{Aad:2011he,atlas-jetcal}. The jet energy scale uncertainty includes uncertainties associated
with the quark--gluon composition of the sample, the heavy-flavour
fraction and pile-up uncertainties. The uncertainties are derived for
$R=0.4$ jets and propagated to all objects and selections
used in the analysis. The 
sources of the jet energy scale uncertainty are treated as correlated between the various  
Standard Model backgrounds as well as with the signal contributions
when setting exclusion limits. The uncertainties on the yields due to
those on the jet energy scale and resolution range typically between 20\% and 30\%. The $b$-tagging efficiency
uncertainties are treated in a similar way when setting limits and
have typical values of $\approx 10\%$. They
are derived from data samples tagged with muons associated with jets,
using techniques described in
refs.~\cite{ATLAS-CONF-2011-102,ATLAS-CONF-2012-097}. 

For the \ttbar{} background, theoretical uncertainties are evaluated by
comparing the particle-level predictions of the nominal \sherpa{}
samples with additional samples in which some of the parameter
settings were varied. These include
variations of the factorisation scale, the matching scale of the matrix element to the parton shower,
the number of partons in the matrix element and the PDFs.
\alpgen\,\cite{Mangano:2002ea} samples are also generated with the renormalisation scale associated with 
$\alpha_{\textrm{S}}\,$ in the matrix element calculation varied up
and down by a factor of two relative to the
 original scale $k_t$ between two partons~\cite{Cooper:2011gk}. Finally, samples with and
without weighting of events initiated by gluon fusion relative to
other processes are used to provide a systematic uncertainty on this procedure. The two
latter sources of systematic uncertainty are the dominant ones with
typical values of 25--30\% each, leading to a total theoretical
uncertainty on the \ttbar{} background of $\approx 40\%$.

Alternative samples are generated similarly for the other smaller backgrounds with different parameters and/or generators
to assess the associated theoretical uncertainties, which are found to
be similar to those for the \ttbar{} background.

\section{Results}\label{sec:results}

Figures~\ref{fig:metsig_8j50_9j50}--\ref{fig:metsig_mj} show the
\metsig{} distributions for all the signal regions of both analysis streams.
In order to check the consistency of the data with the
background-only and signal hypotheses, a simultaneous profile maximum
likelihood fit~\cite{Cowan:2010js} is performed in the control and signal regions, for each of
the analysis streams separately. 
Poisson likelihood functions are used for event counts in signal and control regions. Systematic uncertainties are treated as nuisance parameters.
They are assumed to follow Gaussian distributions and their effect is propagated to the likelihood function.
A control region is taken into
account in the fit if there are at least two expected events associated
with it.
The fits differ significantly between
the two analysis streams, as described in the following sections.

\begin{figure}[h]
\centering
\subfloat[~8j50, no $b$-jets]{\label{fig:metsig_8j50_0bjet}\includegraphics[width=0.45\linewidth]{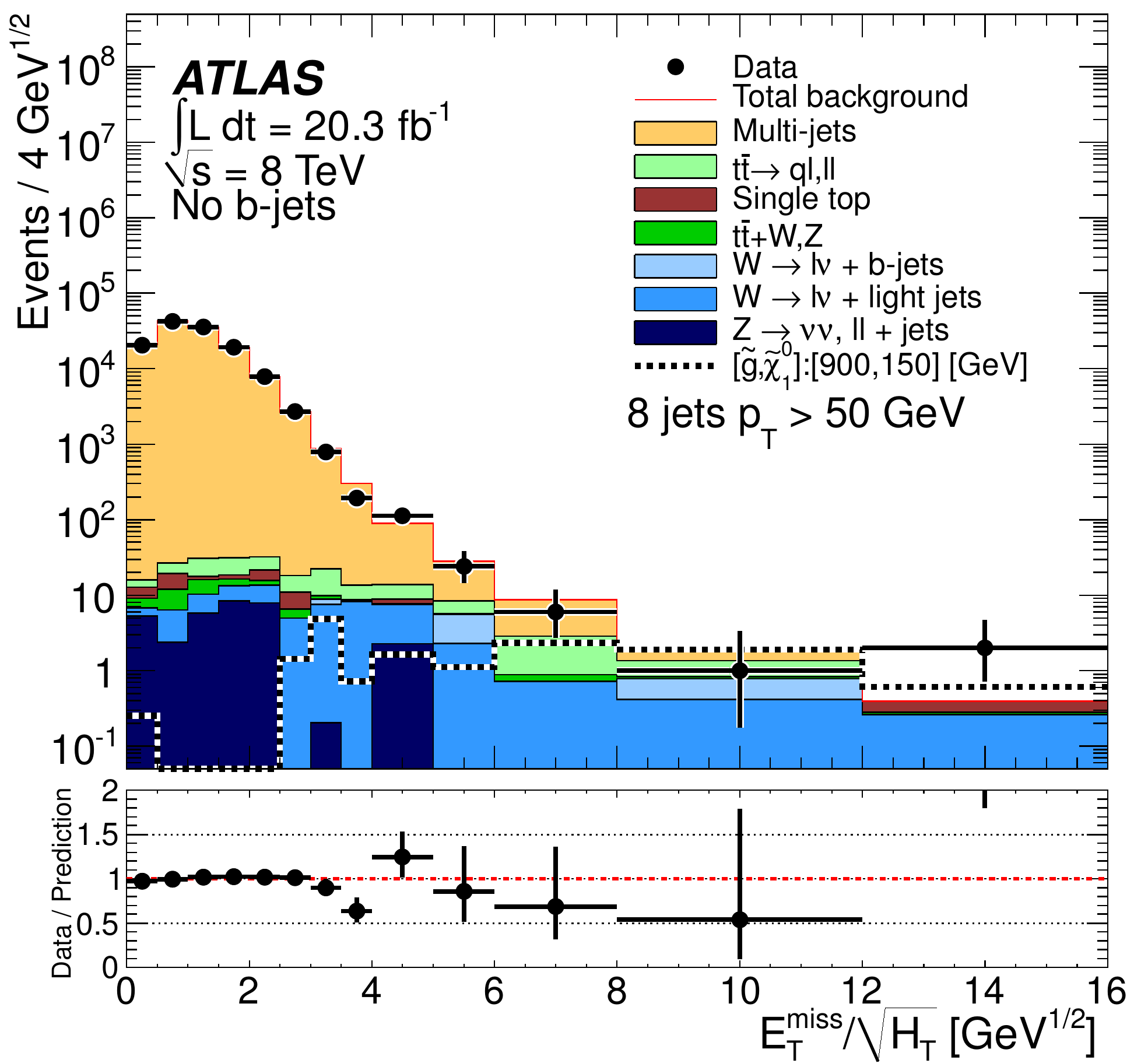}}
\subfloat[~9j50, no $b$-jets]{\label{fig:metsig_9j50_0bjet}\includegraphics[width=0.45\linewidth]{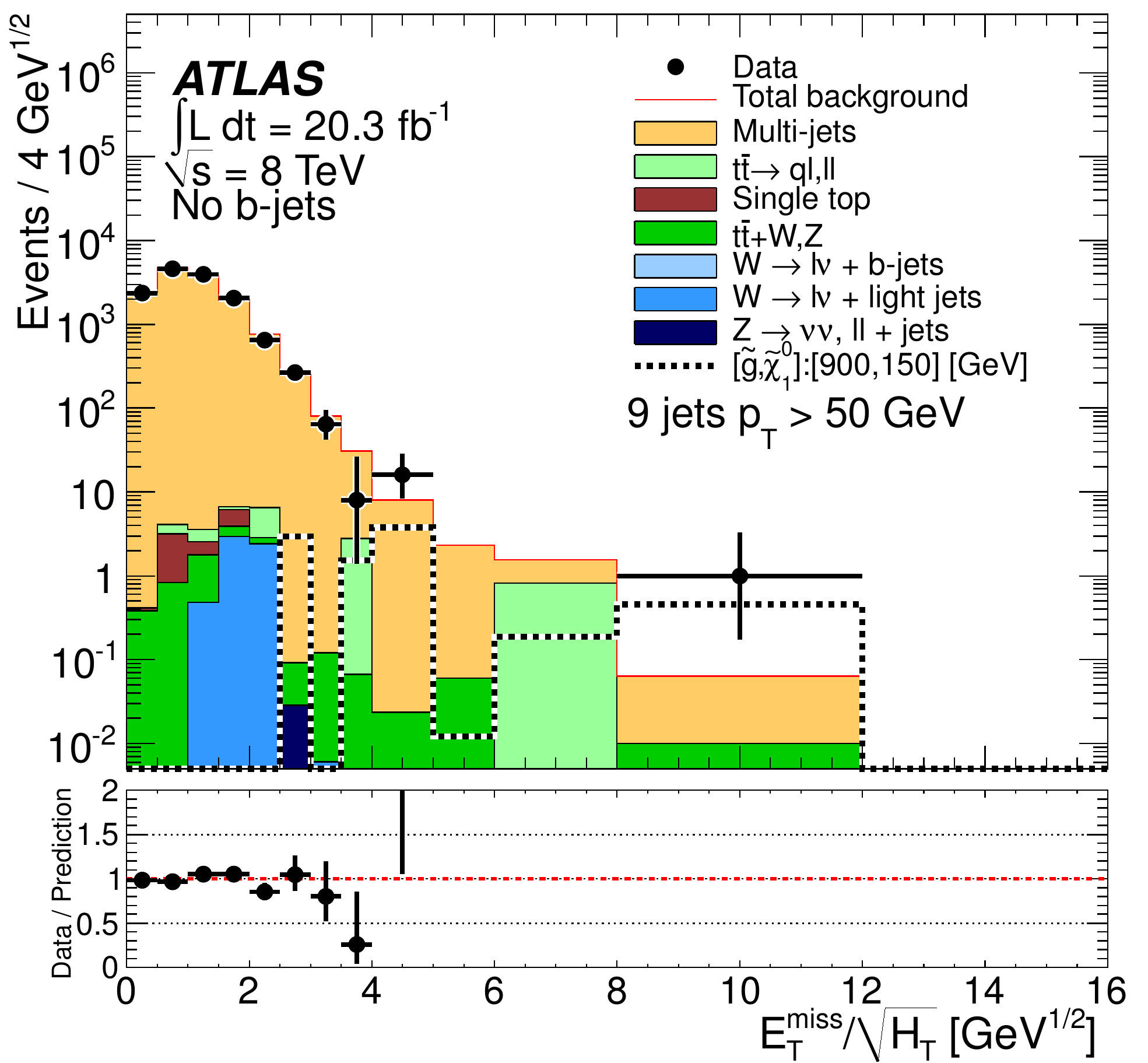}}\\
\subfloat[~8j50, exactly one $b$-jet]{\label{fig:metsig_8j50_1bjet}\includegraphics[width=0.45\linewidth]{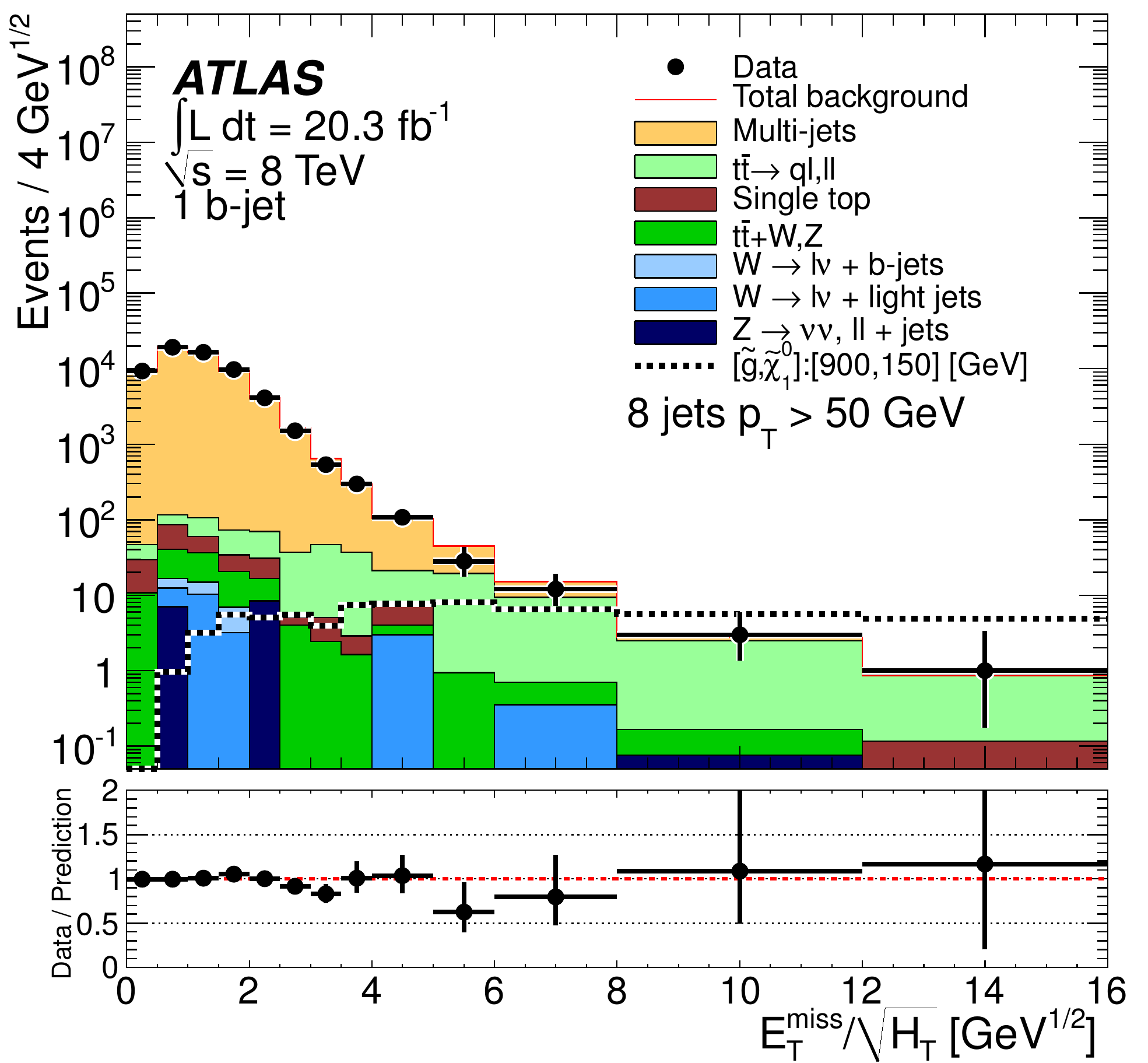}}
\subfloat[~9j50, exactly one $b$-jet]{\label{fig:metsig_9j50_1bjet}\includegraphics[width=0.45\linewidth]{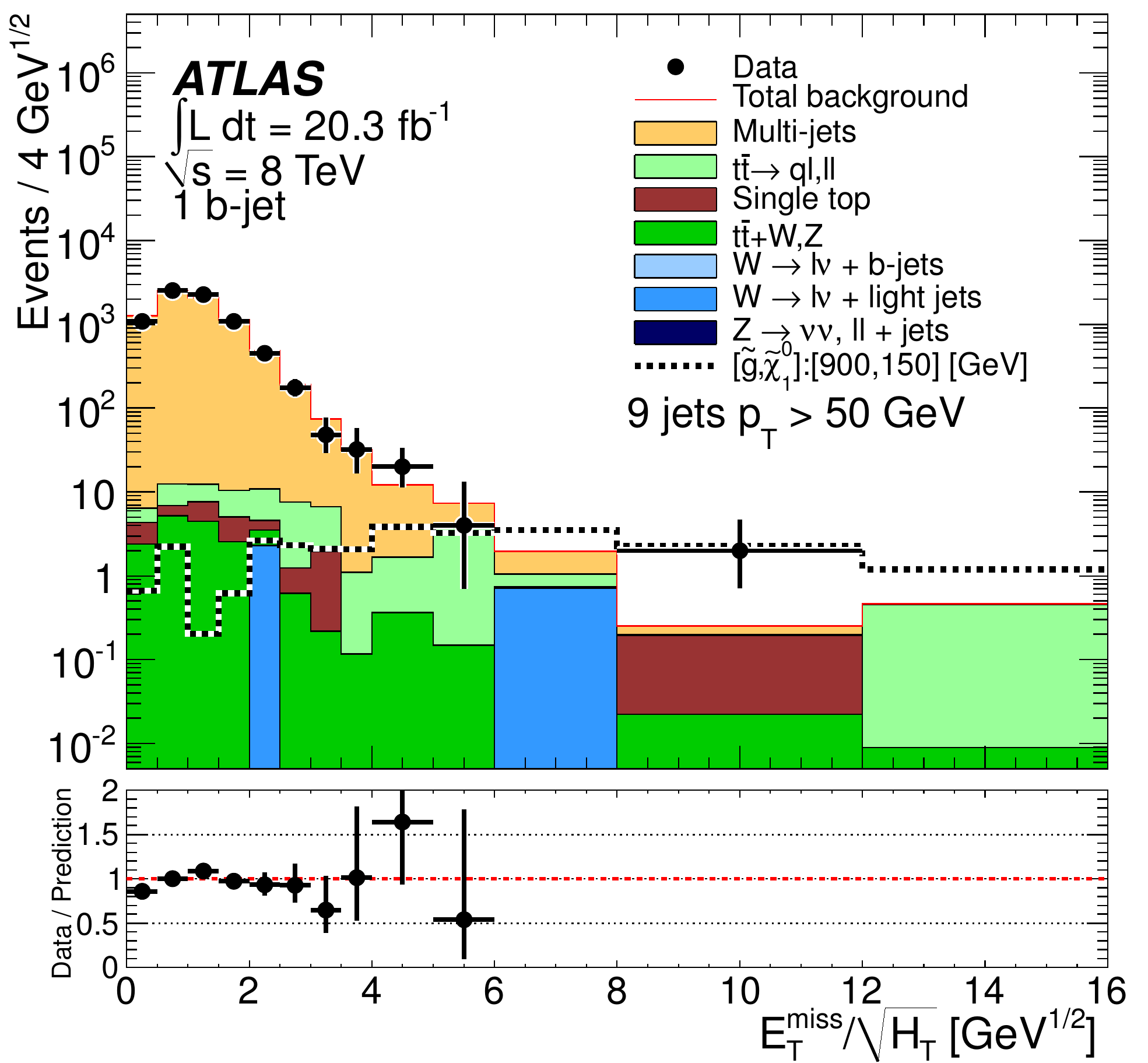}}\\
\subfloat[~8j50, $\geq2$~$b$-jets]{\label{fig:metsig_8j50_2bjet}\includegraphics[width=0.45\linewidth]{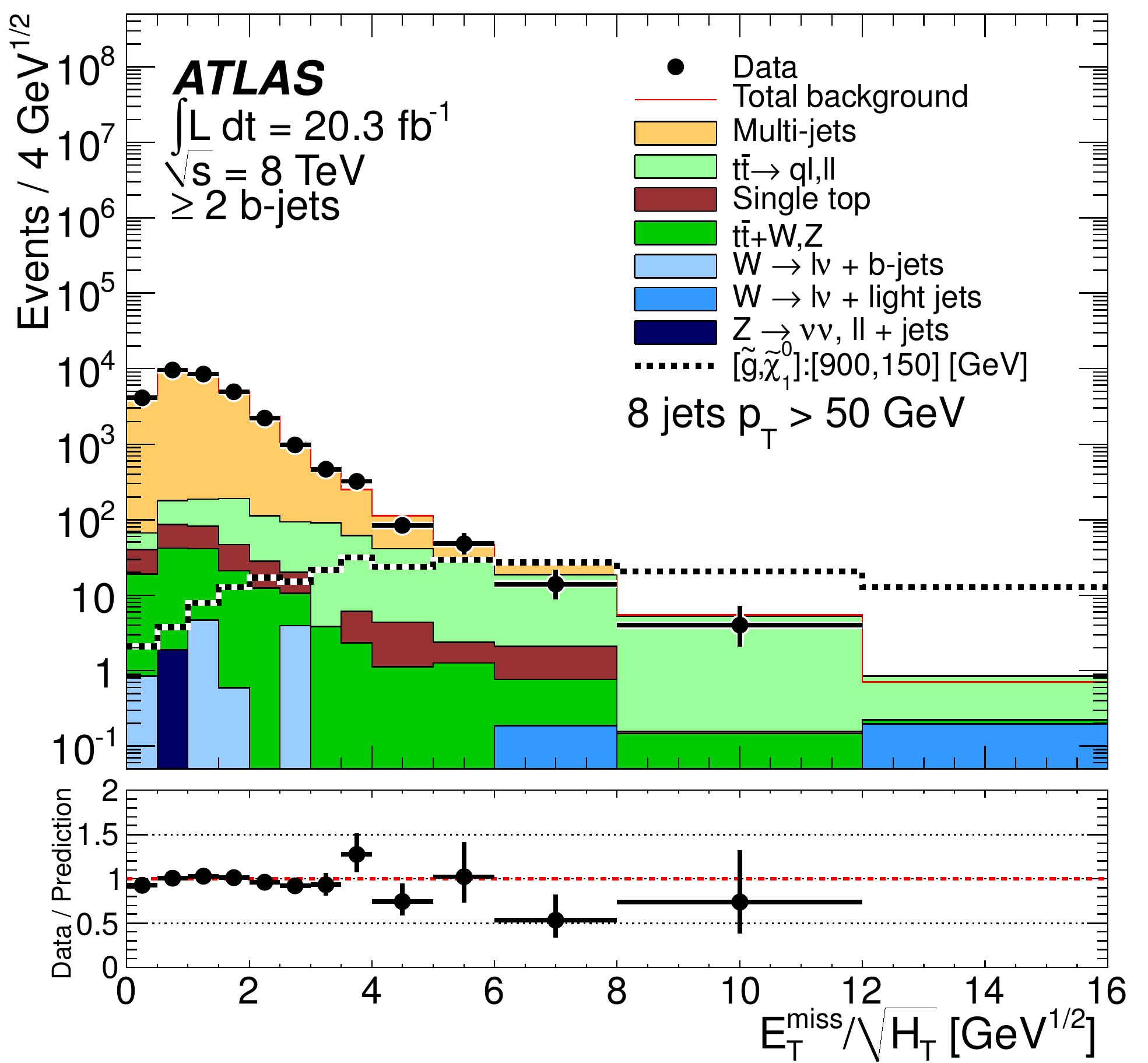}}
\subfloat[~9j50, $\geq2$~$b$-jets]{\label{fig:metsig_9j50_2bjet}\includegraphics[width=0.45\linewidth]{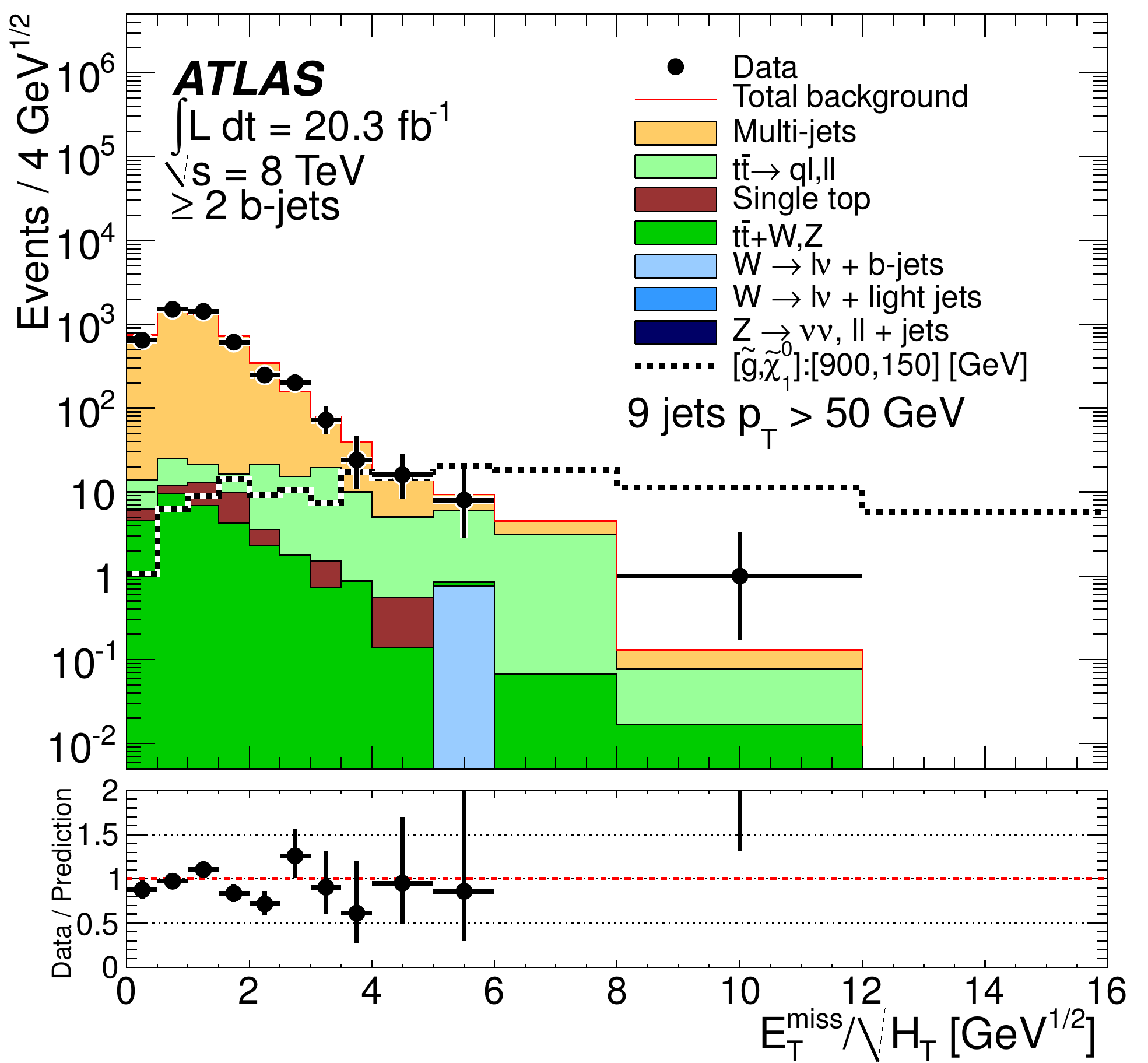}}
\caption{\label{fig:metsig_8j50_9j50}
\metsig{} distributions for the multi-jet + flavour stream with $\pthreshold=50$\,GeV,
and either exactly eight jets (left) or exactly nine jets (right)
with the signal region selection, other than that on \metsig{} itself.
The $b$-jet multiplicity increases from no $b$-jets (top)
to exactly one $b$-jet (middle)
to at least two $b$-jets (bottom). Other details are as
for figure~\ref{fig:metsig_qcd_vr}.
}
\end{figure}

\begin{figure}[h]
\centering
\includegraphics[width=0.55\linewidth]{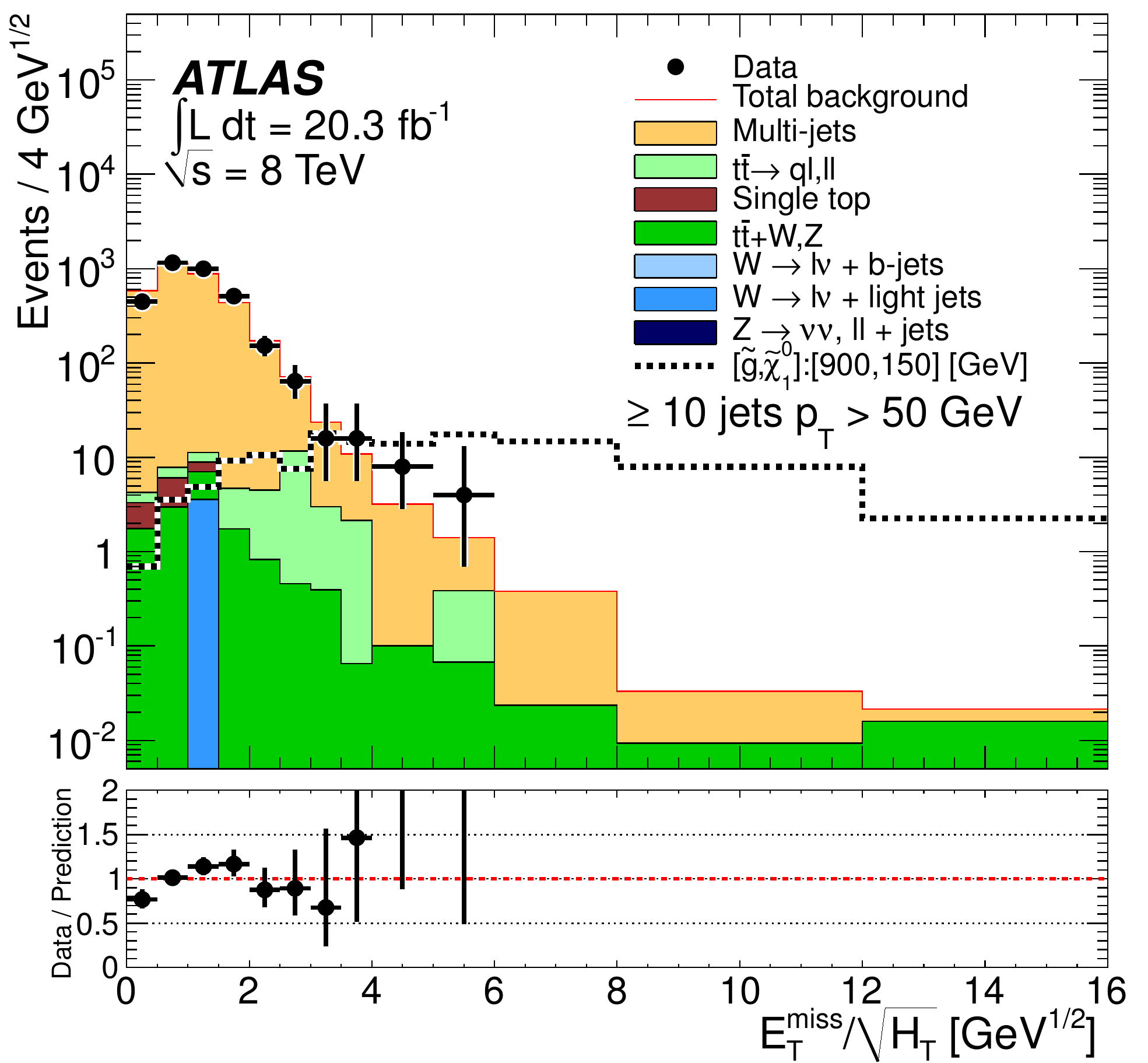}
\caption{\label{fig:metsig_10j50}
\metsig{} distribution for the multi-jet + flavour stream with $\pthreshold=50$\,GeV,
and at least ten jets.
The complete \SR{\geq10j50} selection has been applied, other than
the final \metsig{} requirement. Other details are as
for figure~\ref{fig:metsig_qcd_vr}.
}
\end{figure}

\begin{figure}[h]
\centering
\subfloat[~7j80, no $b$-jets]{\label{fig:metsig_7j80_0bjet}\includegraphics[width=0.45\linewidth]{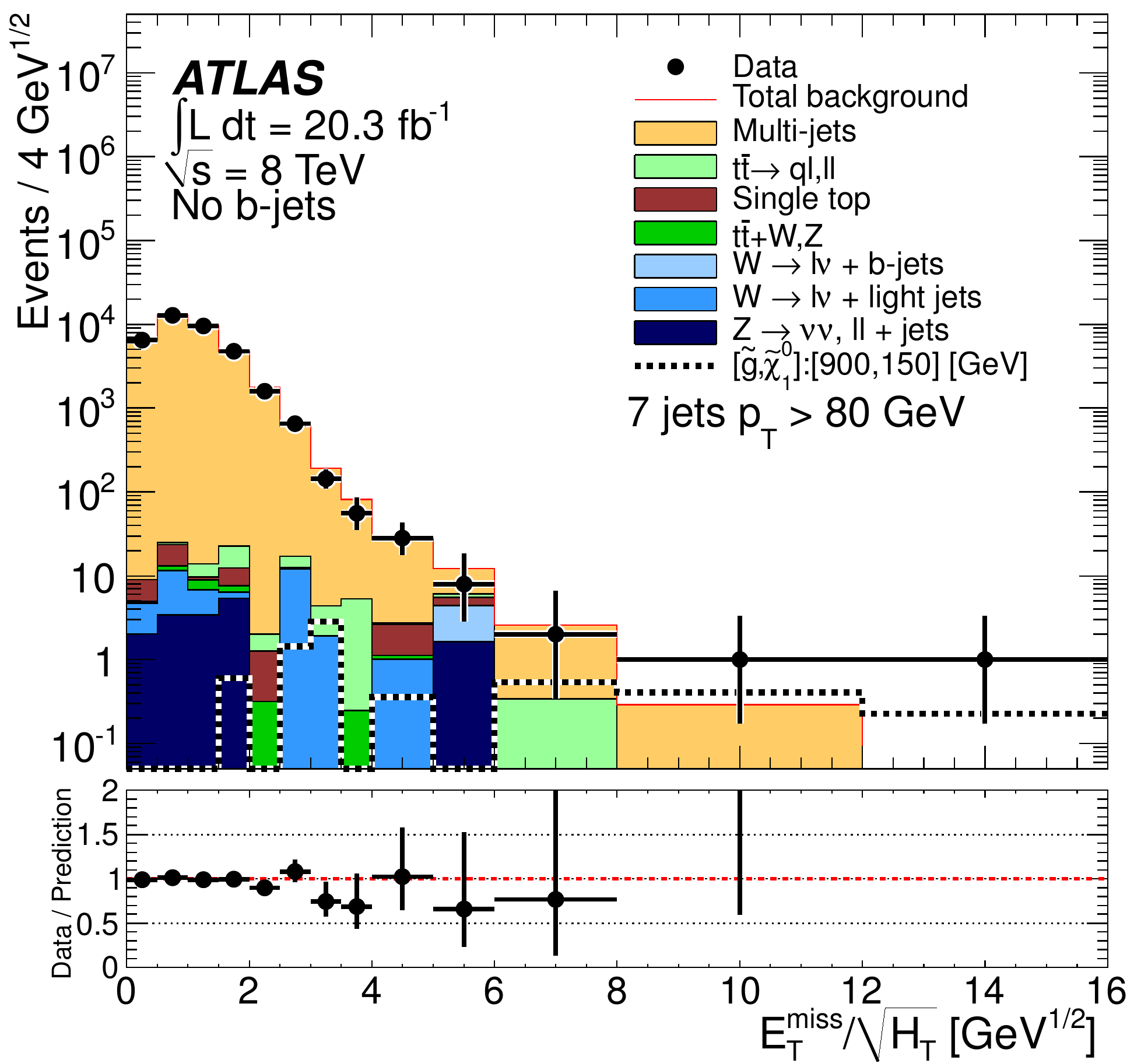}}
\subfloat[~8j80, no $b$-jets]{\label{fig:metsig_8j80_0bjet}\includegraphics[width=0.45\linewidth]{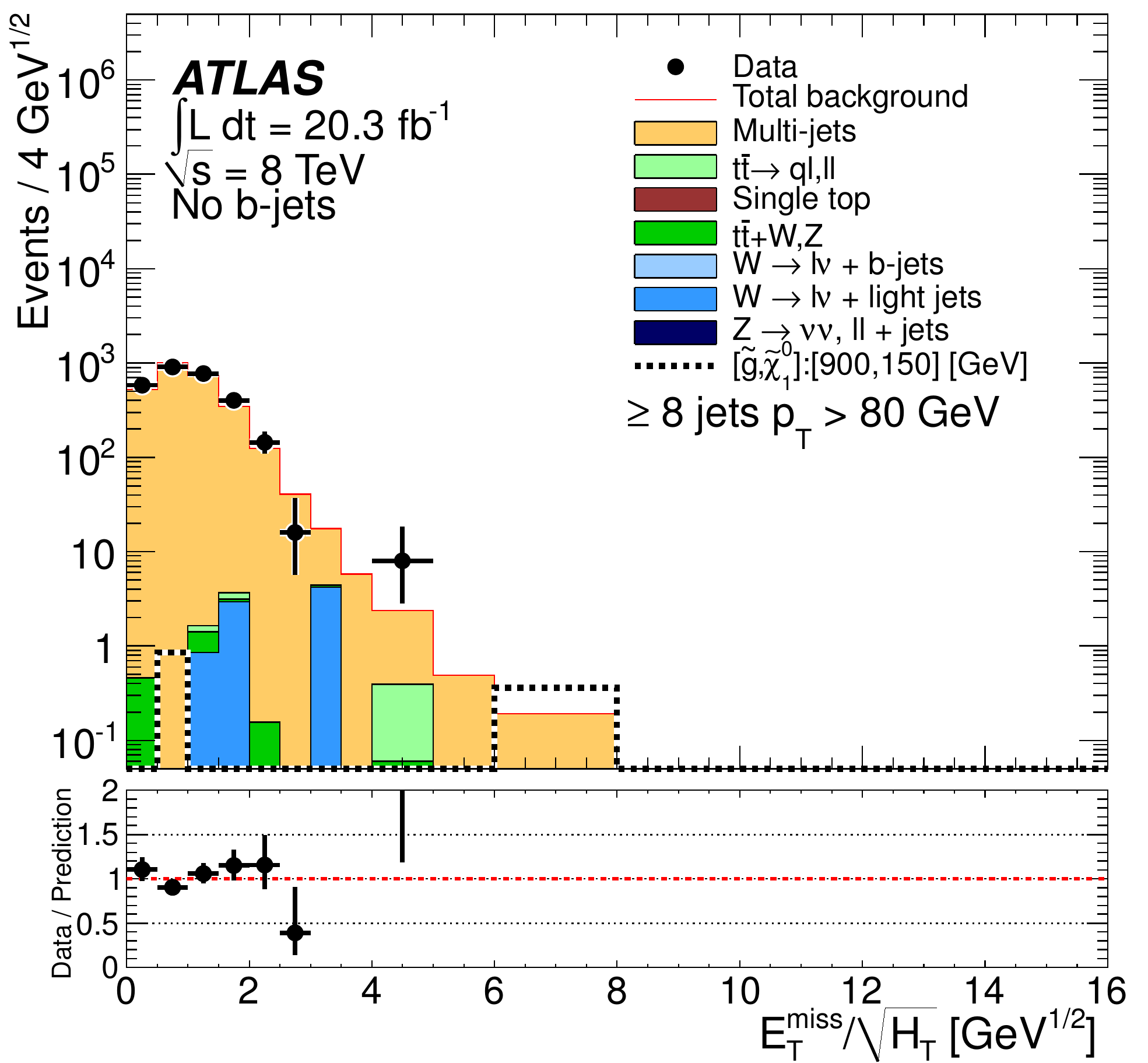}}\\
\subfloat[~7j80, exactly one $b$-jet]{\label{fig:metsig_7j80_1bjet}\includegraphics[width=0.45\linewidth]{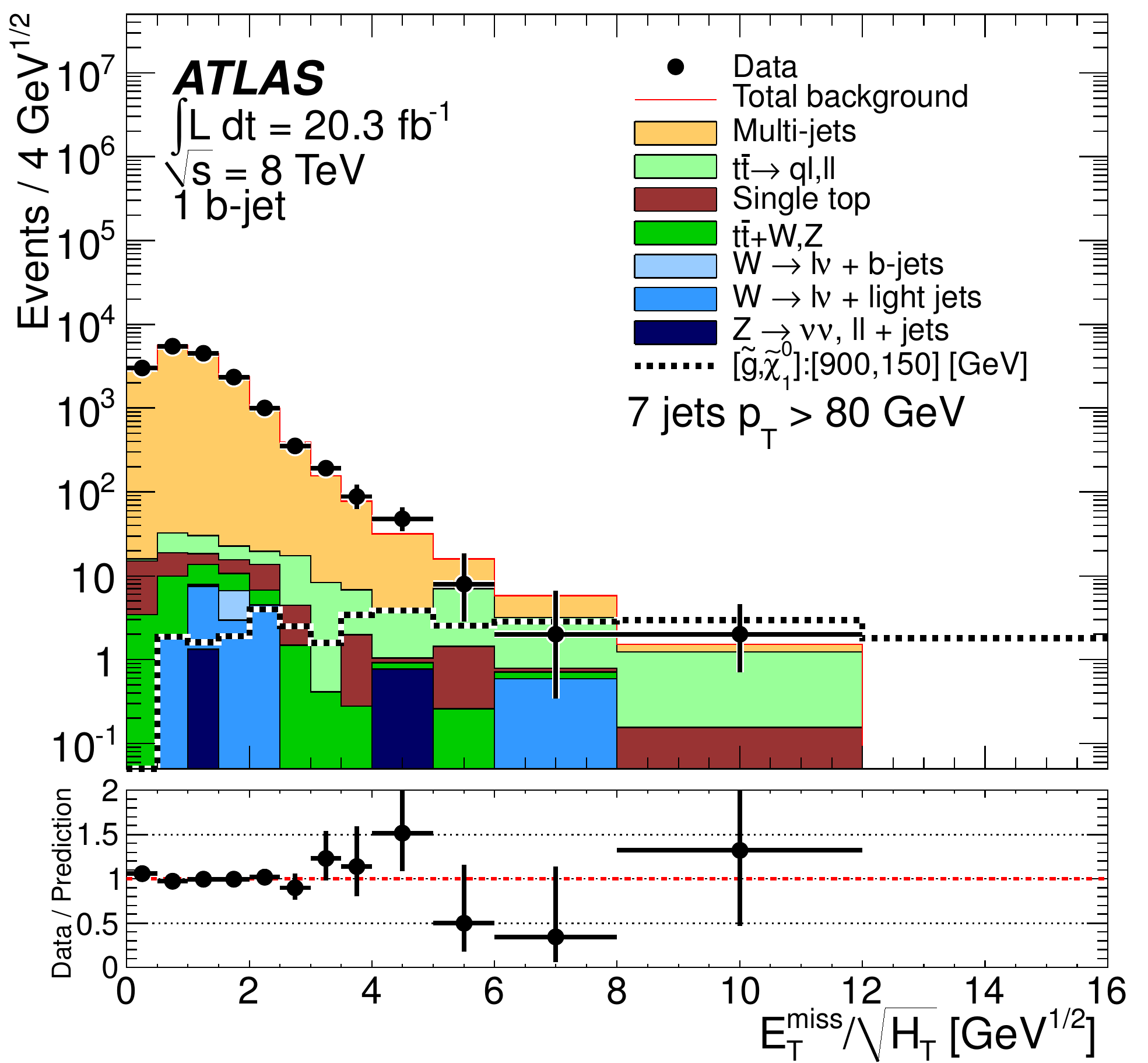}}
\subfloat[~8j80, exactly one $b$-jet]{\label{fig:metsig_8j80_1bjet}\includegraphics[width=0.45\linewidth]{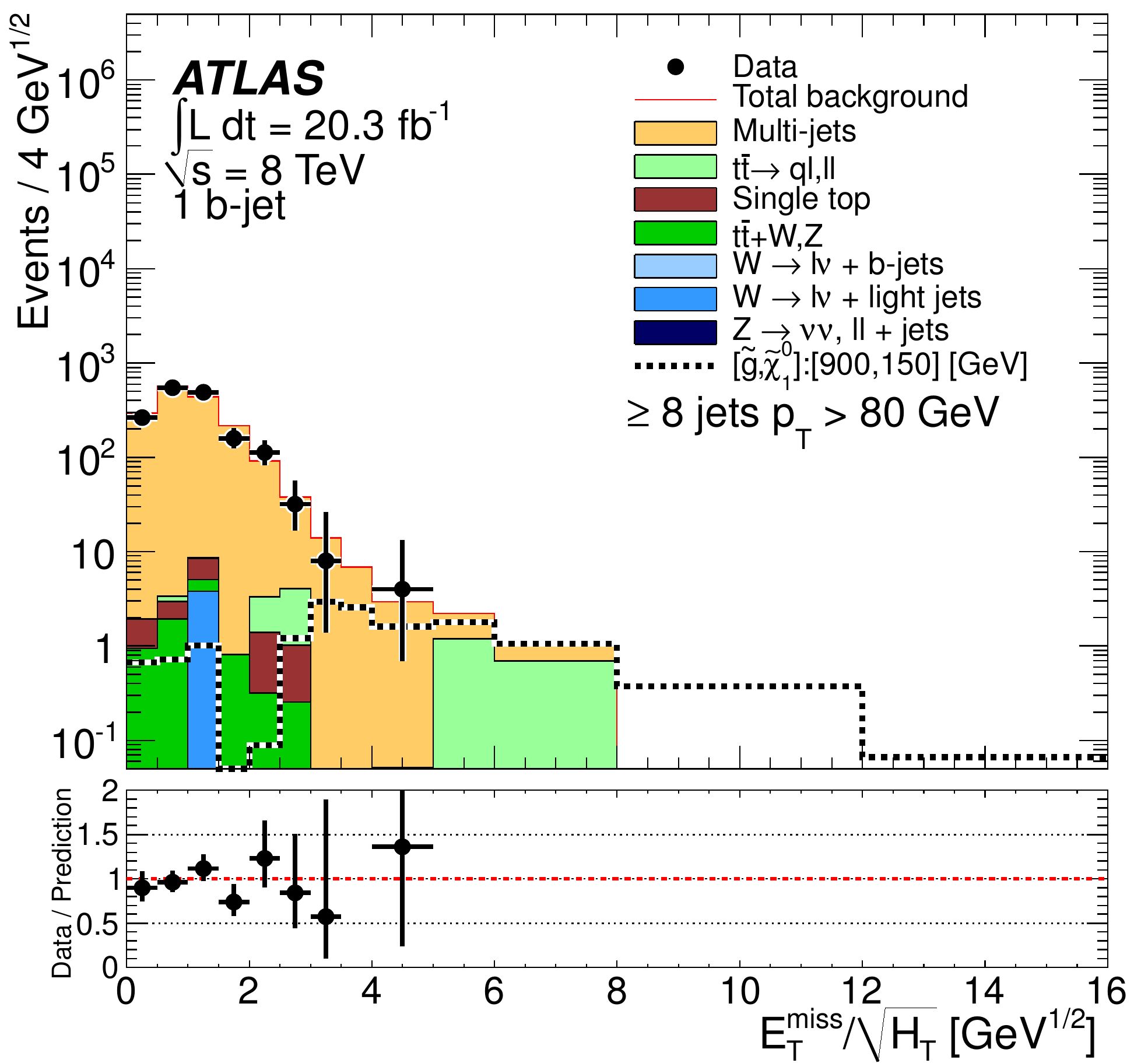}}\\
\subfloat[~7j80, $\geq2$~$b$-jets]{\label{fig:metsig_7j80_2bjet}\includegraphics[width=0.45\linewidth]{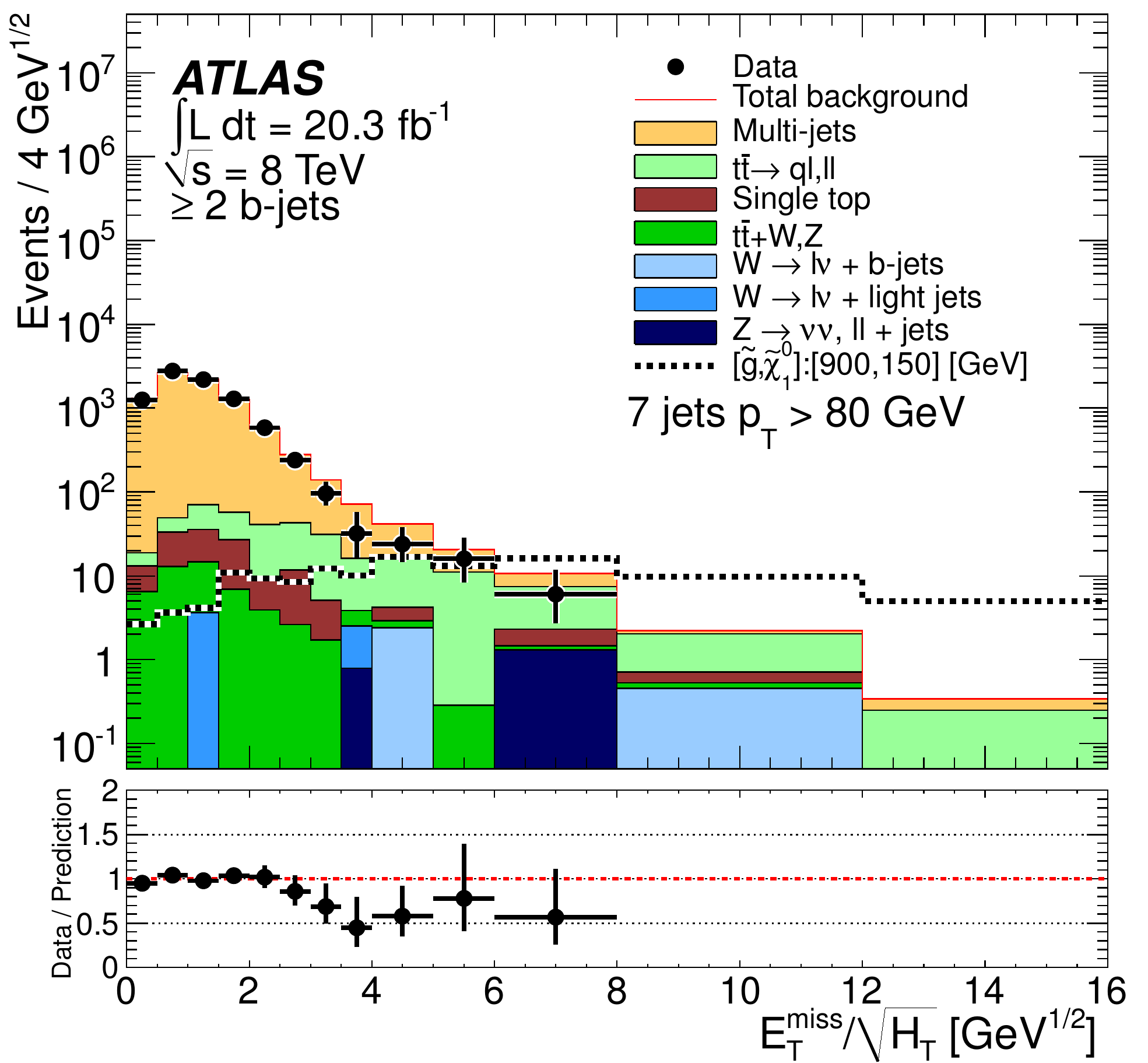}}
\subfloat[~8j80, $\geq2$~$b$-jets]{\label{fig:metsig_8j80_2bjet}\includegraphics[width=0.45\linewidth]{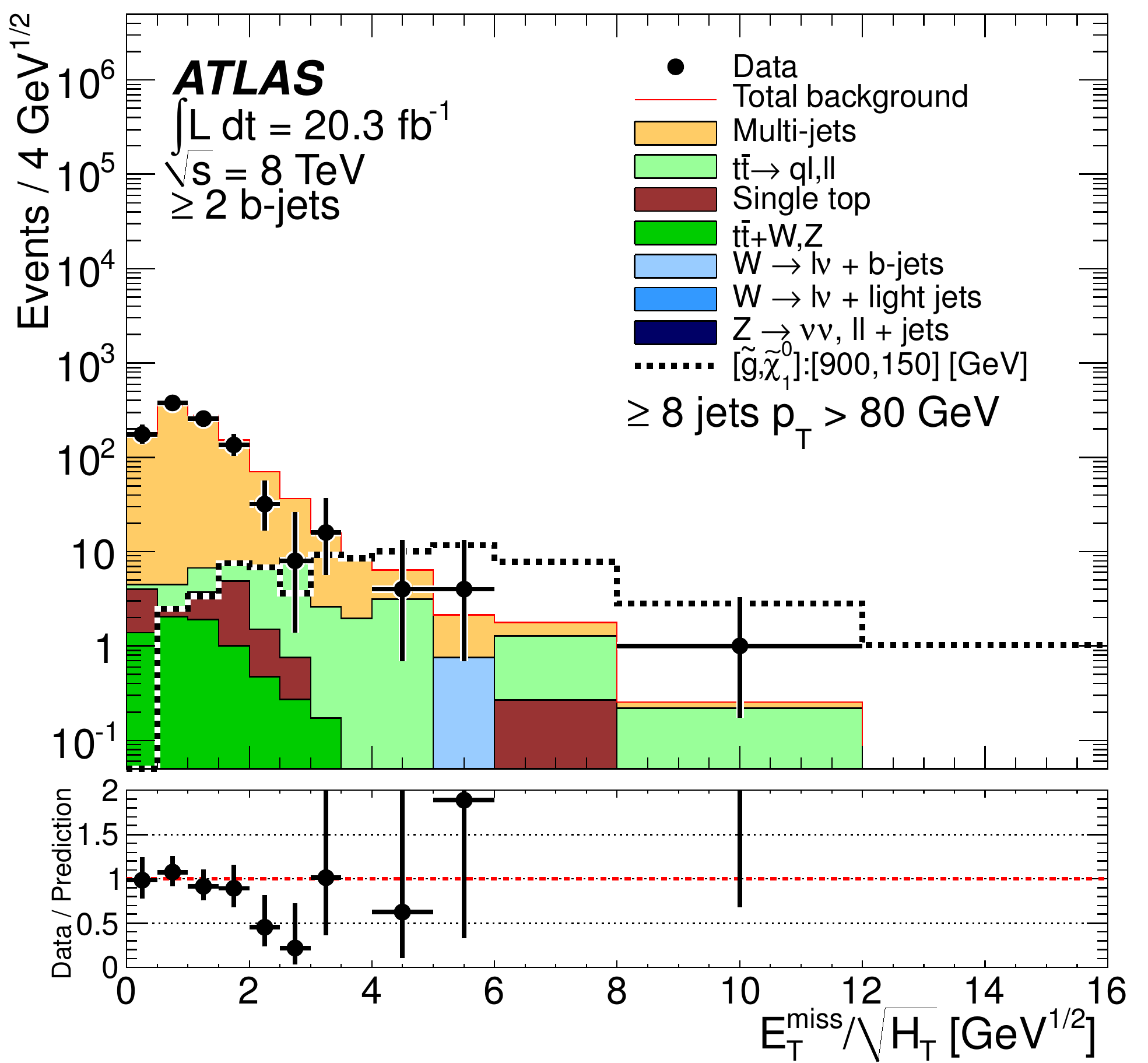}}
\caption{\label{fig:metsig_7j80_8j80}
\metsig{} distributions for the multi-jet + flavour stream with $\pthreshold=80$\,GeV.
The complete signal region selections were applied, other than
the final \metsig{} requirement. Other details are as
for figure~\ref{fig:metsig_qcd_vr}.
}
\end{figure}

\begin{figure}[h]
\centering
{\label{fig:metsig_8j50_mj340}\includegraphics[width=0.45\linewidth]{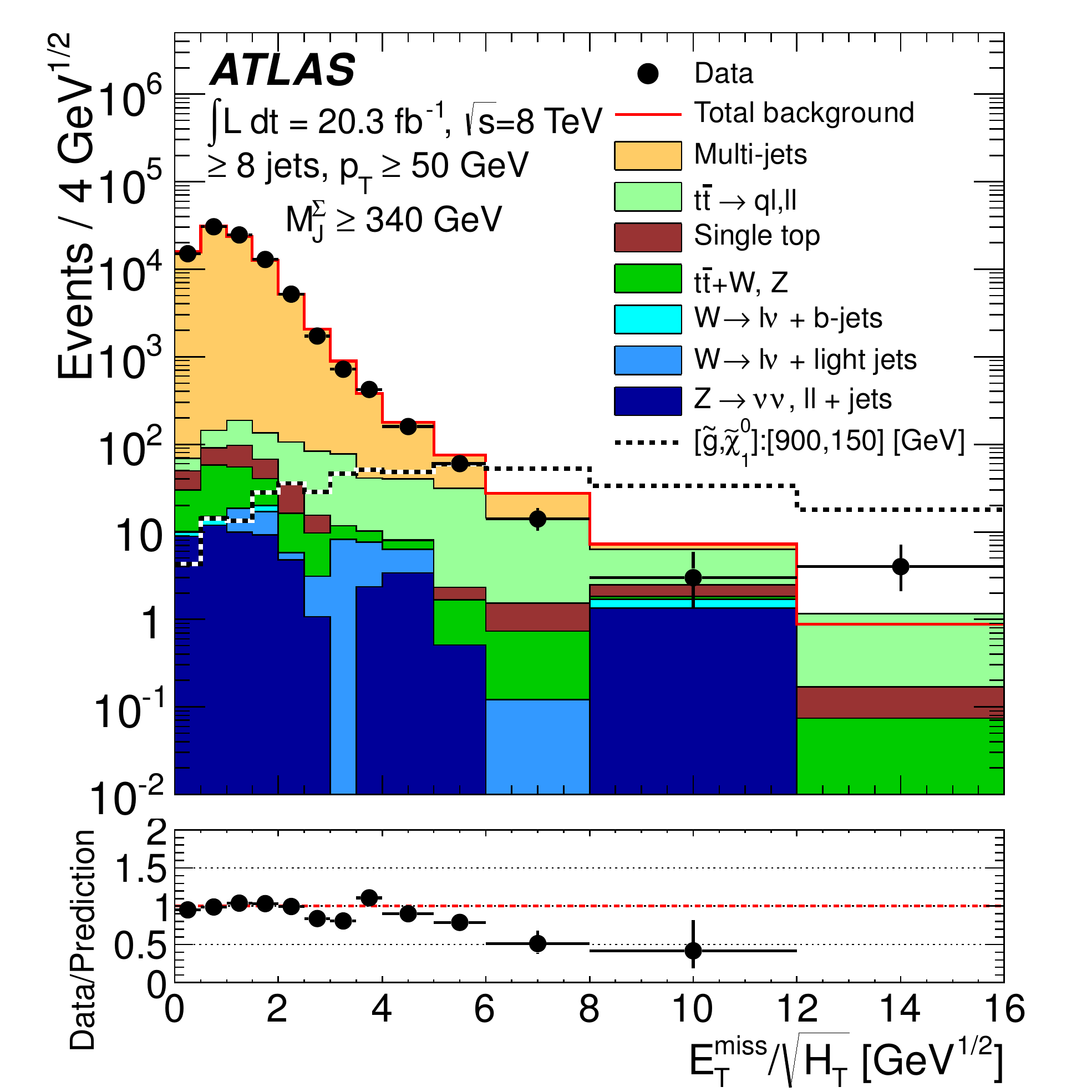}}
{\label{fig:metsig_8j50_mj340}\includegraphics[width=0.45\linewidth]{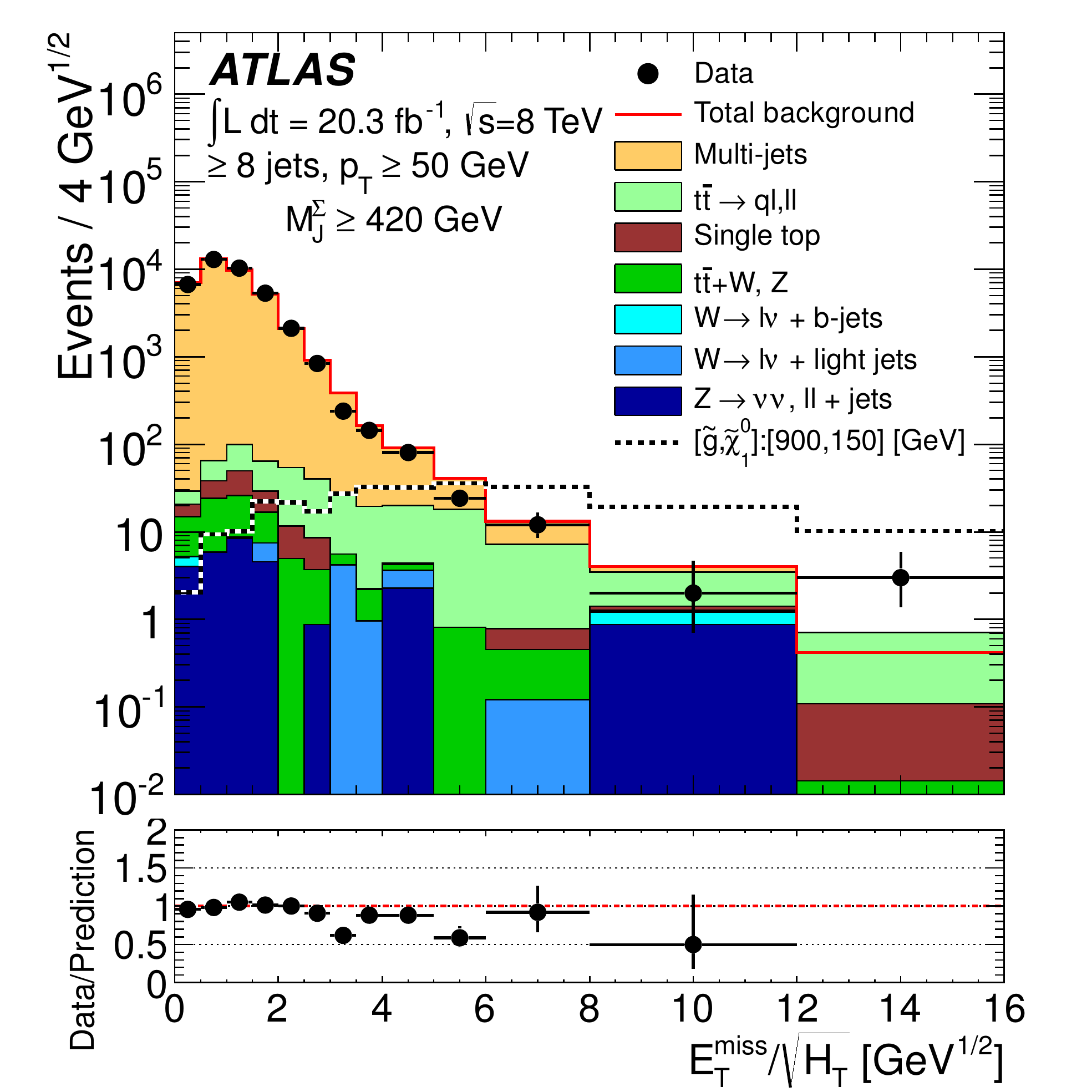}}\\
{\label{fig:metsig_8j50_mj340}\includegraphics[width=0.45\linewidth]{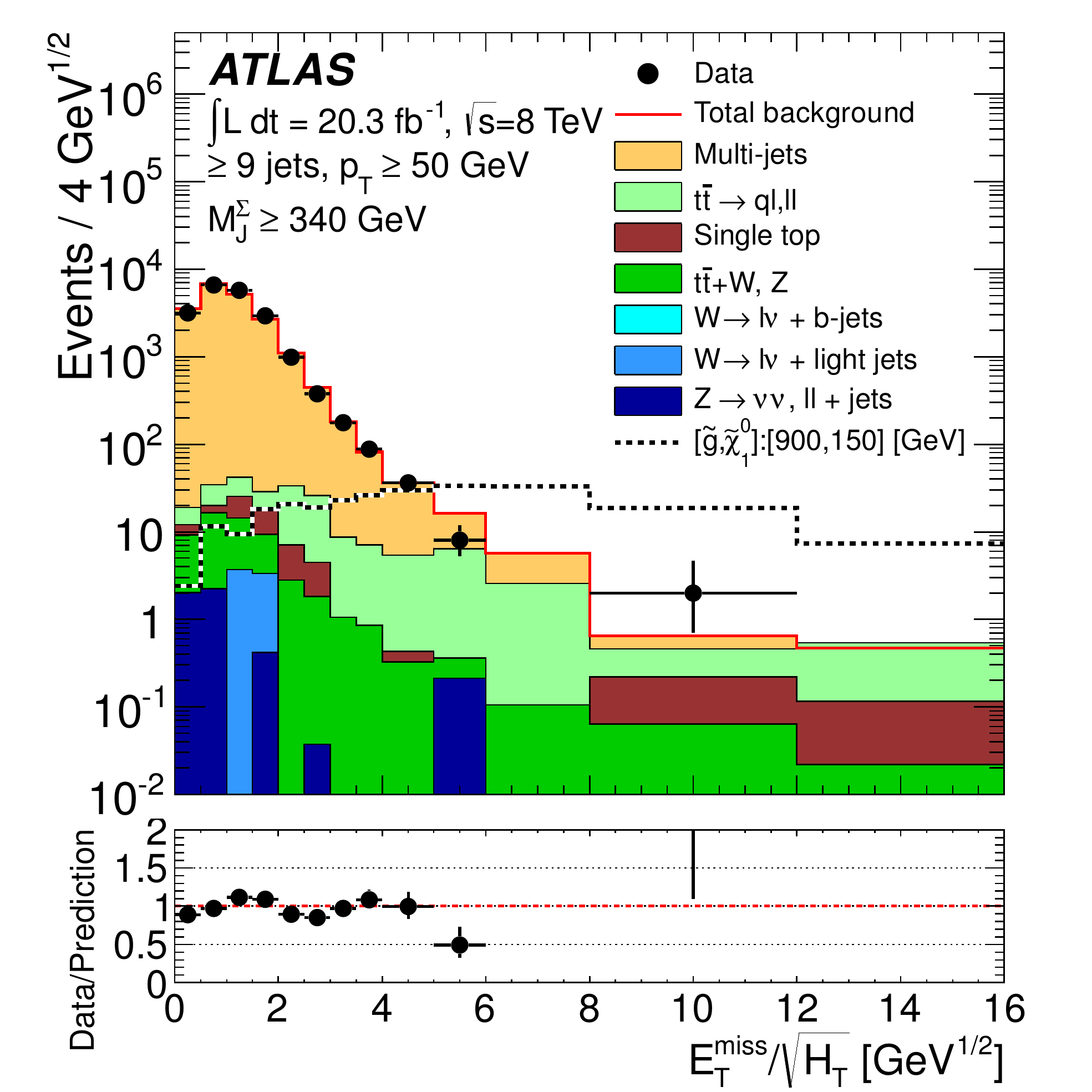}}
{\label{fig:metsig_8j50_mj340}\includegraphics[width=0.45\linewidth]{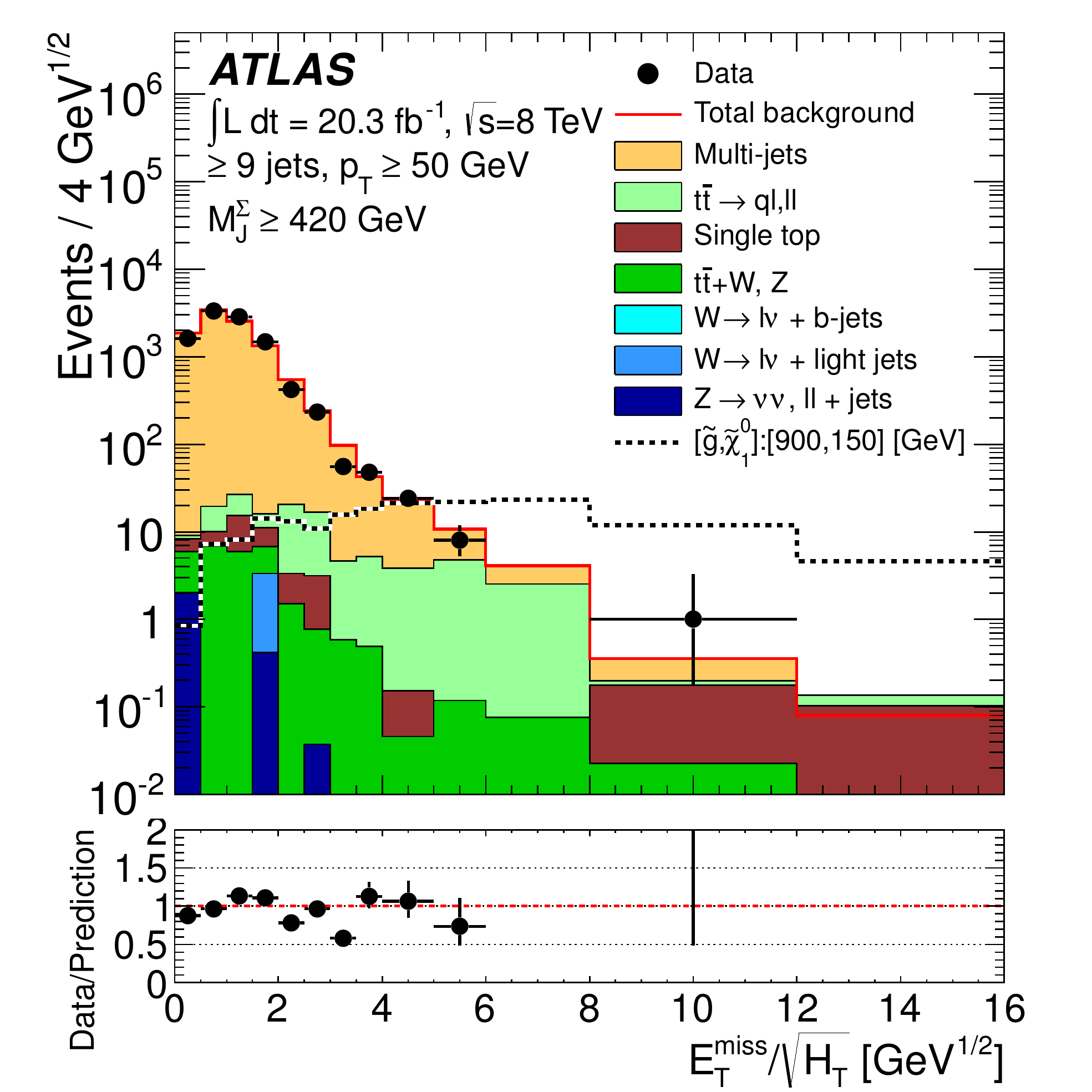}}\\
{\label{fig:metsig_8j50_mj340}\includegraphics[width=0.45\linewidth]{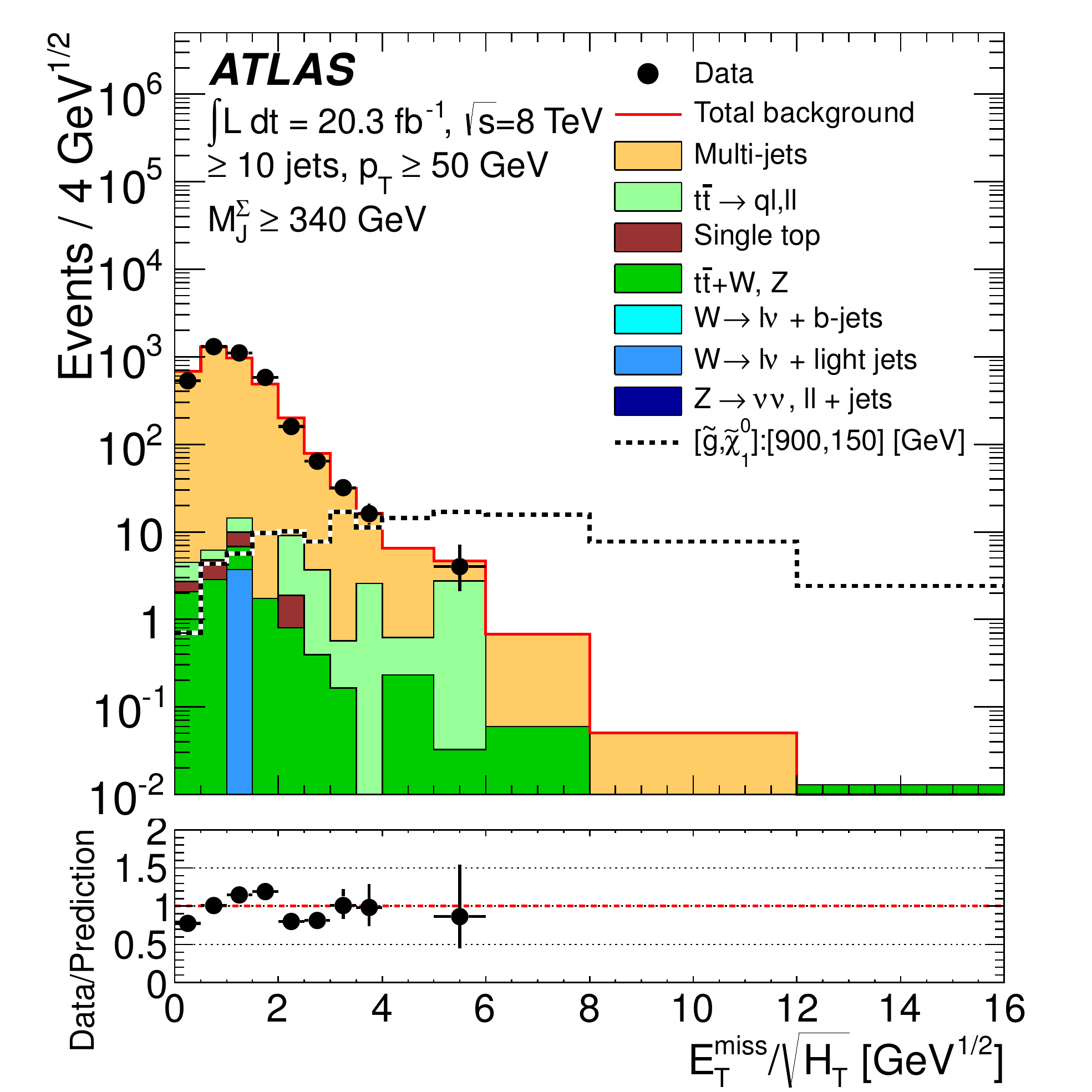}}
{\label{fig:metsig_8j50_mj340}\includegraphics[width=0.45\linewidth]{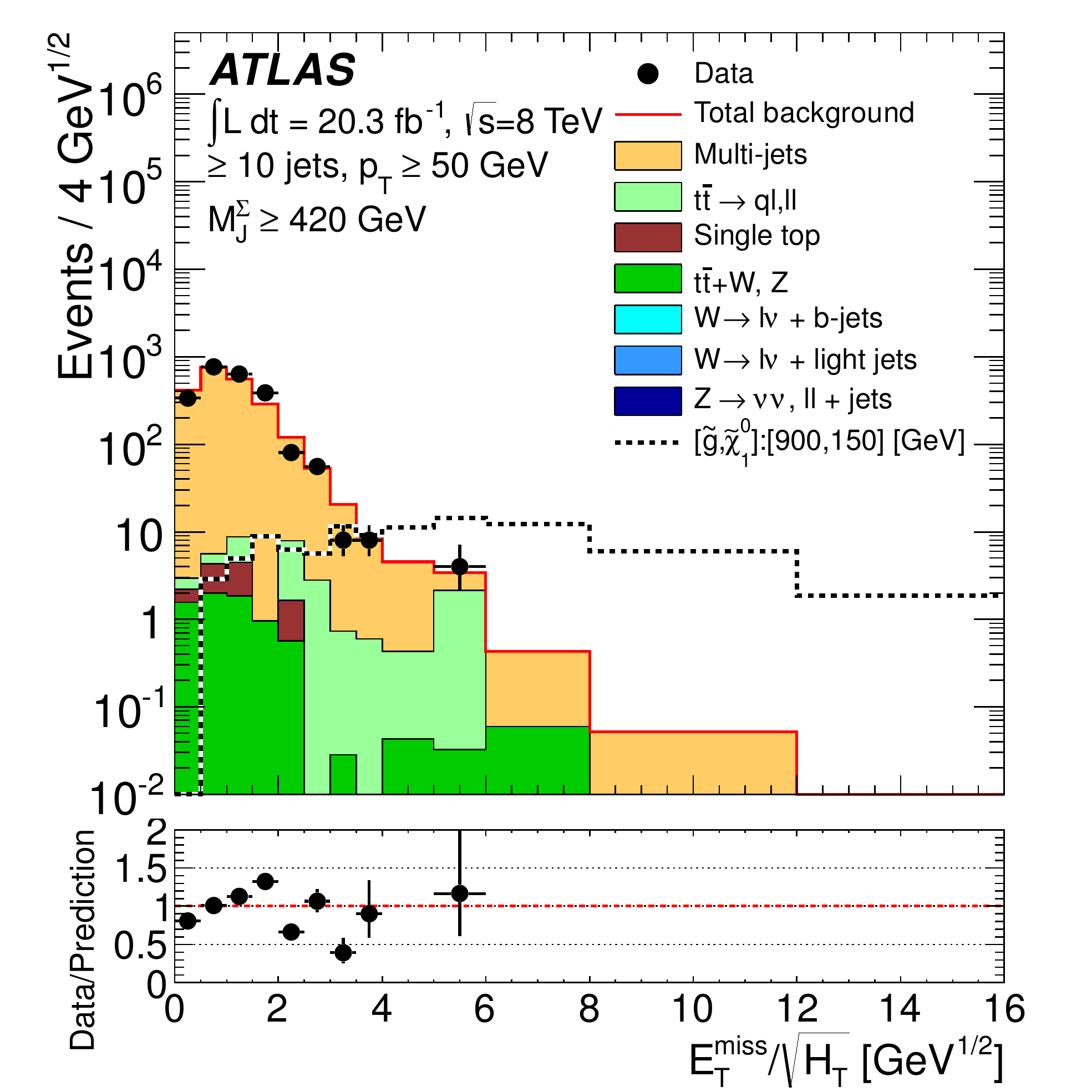}}
\caption{\label{fig:metsig_mj}
\metsig{} distributions for the multi-jet + \MJ{} stream
with the signal region selection, other than the final \metsig{} requirement.
The figures on the left are for events with $\MJ>340\,$GeV, while those on the right 
are for $\MJ>420\,$GeV.
The minimum multiplicity requirement for $\pthreshold=50\,\GeV$, $R=0.4$ jets
increases from eight (top) to nine (middle)
and finally to ten jets (bottom). Other details are as
for figure~\ref{fig:metsig_qcd_vr}.
}
\end{figure}

When evaluating a supersymmetric signal model for exclusion, any
signal contamination in the control regions is taken into
account for each signal point in the control-region fits
performed for each signal hypothesis. Separately,
each signal region (one at a time), along with all control
regions, is also fitted under the background-only hypothesis. 
This fit is used to characterise the agreement in each
signal region with the background-only hypothesis,
and to extract visible cross-section limits and upper limits
on the production of events from new physics. For
these limits, possible signal contamination in the
control regions is neglected. 

\subsection{Simultaneous fit in the multi-jet + flavour stream}
\label{sec:fit:bjet}

The seven $\pthreshold=50$~GeV signal regions (and similarly the six $\pthreshold=80$~GeV signal regions) are
fitted to the background and signal predictions. Correlations from
sample to sample and region to region are taken into account, separately for the $\pthreshold=50$~GeV and
$\pthreshold=80$~GeV signal regions. Systematic
uncertainties arising from the same source are treated as fully
correlated. 

The fit considers several independent background components:
\begin{itemize}
\item \ttbar{} and $W$\,+\,jets. One control region is defined for
  each signal region, as described in table~\ref{tab:one_lepton_cr}; the
  normalisation of each background component is
  allowed to vary freely in the fit. 
\item Less significant backgrounds ($Z$\,+\,jets, $\ttbar$\,+\,$W$,  
$\ttbar$\,+\,$Z$, and single top) are determined using Monte Carlo simulations. These are individually
  allowed to vary within their uncertainties.
\item Multi-jet background. Being data-driven, it is not constrained
  in the fit by any control region. It is constrained in the signal
  regions by its uncertainties, which are described in
  \Secref{sec:backgrounds:multijet:systematic}. 
\end{itemize}

The systematic effects, described in sections~\ref{sec:backgrounds:multijet:systematic} and
\ref{sec:backgrounds:leptonic:systematic}, are treated as nuisance
parameters in the fit. For the signal, the dominant systematic effects
are included in the fit; these are the jet energy scale and
resolution uncertainties, the $b$-tagging efficiency uncertainties, and the
theoretical uncertainties. 

\subsection{Simultaneous fit in the multi-jet + ${\boldsymbol\MJ{}}$ stream}
\label{sec:fit:mj}

For the multi-jet + \MJ{} signal regions, a separate fit is performed
on each signal region to adjust the normalisation of
the \ttbar{} and $W$\,+\,jets backgrounds using control regions, as
defined in table~\ref{tab:one_lepton_cr}. 

The systematic uncertainties affecting the background, described in
sections~\ref{sec:backgrounds:multijet:systematic} and
\ref{sec:backgrounds:leptonic:systematic}, and signal
predictions are treated as nuisance parameters in the fit. The
dominant sources of uncertainty are
considered for the signal predictions: the jet energy scale, the jet energy resolution, and
the theoretical uncertainties.

\subsection{Fit results}
\label{sec:fit:results}

Tables \ref{tab:results:nj50}--\ref{tab:results:mj} summarise the
 fit results; the number of events observed in each of the signal
regions, as well as their Standard Model background expectations, are reported
before and after the fit to the control regions. In each of the signal 
regions, agreement is found between the Standard Model prediction and 
the data. The fit results are checked
for stability and consistency with the background
modelling based on the predictions described in sections~\ref{sec:backgrounds:multijet} and
\ref{sec:backgrounds:leptonic}. There is no indication of a
systematic mis-modelling of any of the major backgrounds; the fitted
values are in all cases consistent with the
Monte Carlo simulation predictions.

In addition to the event yields, the probability
($p_0$-value) that a background-only pseudo-experiment
is more signal-like than the observed data is given for each individual signal
region. To obtain these $p_0$-values, the fit in the signal region proceeds in the same way as the
control-region-only fit, except that the number of events observed in
the signal region is included as an input to the fit. Then, an
additional parameter for the non-Standard-Model signal strength,
constrained to be non-negative, is fitted. The significance ($\sigma$) of
the agreement between data and the Standard Model prediction is
given. No significant deviations from the Standard Model prediction are found.
The 95\% confidence level (CL) upper limit on the number of events
($N\rm{^{95\%}_{BSM}}$) and the cross section
times acceptance times efficiency ($\sigmamaxbsm$) from
non-Standard-Model production are also provided, neglecting in the fit possible
signal contamination in the control regions.

\begin{sidewaystable}[h]
\centering
\footnotesize
\begin{tabular}{| l || c | c | c || c | c | c || c |}
\hline
Signal region  & \multicolumn{3}{| c ||}{\BSR{8j50}} & \multicolumn{3}{| c ||}{\BSR{9j50}} & \BSR{10j50} \\
\hline
$b$-jets       & 0 & 1 & $\geq 2$ & 0 & 1 & $\geq 2$ & --- \\
\hline\hline
        Observed events  		 &  $40$
        &  $44$               		 &  $44$
        &  $5$               		 &  $8$
        &  $7$               		 &  $3$                    \\ \hline\hline
        Total events after fit 		 &  $35 \pm 4$
        &  $40 \pm 10$           		 &  $50 \pm 10$
        &  $3.3 \pm 0.7$           		 &  $6.1 \pm 1.7$
        &  $8.0 \pm 2.7$           		 &  $1.37 \pm 0.35$
        \\ \hline
        Fitted $t\bar{t}$  		 &  $2.7 \pm 0.9$
        &  $11.8 \pm 3.0$           		 &  $23.0 \pm 5.0$
        &  $0.36 \pm 0.18$           		 &  $1.5 \pm 0.5$
        &  $3.2 \pm 1.1$           		 &
        $0.06_{-0.06}^{+0.09}$              \\
        Fitted $W$+jets  		 &  $2.0_{-2.0}^{+2.6}$
        &  $0.62_{-0.62}^{+0.81}$           		 &
        $0.20_{-0.20}^{+0.28}$           		 &  $-$
                   		 &  $0.24_{-0.24}^{+0.65}$
        &  $-$           		 &  $-$
        \\
        Fitted others  			 & $ 2.9 _{- 1.8 }^{+ 1.8
        }$ 			 & $ 1.7 _{- 1.2 }^{+ 1.5 }$
        & $ 2.8 _{- 2.0 }^{+ 2.3 }$ 			 & $ 0.03 \pm
        0.03 $ 			 & $ 0.38 \pm 0.25 $
        & $ 0.40 _{- 0.24 }^{+ 0.60 }$ 			 & $ 0.08 \pm
        0.08 $ \\ \hline \hline
        Total events before fit  	 &  $36  $ 	 &  $48  $
        &  $59  $ 	 &  $3.4  $ 	 &  $6.6$ 	 &  $8.9$
        &  $1.39  $
\\ \hline 

        $t\bar{t}$ before fit  		 &  $3.5$
        &  $15$           		 &  $30$
        &  $0.41$           		 &  $1.8$
        &  $4 $           		 &  $0.08$              \\
        $W$+jets before fit  		 &  $2.9$           		 &  $1.0$           		 &  $0.29$           		 &  $-$           		 &  $0.40$           		 &  $-$          		 &  $-$             \\
        Others before fit 			 & $ 2.4$ 			 & $ 1.8$ 			 & $ 2.8$ 		 & $ 0.03$ 			 & $ 0.34$ 			 & $ 0.4$ 			 & $ 0.08$ \\

\hline \hline
        Multi-jets  			 &  $27 \pm 3$
        &  $30 \pm 10$           			 &  $26 \pm
        10$           			 &  $3.0 \pm 0.6$
        &  $4.0 \pm 1.4$           			 &  $4.4 \pm
        2.2$           			 &  $1.23 \pm 0.32$
        \\ \hline \hline

{  $N\rm{^{95\%}_{BSM}}$ (exp)} &
16 &   
23 & 
26 &    
5 &   
7 &  
8 &  
4 \\ \hline   
{  $N\rm{^{95\%}_{BSM}}$ (obs)} &
20  &  
23  & 
22  & 
7   &  
9   &   
7   &   
6   \\ \hline  \hline
{  $\sigmamaxbsm$ (exp) [fb]} &
0.8 & 
1.2 & 
1.3 & 
0.26 & 
0.36 & 
0.40 & 
0.19  \\ \hline 
{  $\sigmamaxbsm$ (obs) [fb]} & 
0.97 & 
1.1 & 
1.1 & 
0.34 & 
0.43 & 
0.37 & 
0.29 \\ \hline \hline
 $p_0$  & 0.24 & 0.5 & 0.7 & 0.21 & 0.28 &
0.6 & 0.13 \\ \hline
 Significance ($\sigma$) & 0.7   & $-0.02$ & $-0.6$ &
0.8 & 0.6  & $-0.28$ & 1.14 \\ \hline    
\end{tabular}
\caption{\label{tab:results:nj50}
Number of observed and expected (fitted) events for the seven 
$\pthreshold=50\,\GeV$ signal regions of the multi-jet + flavour stream. 
The category indicated by `others' includes the contributions from $Z$\,+\,jets, $\ttbar$\,+\,$W$,  
$\ttbar$\,+\,$Z$, and single top. The table also contains for each
signal region the probability,
$p_0$, that a background-only pseudo-experiment
is more signal-like than the observed data; the significance, $\sigma$, of
the agreement between data and the Standard Model prediction; the 95\%
CL upper limit on the number of events, $N\rm{^{95\%}_{BSM}}$,
originating from sources other than the Standard Model; and the
corresponding cross section
times acceptance times efficiency, $\sigmamaxbsm$.
}
\end{sidewaystable}

\begin{table}[h]
\centering
\footnotesize
\begin{tabular}{| l || c | c | c || c | c | c |}
\hline
Signal region  & \multicolumn{3}{| c ||}{\BSR{7j80}} & \multicolumn{3}{| c |}{\BSR{8j80}} \\
\hline
$b$-jets       & 0 & 1 & $\geq 2$ & 0 & 1 & $\geq 2$ \\
\hline\hline
        Observed events  		 &  $12$
        &  $17$               		 &  $13$
        &  $2$               		 &  $1$
        &  $3$                    \\ \hline \hline
        Total fitted events  		 &  $11.0 \pm 2.2$
        &  $17 \pm 6$           		 &  $25 \pm 10$
        &  $0.9 \pm 0.6$           		 &  $1.5 \pm 0.9$
        &  $3.3 \pm 2.2$              \\ \hline
        Fitted $t\bar{t}$  		 &  $0.00_{-0.00}^{+0.26}$
        &  $5.0 \pm 4.0$           		 &  $12 \pm 9$
        &  $0.10_{-0.10}^{+0.14}$           		 &
        $0.32_{-0.32}^{+0.67}$           		 &
        $1.5_{-1.5}^{+1.9}$              \\
        Fitted $W$+jets  		 &  $0.07_{-0.07}^{+0.38}$
        &  $0.29_{-0.29}^{+0.37}$           		 &  $-$
                  		 &  $-$
        &  $-$           &  $-$
        \\
        Fitted others  			 & $ 1.9 _{- 0.9 }^{+ 1.1 }$
        & $ 0.71 _{- 0.25 }^{+ 0.31 }$ 			 & $ 2.6 _{-
          1.1 }^{+ 1.7 }$ 			 & $ 0.02 \pm 0.02 $
        & $ 0.02 \pm 0.02 $ 			 & $ 0.32 _{- 0.21
        }^{+ 0.36 }$ \\ \hline \hline
        Total events before fit  	 &  $11.7  $ 	 &  $16  $
        &  $23  $ 	 &  $0.8  $ 	 &  $1.8  $ 	 &  $3.3  $
        \\ \hline 
        $t\bar{t}$ before fit 		 &  $0.34$           		 &  $4$           		 &  $10$           		 &  $0.08$           		 &  $0.6$           		 &  $1.5$              \\
        $W$+jets before fit 		 &  $0.46$
        &  $0.29$           		 &  $-$
        &  $-$           		 &  $-$          		 &  $-$              \\
        Others before fit  			 & $ 1.8$
        & $ 0.89$ 			 & $ 3.0$
        & $ 0.02$ 			 & $ 0.02$
        & $ 0.35$ \\ \hline \hline
        Multi-jets  			 &  $9.1 \pm 1.6$
        &  $11 \pm 4$           			 &  $10 \pm
        4$           			 &  $0.75 \pm 0.56$
        &  $1.2 \pm 0.5$           			 &  $1.4 \pm
        1.0$              \\ \hline \hline
{  $N\rm{^{95}_{BSM}}$ (exp)} &
10    & 
17   & 
14  & 
4    &
4    &
6    \\ \hline 
{  $N\rm{^{95}_{BSM}}$ (obs)} &
10   &  
16  & 
12  & 
5   & 
3.5   &   
6     \\ \hline \hline
{  $\sigmamaxbsm $ (exp) [fb]} &
0.5   & 
0.8   & 
0.7   & 
0.18   & 
0.18   & 
0.31  \\ \hline
{  $\sigmamaxbsm $ (obs) [fb]} & 
0.5   & 
0.8   & 
0.6   & 
0.24   & 
0.17   & 
0.31  \\ \hline \hline
 $p_0$  & 0.5 & 0.6 & 0.8 & 0.19 & 0.6 &
0.5 \\ \hline
 Significance ($\sigma$) & 0.05 & $-0.14$ & $-1.0$ & 0.9
 & $-0.28$ & $-0.06$ \\ \hline
\end{tabular}
\caption{\label{tab:results:nj80}
As for \tabref{tab:results:nj50} but for the six signal regions 
for which $\pthreshold=80\,\GeV$.
}
\end{table}

\begin{table}[h]
\centering
\renewcommand\arraystretch{1.2}
\footnotesize
\begin{tabular}{| l || c | c |}
\hline
Signal region  & \multicolumn{2}{| c |}{\BSR{8j50}}  \\
\hline
\MJ{}~[GeV]         & 340 & 420  \\
\hline\hline
       Observed events  		 &  $69$                         &  $37$                          \\ \hline\hline
        Total events after fit           &  $75 \pm 19$            &  $45 \pm 14$             \\ \hline
        Fitted $t\bar{t}$  		 &  $17 \pm 11$            &  $16 \pm 13$             \\ 
        Fitted $W$+jets  		 &  $0.8_{-0.8}^{+1.3}$       &  $0.4_{-0.4}^{+0.7}$        \\ 
        Fitted others  			 & $ 5.2 _{- 2.5 }^{+ 4.0 }$
        & $ 2.8 _{- 1.6 }^{+ 2.9 }$   \\ \hline \hline
        Total events before fit  	 &  $85  $
        &  $44  $                     \\ \hline
        $t\bar{t}$ before fit 		 &  $27$            &  $14$             \\ 
        $W$+jets before fit 		 &  $0.8$       &  $0.4$        \\ 
        Others before fit 			 & $ 5 $
        & $ 2.8$   \\ \hline \hline

        Multi-jets  			 &  $52 \pm 15$            &
        $27 \pm 7$              \\ \hline \hline

{ $N\rm{^{95\%}_{BSM}}$ (exp)}    &
40 &
23 \\ \hline 
{ $N\rm{^{95\%}_{BSM}}$ (obs)}  &
35 &
20 \\ \hline \hline
{ $\sigmamaxbsm $ (exp) [fb]}  &
1.9  & 
1.1  \\ \hline
{ $\sigmamaxbsm $ (obs) [fb]}  &
1.7  & 
1.0  \\ \hline \hline
 $p_0$  &  0.60 & 0.7  \\ \hline
 Significance ($\sigma$)  &
 $-0.27$ & $-0.6$  \\ \hline
\end{tabular}
\caption{\label{tab:results:mj}
As for \tabref{tab:results:nj50} but for the signal regions in the
multi-jet + \MJ{} stream for which the number of events in the control regions
allowed background determination using a fit.
}
\end{table}

\begin{table}[h]
\centering
\renewcommand\arraystretch{1.2}
\footnotesize
\begin{tabular}{| l || c | c || c | c |}
\hline
Signal region   & \multicolumn{2}{| c ||}{\BSR{9j50}} & 
                 \multicolumn{2}{| c |}{\BSR{10j50}} \\
\hline
\MJ{}~[GeV]       & 340 & 420 & 340 & 420 \\
\hline\hline
        Observed events  		 &  $13$
        &  $9$                           &  $1$                    &
        $1$                     \\ \hline \hline
        Total events     		 &  $17 \pm 7$                &  $11 \pm 5$              &  $3.2_{-3.2}^{+3.7}$ &  $2.2 \pm 2.0$         \\ \hline 
        $t\bar{t}$       		 &  $5 \pm 4$                 &  $3.4_{-3.4}^{+3.6}$        &  $0.8_{-0.8}^{+0.8}$ &  $0.6_{-0.6}^{+0.9}$  \\  
        $W$+jets        		 &  $-$                 &  $-$               &  $-$        &  $-$         \\  
        Others  			 & $ 0.58 _{- 0.33 }^{+ 0.54 }$     & $ 0.39 _{- 0.30 }^{+ 0.32 }$    & $ 0.12 \pm 0.12 $       & $ 0.06 \pm 0.06 $        \\  
\hline\hline
        Multi-jets  			 &  $12 \pm 4$
        &  $7.0 \pm 2.3$               &  $2.3_{-2.3}^{+3.6}$ &
        $1.6_{-1.6}^{+1.8}$  \\ \hline \hline
{ $N\rm{^{95\%}_{BSM}}$ (exp)}    &
13 &
11 &
5  &
5  \\ \hline 
{ $N\rm{^{95\%}_{BSM}}$ (obs)}  &
11 &
10  &
4  &
4 \\ \hline \hline
{ $\sigmamaxbsm $ (exp) [fb]}  &
0.7  & 
0.5  & 
0.23  & 
0.23  \\ \hline
{ $\sigmamaxbsm $ (obs) [fb]}  & 
0.5  & 
0.5  & 
0.2  & 
0.2  \\ \hline \hline
 $p_0$    & 0.7 &
0.6 & 0.8 & 0.7 \\ \hline
 Significance ($\sigma$)  &
$-0.6$  & $-0.34$ & $-0.8$ & $-0.6$ \\ \hline
\end{tabular}
\caption{\label{tab:results:mj}
As for \tabref{tab:results:nj50} but for the signal regions in the
multi-jet + \MJ{} stream for which the number of events in the control regions
did not allow background determination using a fit and therefore the leptonic background is extracted directly from
Monte Carlo simulations.
}
\end{table}

\section{Interpretation}\label{sec:interpretation}

\input{interpretations}

\section{Conclusion}\label{sec:conclusion}
\input{Conclusions.tex}

%% file: interpretations.tex
In the absence of significant discrepancies, exclusion limits at 95\%
CL are set
in the context of several simplified
supersymmetric models and an \mSUGRA{} model, all described in \Secref{intro}.
Theoretical uncertainties on the SUSY signals are estimated as described in section~\ref{sec:mc}.
Combined experimental systematic uncertainties on the signal yield
from the jet energy scale, resolution, and $b$-tagging efficiency in the
case of the flavour stream, range from 15\% to 25\%. Acceptance and
efficiency values, uncertainties and other information per signal
region are tabulated in HepData~\cite{multijet_2013_auxiliary}.

The limit for each signal region is obtained by comparing the observed event count with
that expected from Standard Model background plus SUSY signal
processes.
All uncertainties on the Standard Model expectation are used, including those which
are correlated between signal and background (for instance jet energy
scale uncertainties) and all but theoretical cross-section
uncertainties (PDF and scale) on the signal expectation.
The resulting exclusion regions are
obtained using the $CL_s$ prescription~\cite{clsread}. For the
multi-jet + flavour stream a simultaneous fit is performed in all the signal
regions for each of the two values of \pthreshold, and the two fit results are combined using the better expected
limit per point in the parameter space, as described in \Secref{sec:fit:bjet}. For the multi-jet + \MJ{}
stream the signal region with the best expected limit
at each point in parameter space is used. The stream
with the better expected limit
at each point in parameter space is chosen when
combining the two streams.  
The multi-jet + flavour stream typically has stronger expected
exclusion limits than the multi-jet + \MJ{} stream. However, in models
with large numbers of objects in the final state, and more so in
boosted topologies, the multi-jet + \MJ{} stream becomes competitive.
Limits on sparticle masses quoted in the text are
 those from the lower edge of the $1\sigma$ signal cross-section band
 rather than the central value of the observed limit.

As shown in the rest of this section, the analysis substantially extends previous published exclusion
limits on various models, from ATLAS~\cite{Aad:2012hm,Aad:2012pq} and
CMS~\cite{Chatrchyan:1527115,Chatrchyan:2013sza}.

\subsection*{`Gluino--stop (off-shell)' model}

\begin{figure}
\centering
{\label{fig:exclusion_combined}\includegraphics[width=0.6\linewidth]{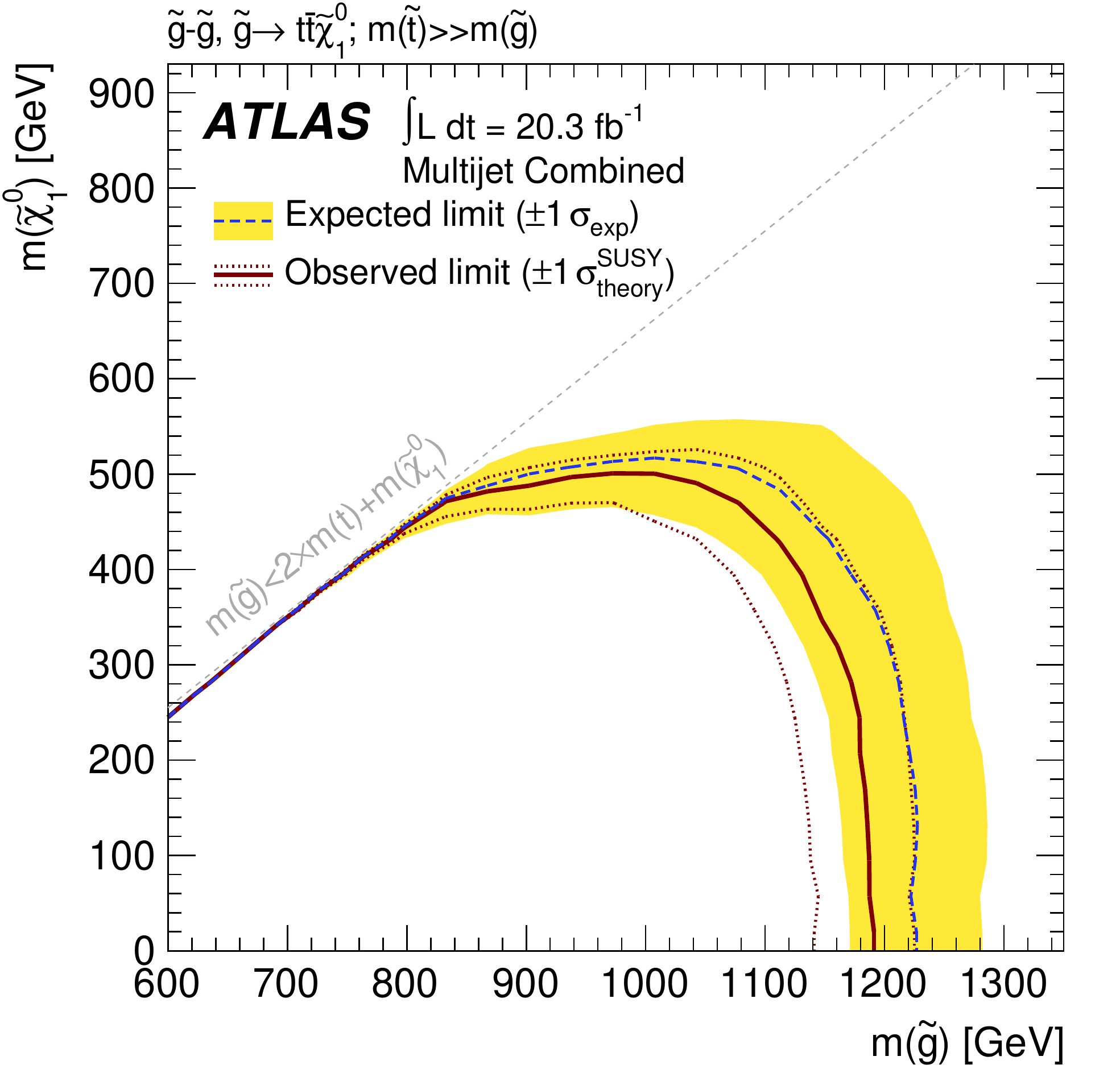}}

\caption{\label{fig:exclusion}
95\% CL exclusion curve for the simplified gluino--stop (off-shell)
  model. The dashed grey and solid red curves show the
  95\% CL expected and observed limits, respectively,
  including all uncertainties except the theoretical signal
  cross-section uncertainty (PDF and scale). The shaded yellow band around
  the expected limit shows the $\pm1\sigma$ result. The $\pm1\sigma$ lines around the observed limit represent the
  result produced when moving the signal cross section by $\pm1\sigma$
  (as defined by the PDF and scale uncertainties). The diagonal dashed line is the kinematic limit for this decay channel.
}
\end{figure}

The analysis result is interpreted in a simplified model
that contains only a gluino octet and a neutralino $\ninoone$ within kinematic reach,
and decaying with unit probability according to
Eq.~\ref{eq:gtt}, via an off-shell \stop-squark. 
 The results are presented in the ($m_{\tilde{g}},m_{\ninoone}$) plane
 in figure~\ref{fig:exclusion}, which shows the combined exclusion. Within the
 context of this simplified model, the 95\% CL exclusion
 bound
on the gluino mass is
 \mgluinoexcluded{} for the lightest neutralino mass up
 to \mlspexcluded{}.

\subsection*{`Gluino--stop (on-shell)' model}

\begin{figure}
\centering
\label{fig:exclusion_combined_onshell11}\includegraphics[width=0.6\linewidth]{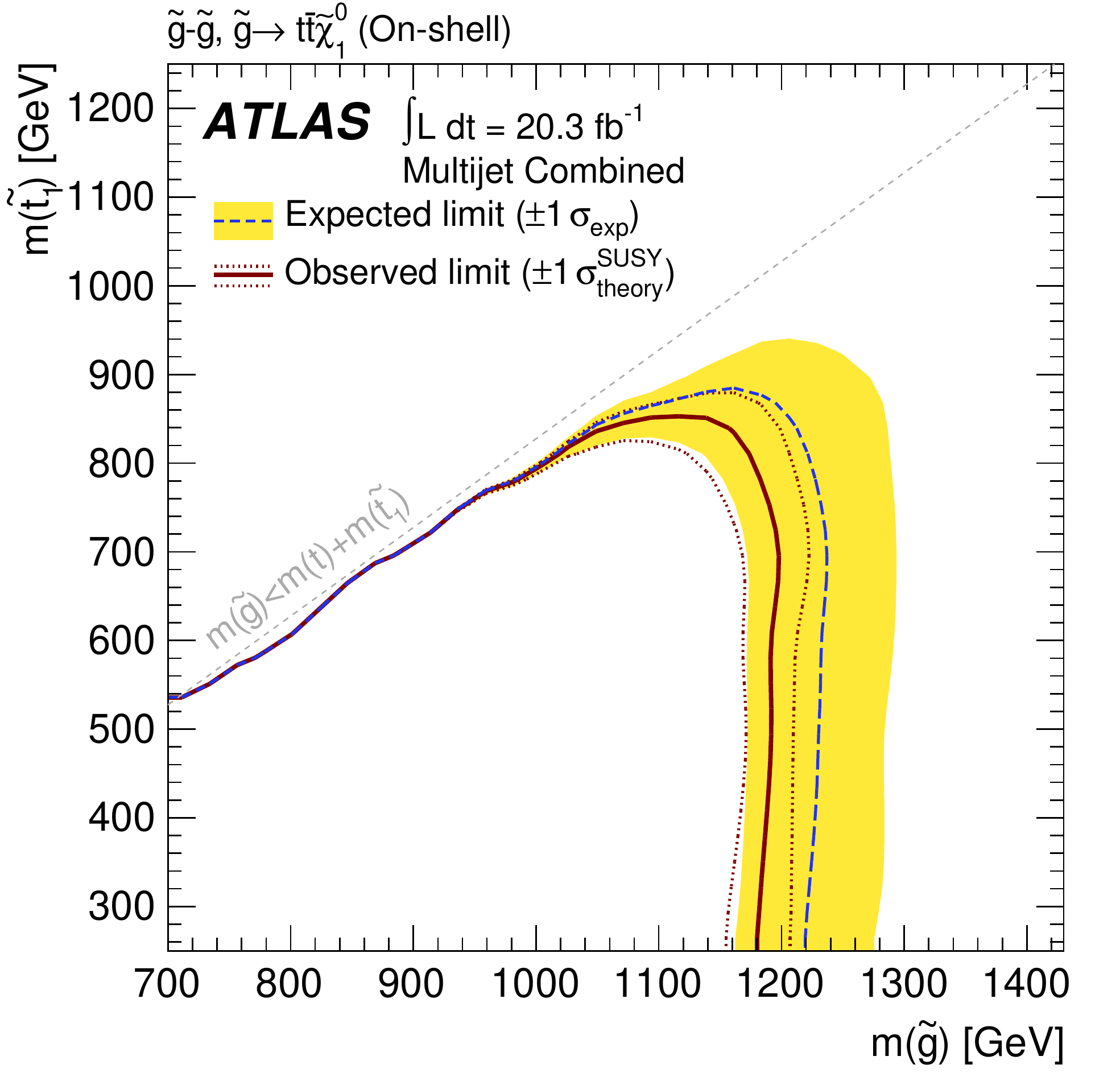}
\caption{\label{fig:exclusion_combined_onshell}
95\% CL exclusion curve for the simplified gluino--stop (on-shell)
 model, where the gluino decays as $\gluino
 ~\to~ \stop + \tbar$ and the stop as $\stop \to \ninoone + t$, with $m_{\ninoone{}}=60$~GeV. Other
 details are as in figure~\ref{fig:exclusion}. 
}
\end{figure}

In this simplified model, each gluino of a pair decays as $\gluino
 ~\to~ \stop + \tbar$; $\stop \to \ninoone + t$. The mass of \ninoone{}
is fixed to 60~GeV. The
results are presented in the ($m_{\tilde{g}},m_{\stop}$) plane in 
figure~\ref{fig:exclusion_combined_onshell} which 
shows the combined exclusion limits. Within the context of this simplified model, the 95\% CL exclusion bound
on the gluino mass is
1.15~TeV for stop masses up to 750~GeV.

\subsection*{`Gluino--squark (via
 $\boldsymbol{ \tilde{\chi}^{\pm}_{1} }$)' model}

\begin{figure}
\centering
\subfloat[{$\rm{x}=1/2$}]{\label{fig:exclusion_combined_1step11}\includegraphics[width=0.48\linewidth]{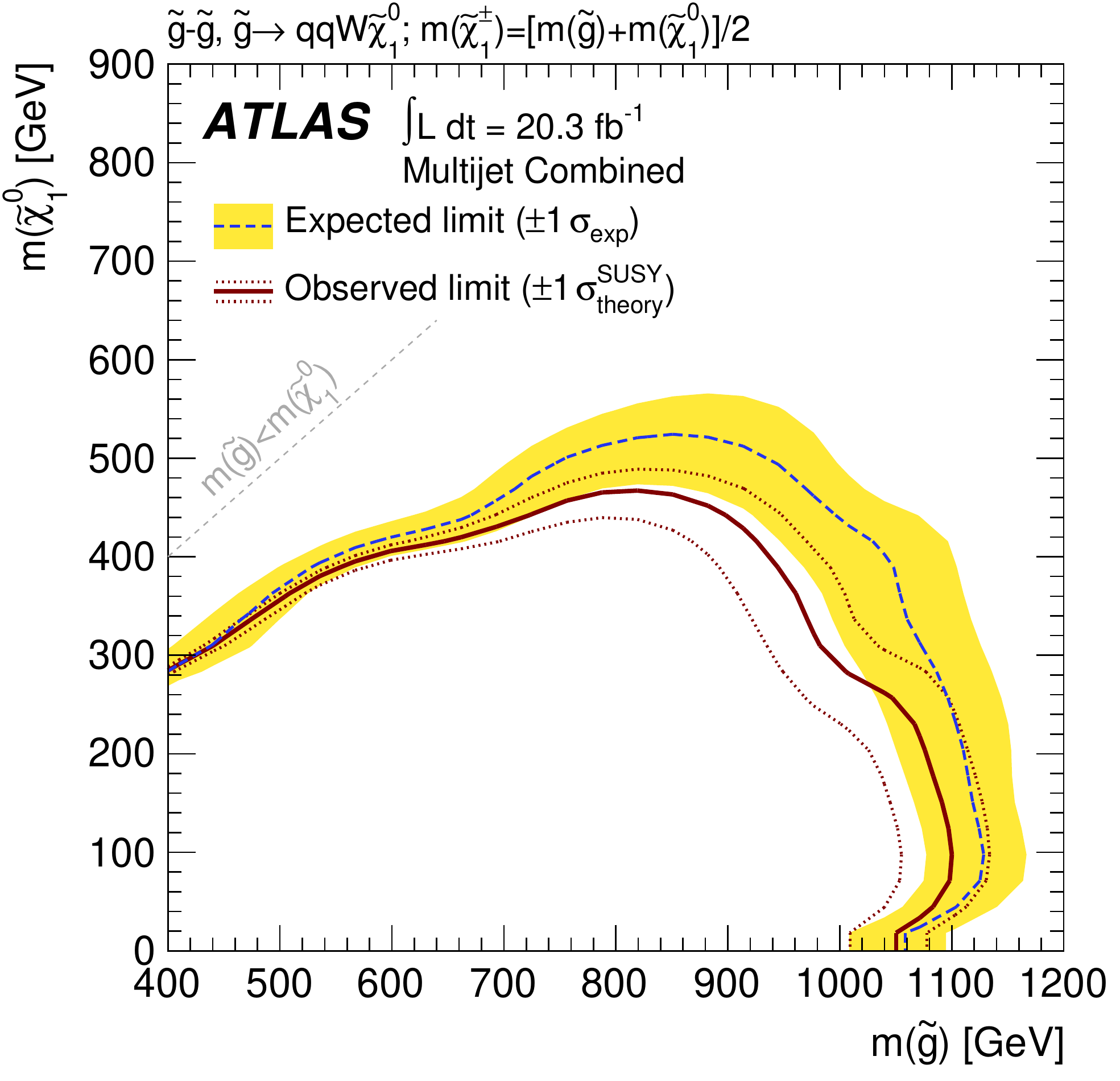}}
\subfloat[$m_{\ninoone}=60$~GeV]{\label{fig:exclusion_combined_1stepx11}\includegraphics[width=0.48\linewidth]{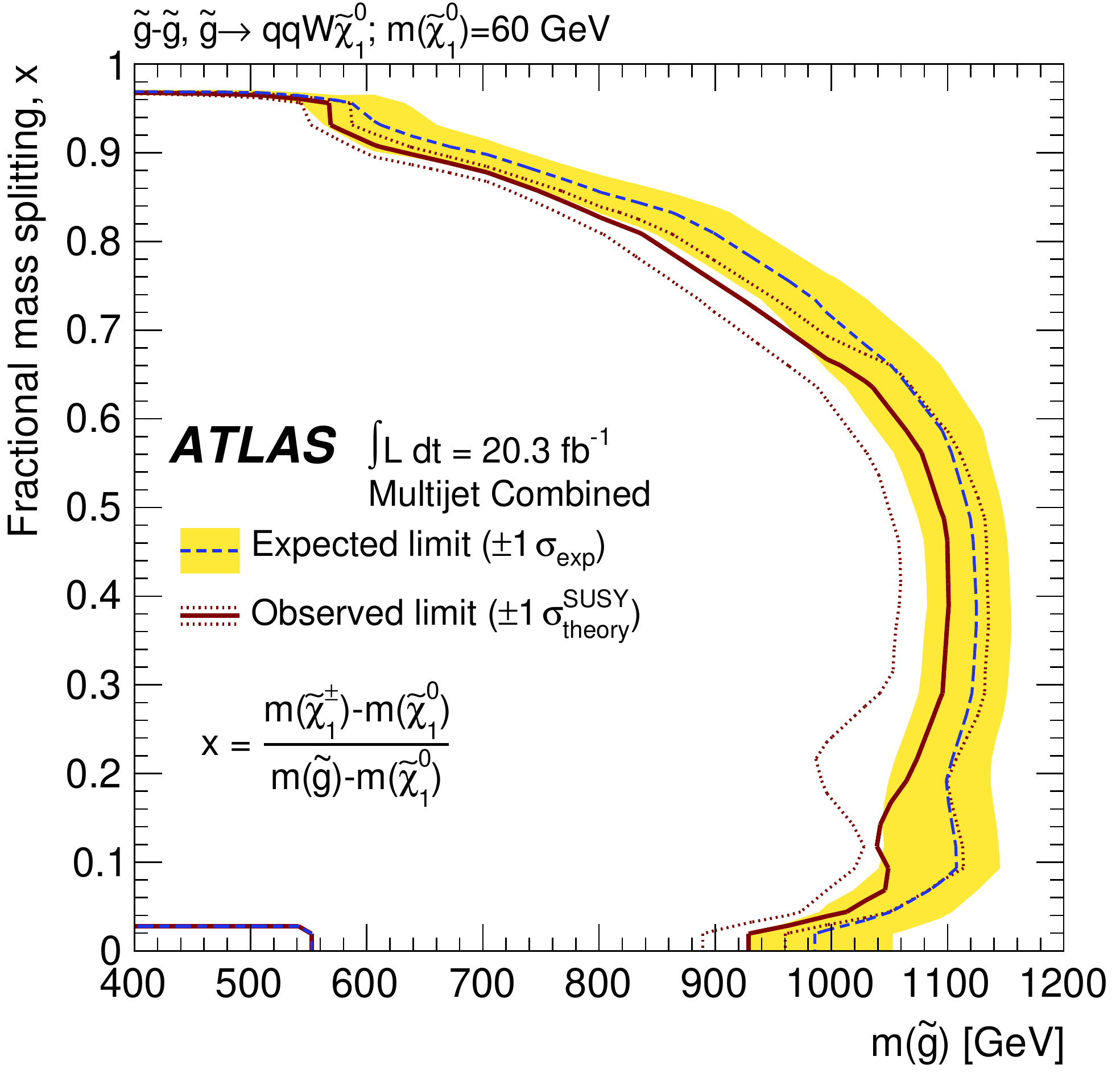}}
\caption{\label{fig:exclusion_combined_1step}
  95\% CL exclusion curve for the simplified gluino--squark (via $\tilde{\chi}^{\pm}_{1}$)
  model, for the two versions of the model; fixed $\rm{x}=1/2$, where
  $\rm{x} = (m_{\tilde{\chi}^{\pm}_{1}} -
  m_{\ninoone}) / (m_{\tilde{g}} - m_{\ninoone})$, and
  varying \ninoone{} mass on the left, and \ninoone{} mass fixed to
  60~GeV and varying $\rm{x}$ on the right. 
  The region with gluino masses between 400~GeV and 550~GeV at small ${\rm x}$ has no signal Monte Carlo simulation.
  Other details are as in figure~\ref{fig:exclusion}.
}
\end{figure}

In this simplified model, each gluino of a pair decays promptly via an
off-shell squark as $\tilde{g}
\to \bar{q} + q' +
\tilde{\chi}^{\pm}_{1} \to \bar{q} + q' + W^{\pm} + \ninoone$. Two versions of this
model are evaluated, and the combined exclusion results are shown in 
figure~\ref{fig:exclusion_combined_1step}. 
In figure~\ref{fig:exclusion_combined_1step}a, the fractional mass splitting,
$\rm{x}$, defined as $\rm{x} = (m_{\tilde{\chi}^{\pm}_{1}} -
  m_{\ninoone}) / (m_{\tilde{g}} - m_{\ninoone})$, is set to 1/2, while the
\ninoone{} mass varies, and the results are shown in the 
($m_{\tilde{g}},m_{\ninoone}$) plane. In the
second case, the \ninoone{} mass is fixed to 60~GeV while $\rm{x}$ varies,
and the results are presented in the ($m_{\tilde{g}},{\rm x}$) plane.
Gluino masses are excluded below 1~TeV at 95\% CL, for
$\ninoone$ masses below 200~GeV, in the case of $\rm{x}=1/2$.

\subsection*{`Gluino--squark (via
  $\boldsymbol{ \tilde{\chi}^{\pm}_{1}}$ and $\boldsymbol{ \ninotwo}$)' model}

\begin{figure}
\centering
\label{fig:exclusion_combined_2step11}\includegraphics[width=0.6\linewidth]{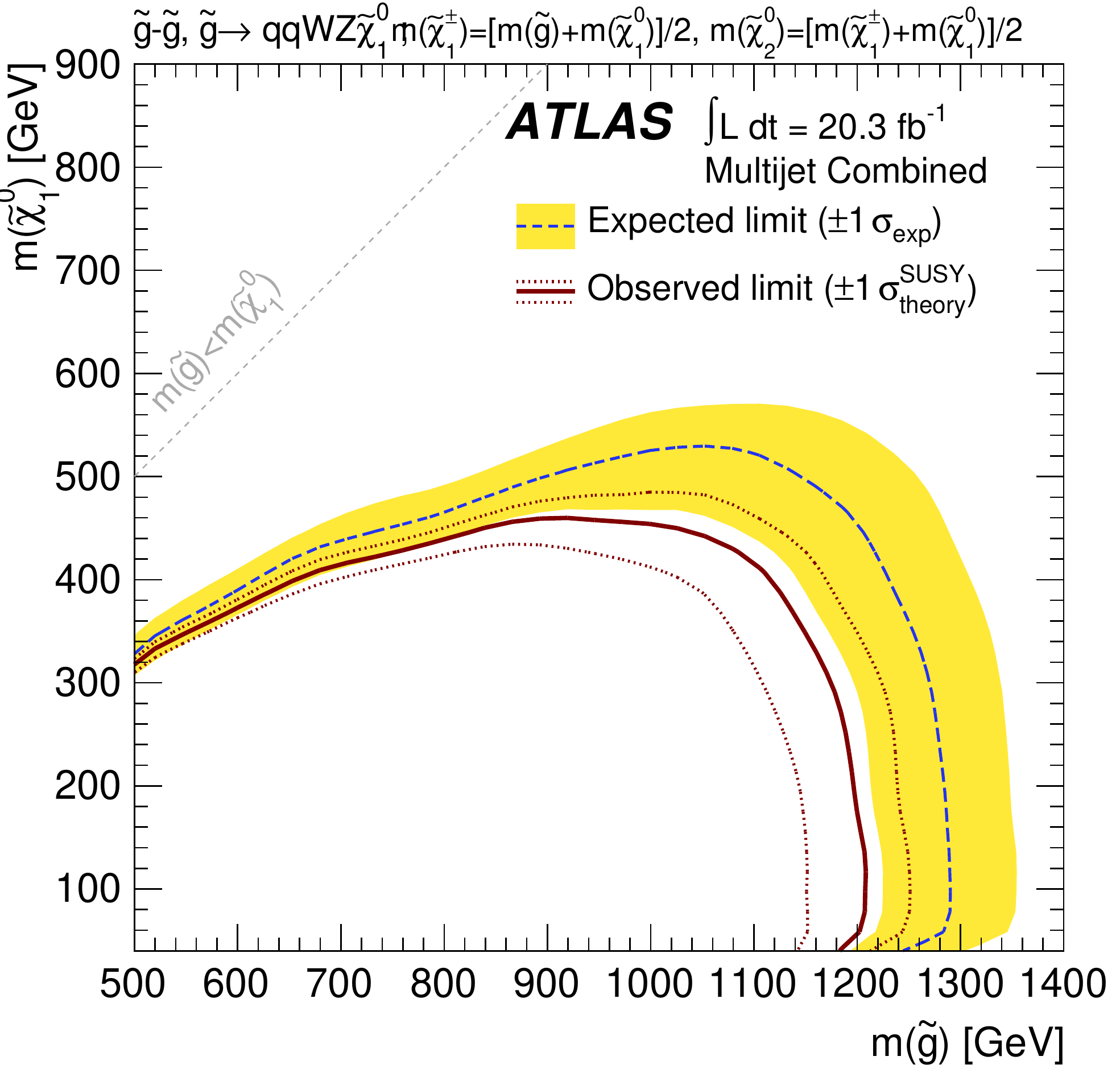}
\caption{\label{fig:exclusion_combined_2step}
95\% CL exclusion curve for the simplified gluino--squark (via
  $\tilde{\chi}^{\pm}_{1}$ and \ninotwo)
  model. Other
  details are as in figure~\ref{fig:exclusion}.
}
\end{figure}

In this simplified model, each gluino of a pair decays promptly via an
off-shell squark as $\tilde{g}
\to \bar{q} + q' +
\tilde{\chi}^{\pm}_{1} \to \bar{q} + q' + W^{\pm} + \ninotwo \to \bar{q} + q' + W^{\pm} + Z^{0}
+ \ninoone$. The intermediate particle masses,
$m_{\tilde{\chi}^{\pm}_{1}}$ and $m_{\ninotwo}$, are set to $(m_{\tilde{g}} +
m_{\ninoone})/2$ and $(m_{\tilde{\chi}^{\pm}_{1}} +
m_{\ninoone})/2$, respectively. 
The results are presented in the ($m_{\tilde{g}},m_{\ninoone}$) plane in figure~\ref{fig:exclusion_combined_2step}, 
which shows the combined exclusion limits for this model.
Gluino masses are excluded below 1.1~TeV at 95\% CL, for
$\ninoone$ masses below 300~GeV.

\subsection*{\mSUGRA}

\begin{figure}
\centering
\label{fig:exclusion_combined_msu}\includegraphics[width=0.6\linewidth]{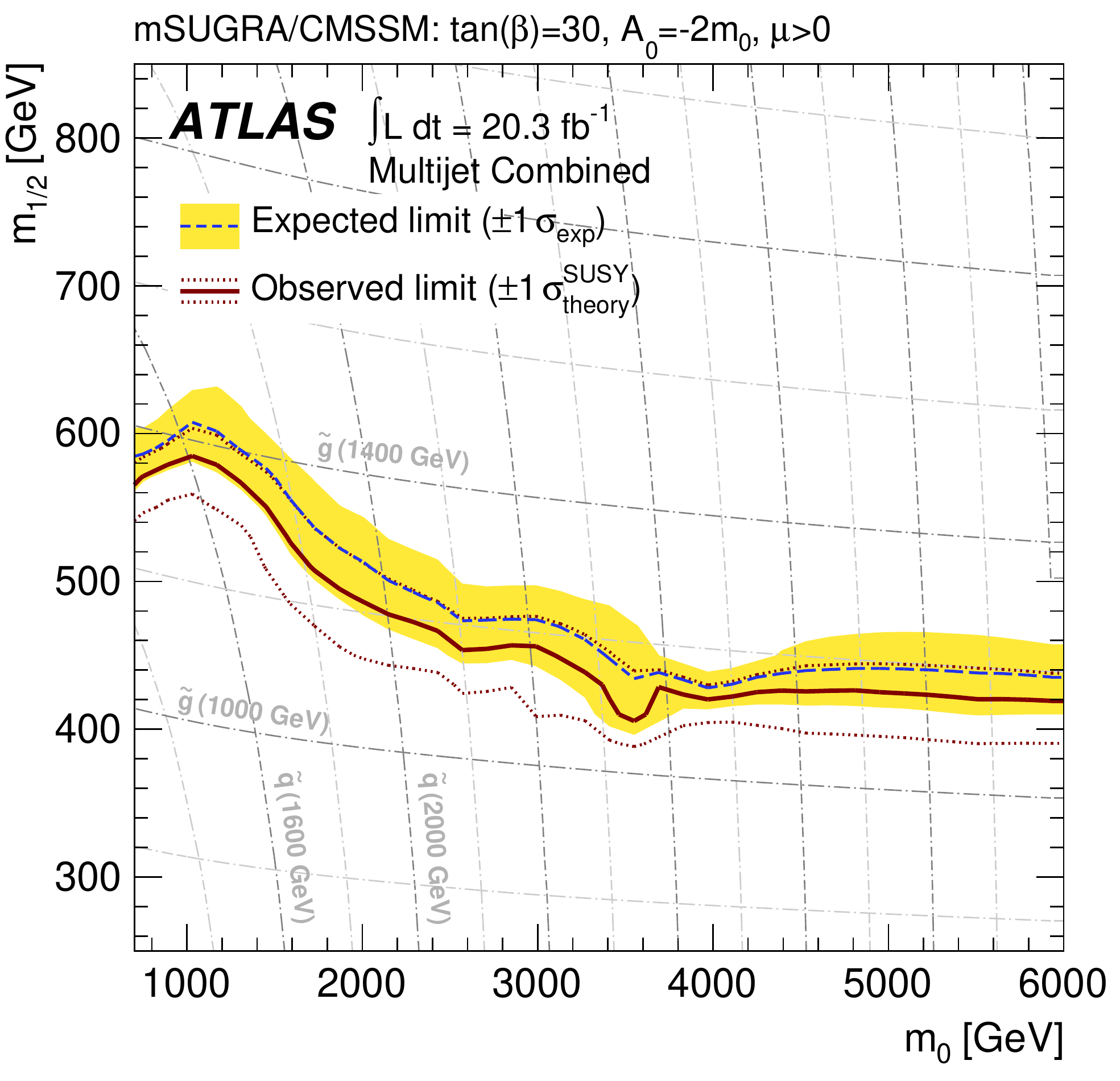}
\caption{\label{fig:exclusion_msugra}
95\% CL exclusion curve for the \msugra{} model, generated with
parameters $\tan{\beta}=30$, $A_0=-2m_0$ and
$\mu>0$. Other
  details are as in figure~\ref{fig:exclusion}. 
}
\end{figure}

An \mSUGRA{} model with parameters $\tan{\beta}=30$, $A_0=-2m_0$ and
$\mu>0$ is also used
to interpret the analysis results. The
exclusion limits are presented in the ($m_{0},m_{1/2}$)
plane in figure~\ref{fig:exclusion_msugra}. 
For large universal scalar mass $m_0$, gluino
masses smaller than 1.1~TeV are excluded at 95\% CL.

\subsection*{`Gluino--stop (RPV)' model}

\begin{figure}
\centering
\label{fig:exclusion_combined_RPVUDD}\includegraphics[width=0.6\linewidth]{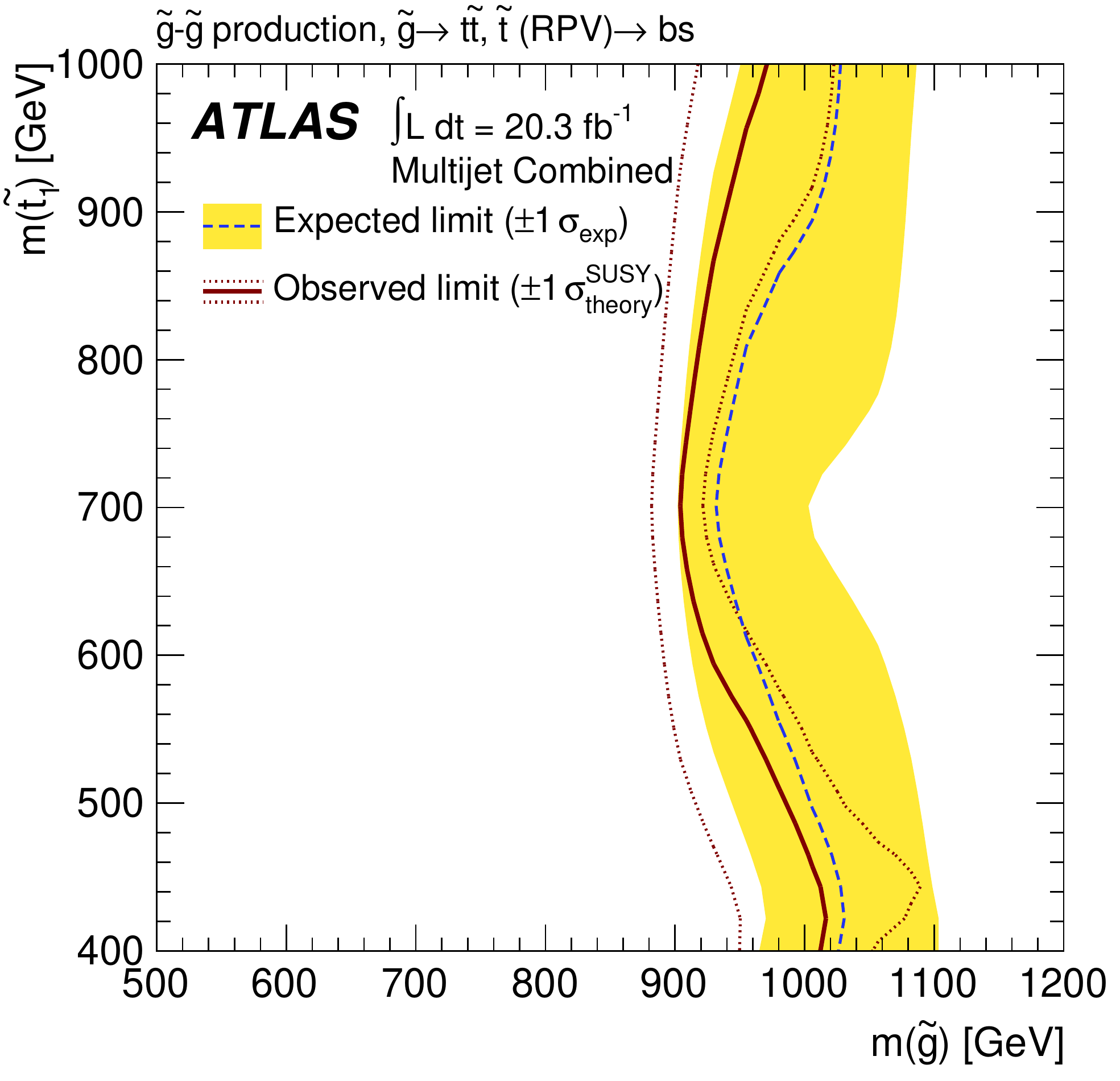}
\caption{\label{fig:exclusion_RPVUDD}
95\% CL exclusion curve for the simplified gluino--stop (RPV)
  model. Other
  details are as in figure~\ref{fig:exclusion}.}

\end{figure}

In this simplified model, each gluino of a pair decays as $\gluino
  ~\to~ \stop + \tbar$; and the \stop-squark decays via the
R-parity- and baryon-number-violating decay $\stop \to \bar{s} + \bar{b}$. The
results are presented in the ($m_{\tilde{g}},m_{\stop}$) plane
 in figure~\ref{fig:exclusion_RPVUDD}. 
 Within the context of this simplified model, the 95\% CL exclusion bound
on the gluino mass is
 900~GeV for \stop-squark masses ranging from 400~GeV to 1~TeV.

%% file: Conclusions.tex
A search is presented for new phenomena with large jet multiplicities (from 7 to $\ge$10) and missing transverse momentum using
\ourintlumi{} of 8~TeV $pp$ collision data collected by the ATLAS experiment 
at the Large Hadron Collider. The sensitivity to new physics is enhanced by considering the number of $b$-tagged jets 
and the scalar sum of masses of radius $R=1.0$ jets in the event,
reconstructed using the anti-$k_t$ clustering algorithm. The
Standard Model predictions are found to be consistent with the data. 
The results are interpreted in the
context of an \mSUGRA{} model and various simplified models resulting
in final states with large jet multiplicity and \met. The exclusion
limits substantially extend previous results. 
For example, in a model where both of the pair-produced gluinos
decay via 
$\gluino \rightarrow t + \bar{t} + \ninoone$, gluino
masses smaller than \mgluinoexcluded{} are excluded for neutralino masses below \mlspexcluded{}.

%% file: acknowledgements.tex

We thank CERN for the very successful operation of the LHC, as well as the
support staff from our institutions without whom ATLAS could not be
operated efficiently.

We acknowledge the support of ANPCyT, Argentina; YerPhI, Armenia; ARC,
Australia; BMWF and FWF, Austria; ANAS, Azerbaijan; SSTC, Belarus; CNPq and FAPESP,
Brazil; NSERC, NRC and CFI, Canada; CERN; CONICYT, Chile; CAS, MOST and NSFC,
China; COLCIENCIAS, Colombia; MSMT CR, MPO CR and VSC CR, Czech Republic;
DNRF, DNSRC and Lundbeck Foundation, Denmark; EPLANET, ERC and NSRF, European Union;
IN2P3-CNRS, CEA-DSM/IRFU, France; GNSF, Georgia; BMBF, DFG, HGF, MPG and AvH
Foundation, Germany; GSRT and NSRF, Greece; ISF, MINERVA, GIF, DIP and Benoziyo Center,
Israel; INFN, Italy; MEXT and JSPS, Japan; CNRST, Morocco; FOM and NWO,
Netherlands; BRF and RCN, Norway; MNiSW, Poland; GRICES and FCT, Portugal; MERYS
(MECTS), Romania; MES of Russia and ROSATOM, Russian Federation; JINR; MSTD,
Serbia; MSSR, Slovakia; ARRS and MIZ\v{S}, Slovenia; DST/NRF, South Africa;
MICINN, Spain; SRC and Wallenberg Foundation, Sweden; SER, SNSF and Cantons of
Bern and Geneva, Switzerland; NSC, Taiwan; TAEK, Turkey; STFC, the Royal
Society and Leverhulme Trust, United Kingdom; DOE and NSF, United States of
America.

The crucial computing support from all WLCG partners is acknowledged
gratefully, in particular from CERN and the ATLAS Tier-1 facilities at
TRIUMF (Canada), NDGF (Denmark, Norway, Sweden), CC-IN2P3 (France),
KIT/GridKA (Germany), INFN-CNAF (Italy), NL-T1 (Netherlands), PIC (Spain),
ASGC (Taiwan), RAL (UK) and BNL (USA) and in the Tier-2 facilities
worldwide.

%% file: atlas_authlist.tex
\begin{flushleft}
{\Large The ATLAS Collaboration}

\bigskip

G.~Aad$^{\rm 48}$,
T.~Abajyan$^{\rm 21}$,
B.~Abbott$^{\rm 112}$,
J.~Abdallah$^{\rm 12}$,
S.~Abdel~Khalek$^{\rm 116}$,
O.~Abdinov$^{\rm 11}$,
R.~Aben$^{\rm 106}$,
B.~Abi$^{\rm 113}$,
M.~Abolins$^{\rm 89}$,
O.S.~AbouZeid$^{\rm 159}$
% ...
\bigskip
\\
% ...
$^{11}$ Institute of Physics, Azerbaijan Academy of Sciences, Baku, Azerbaijan\\
$^{12}$ Institut de F{\'\i}sica d'Altes Energies and Departament de F{\'\i}sica de la Universitat Aut{\`o}noma de Barcelona, Barcelona, Spain\\
% ...
$^{21}$ Physikalisches Institut, University of Bonn, Bonn, Germany\\
% ...
$^{48}$ Fakult{\"a}t f{\"u}r Mathematik und Physik, Albert-Ludwigs-Universit{\"a}t, Freiburg, Germany\\
% ...
$^{89}$ Department of Physics and Astronomy, Michigan State University, East Lansing MI, United States of America\\
% ...
$^{106}$ Nikhef National Institute for Subatomic Physics and University of Amsterdam, Amsterdam, Netherlands\\
% ...
$^{112}$ Homer L. Dodge Department of Physics and Astronomy, University of Oklahoma, Norman OK, United States of America\\
$^{113}$ Department of Physics, Oklahoma State University, Stillwater OK, United States of America\\
% ...
$^{116}$ LAL, Universit{\'e} Paris-Sud and CNRS/IN2P3, Orsay, France\\
% ...
$^{159}$ Department of Physics, University of Toronto, Toronto ON, Canada\\
% ...
\end{flushleft}
